\documentclass[iop]{emulateapj}
\usepackage{amsmath}
\usepackage{amssymb}
\usepackage{verbatim}
\usepackage{graphicx}
\usepackage{morefloats}
\usepackage{float}
\usepackage{lipsum}
\usepackage{diagbox}
\usepackage{subfigure}
\usepackage{longtable}
\usepackage{lipsum} 
\usepackage{rotating}
\usepackage{wasysym}
\usepackage{natbib}
\usepackage{ulem}

\begin{document}

\title{Analyzing the Largest Spectroscopic Dataset of Stripped Supernovae to Improve Their Identifications and Constrain Their Progenitors}

\author{Yu-Qian Liu\altaffilmark{1}, Maryam Modjaz\altaffilmark{1}, Federica B. Bianco\altaffilmark{1,2}, Or Graur\altaffilmark{1,3}}
\altaffiltext{1}{Center for Cosmology and Particle Physics, New York University, 4 Washington Place, New York, NY 10003-6603, USA; YL1260@nyu.edu}
\altaffiltext{2}{Center for Urban Science and Progress, New York University,
1 MetroTech Center, Brooklyn, NY 11201, USA}
\altaffiltext{3}{Department of Astrophysics, American Museum of Natural History, Central Park West and 79th Street, New York, NY 10024-5192, USA}

\slugcomment{Accepted by ApJ}

\begin{abstract}
Using the largest spectroscopic dataset of stripped-envelope core-collapse supernovae (stripped SNe), we present a systematic investigation of spectral properties of Type IIb SNe (SNe IIb), Type Ib SNe (SNe Ib), and Type Ic SNe (SNe Ic). Prior studies have been based on individual objects or small samples. Here, we analyze 242 spectra of 14 SNe IIb, 262 spectra of 21 SNe Ib, and 207 spectra of 17 SNe Ic based on the stripped SN dataset of \citet{modjaz14} and other published spectra of individual SNe. Each SN in our sample has a secure spectroscopic ID, a date of $V$-band maximum light, and most have multiple spectra at different phases. We analyze these spectra as a function of subtype and phase in order to improve the SN identification scheme and constrain the progenitors of different kinds of stripped SNe. By comparing spectra of SNe IIb with those of SNe Ib, we find that the strength of H$\alpha$ can be used to quantitatively differentiate between these two subtypes at all epochs. Moreover, we find a continuum in observational properties between SNe IIb and Ib. We address the question of hidden He in SNe Ic by comparing our observations with predictions from various models that either include hidden He or in which He has been burnt. Our results favor the He-free progenitor models for SNe Ic. Finally, we construct continuum-divided average spectra as a function of subtype and phase to quantify the spectral diversity of the different types of stripped SNe.
% and to compare them with the often-used \citet{nugent02} templates. We suggest that instead of the \citet{nugent02} templates, the mean spectra and the corresponding standard deviations released in this study should be used in cases where only the SN absorption features are important, since our mean spectra have no information about the SN continuum, however, they include many more SNe than Nugent included, cover a larger wavelength range at some phases, and account for the spectral diversity of the SNe.
\end{abstract}
\keywords{supernovae: general---supernovae: individual (SNe 1993J, 1998dt, 1999ex, 2005bf, 2005E, 2006el, 2007Y, 2009mg, 2011dh, 2011ei)---methods: data analysis}

\section{Introduction}
\label{intro}

Supernovae (SNe) mark the diverse deaths of stars and contribute to the production and release of heavy elements in the universe. In particular, stripped-envelope core-collapse SNe \citep[stripped SNe; e.g.,][]{clocchiatti97, filippenko97_review, modjaz14} are the deaths of massive stars ($\gtrsim8~\text{M}_{\astrosun}$) that have lost some, if not all, of their outer hydrogen and helium envelopes through strong winds \citep{Woosley93}, binary interactions \citep{Nomoto95, Podsiadlowski04}, or enhanced mixing \citep{frey13}. Unlike Type II SNe (SNe II), whose spectra show strong H lines during all photospheric phases, stripped SNe have spectra with no or weak H lines, or strong H lines only at early phases. In contrast to Type Ia SNe (SNe Ia), which are considered to be the outcome of thermonuclear explosions of carbon-oxygen white dwarfs \citep{nugent11,bloom12,maoz14}, stripped SNe are thought to explode due to the core-collapse of their massive progenitors \citep[$\gtrsim8~\text{M}_{\astrosun}$;][]{woosley02,burrows13}. Following the empirical classification based on the presence or absence of certain lines in SN spectra \citep{filippenko97_review}, stripped SNe can be divided into several subtypes: Type IIb SNe (SNe IIb) initially show strong H lines, but over time the H lines become weaker whereas the He I lines grow stronger \citep{filippenko93}; Type Ib SNe (SNe Ib), which show conspicuous He I lines; Type Ic SNe (SNe Ic), which do not show prominent H lines or He I lines; and broad-lined SNe Ic (SNe Ic-bl), which are similar to SNe Ic, but exhibit much broader lines \citep[by $\sim9000$ km s$^{-1}$ around maximum light;][]{modjaz15}.

Dozens of stripped SNe are discovered every year, but only a few of them have a large amount of spectroscopic and photometric data and have been studied in detail. For example, the best-studied SN IIb and SN Ic are SN 1993J  \citep{filippenko93, matheson00_93jdetail, matheson00_93j} and SN 1994I \citep{filippenko95, richmond96}, respectively. However, statistical analyses of large SN samples are needed to characterize spectra of different SN subtypes. This will help to explore whether there is an observational continuum between SN subtypes, as well as to correctly classify SNe, which is vital in matching various progenitor models to different SN subtypes. In addition, it will help to sharpen the use of SNe Ia for high-precision cosmology since SNe Ic, which are potential contaminants in high-redshift SN Ia surveys \citep[e.g.,][]{Clocchiatti00, Homeier05, graur13, jones13, rodney15}, can be better distinguished. Finally, statistical studies will help to assess whether there is hidden He in SN Ic spectra via comparisons between the bulk properties of observed spectra and predicted spectroscopic properties based on various models. The most recent work that statistically compared spectra of different stripped SN subtypes was conducted by \citet{matheson01}, who used 84 spectra of 28 stripped SNe, many of which did not have light curves to determine the phases of the spectra. We discuss this paper in detail in Sections \ref{sec_He_IbIIb_pEW} and \ref{sec_OI7774}.

Recently, \citet{modjaz14} published optical spectra of 73 stripped SNe collected at the Harvard-Smithsonian Center for Astrophysics (CfA; the M14 sample hereafter), doubling the supply of well-observed stripped SNe. Forty-four of these 73 stripped SNe have a date of maximum light. Besides the above data, which comprise more than half of our SN sample, we have collected the spectra of all available stripped SNe from the literature until September of 2014. Thus, we analyze the optical spectra of a sample of 71 stripped SNe with a well-defined date of maximum light and type. 

This study will focus on spectroscopic comparisons between SNe IIb, SNe Ib, and SNe Ic. A spectroscopic comparison between SNe Ic, SNe Ic-bl without Gamma-Ray Bursts (GRBs), and SNe Ic-bl connected with GRBs is presented in a companion paper \citep{modjaz15}. In Section \ref{data}, we summarize the phases and references of the SN spectral sample used in this paper. In Section \ref{method}, we describe the velocity and strength measurements we make based on the SN spectra. In Section \ref{classify}, we conduct spectroscopic comparisons between SNe IIb and SNe Ib to see whether there is an observational continuum between the two subtypes, as well as to better characterize them. In Section \ref{He_pro}, we discuss whether there is hidden He in progenitors of SNe Ic using the strength and velocity of the O I $\lambda$7774 line and the velocity of the Fe II $\lambda$5169 line as indicators to test two competing models. We also explore the spectral diversity within each SN subtype using mean spectra and their corresponding standard deviations in Section \ref{sec_meanspec}. Finally, we summarize our conclusions in Section \ref{conc}.

\section{SN Spectral Samples}
\label{data}

We list our SN IIb and Ib samples in Tables \ref{table_IIb} and \ref{table_Ib-norm}, respectively. Our SN Ic sample is the same as in \citet{modjaz15}. To improve both the accuracy and the precision of our analyses, we consistently use the date of maximum light in the $V$-band since it is the best-sampled light curve for current SN photometric datasets \citep{bianco14}. For SNe with $V$-band light curves in the literature but for which the authors did not explicitly report the date of maximum light \citep[e.g., SN 2004ff;][]{drout11}, we calculated the date of $V$-band maximum using a polynomial fit as described in \citet{bianco14}. For SNe without $V$-band light curves but with a date of maximum light in either the $R$-band or $B$-band \citep[e.g., SN 1998fa;][]{matheson01}, we converted the known date to the date of $V$-band maximum light using the relationship listed in \citet{bianco14}: $\mathrm{JD}_{Vmax}=\mathrm{JD}_{Rmax}-1.8~\mathrm{days}=\mathrm{JD}_{Bmax}+2.3$ days, where JD stands for Julian date. 

Since we want to analyze SN spectra as a function of SN type and phase in a statistical manner, we included as many stripped SNe that have a secure ID and a date of maximum light as possible. The classifications of these SNe are taken from the literature as indicated in Tables 1 and 2. Authors usually classifies a SN using SNID, GELATO, or other SN classification codes. Note that we adopt the changes in classification as discussed in \citet{modjaz14}. We only include normal SNe IIb, SNe Ib and SNe Ic, i.e., we exclude SNe Ib-n \citep[e.g., SN 2006jc;][]{pastorello07, modjaz14} , SNe Ib-pec \citep[e.g., SNe 2002bj, 2007uy, and 2009er; ][]{poznanski10, modjaz14}, SNe Ib-Ca \citep[e.g., SN 2005E;][]{perets10}, super-luminous SNe Ic \citep[e.g., SN 2010gx;][]{quimby11, stoll11, gal-yam12}, and SNe that transition from one type to another during their evolution \citep[e.g., ASASSN-15ed;][]{pastorello15}. Our SN spectroscopic sample consists of relevant spectra (that satisfy the above requirements) in the database templates-2.0 of the SuperNova IDentification code \citep[SNID;\footnote{http://people.lam.fr/blondin.stephane/software/snid/index.html}][]{Blondin2007}, the M14 sample, and relevant spectra from the literature available before September of 2014. These spectra were drawn from SNID, public archives such as the CfA Supernova Archive\footnote{http://www.cfa.harvard.edu/supernova/SNarchive.html} and Weizmann Interactive Supernova data REPository  \citep[WISeREP;\footnote{http://wiserep.weizmann.ac.il/spectra/list}][]{Yaron2012WISeREP}, or requested from the authors.   

In summary, we have 242 optical spectra of 14 SNe IIb, 262 optical spectra of 21 SNe Ib, and 207 optical spectra of 17 SNe Ic. These SNe have low redshifts ($z<0.027$) with mean and median values of 0.011 and 0.009, respectively.

\begin{deluxetable*}{lcl}
\tablecolumns{3}
\singlespace
\tablecaption{Spectral sample of SNe IIb}
\tablehead{
\colhead{SN Name} &
\colhead{Phases\tablenotemark{a} of Spectra}  &
\colhead{References\tablenotemark{b}}
}
\startdata
SN 1993J &  $-$18, $-18$, $-$17, $-$16, $-$11, $-$5, $-$4, $-$3, $-$2, $-2$, $-2$, 1, 3, 3, 4, 4, 5, 6, 7, 11, 12, & SNID (J94, B95, M00, F05), M14\\ [1ex]
 &   13, 17, 20, 24, 29, 30, 32, 35, 35, 36, 36, 37, 38, 39, 62, 64, 68, 70, 72, 88, 90+(32)&  \\ [1ex]
SN 1996cb &  $-$20, $-$19, $-$4, $-$1, 3, 4, 5, 6, 25, 27, 31, 35, 39, 55, 66,
90+(7) & SNID (M01), M14\\ [1ex]
SN 1998fa\tablenotemark{c} &  4, $-$3,
18  & M14 \\ [1ex]
SN 2000H &  $-$2, $-$1, 0, 1, 2, 3, 5, 17, 21, 32, 47, 57 
 & SNID (B02), M14\\ [1ex]
SN 2003bg\tablenotemark{d} &  $-$19, $-$17, $-$15, $-$7, 12, 16, 20, 21,
90+(6) & H09\\ [1ex]
SN 2004ff\tablenotemark{e} &  $-$1
 & M14 \\ [1ex]
SN 2006el &  $-$4, 10, 11, 12, 13, 16,
17  & M14 \\ [1ex]
SN 2006T &  $-$14, $-$12, 6, 7, 14,
90+(2)  &  S12, M14 \\ [1ex]
SN 2008ax &  $-$20, $-$19, $-$18, $-$16, $-$15, $-$13, $-$12, $-$11, $-$10, $-$9, $-$8, 6, 8, 9, 9, 12, 15, 16,  & M10, M14\\ [1ex]
& 
19, 23, 38, 45, 67, 74, 80,
90+(1) 
& \\[1ex]
SN 2008bo &  $-$10, $-$7, $-$3, $-$1, 16, 21, 25, 31, 45, 54, 54,
90+(1)  & M10, M14\\ [1ex]
SN 2009mg &  $-$4, $-$1, 3, 3, 12, 13, 14, 24, 39, 39, 88
  & O12\\ [1ex]
SN 2011dh\tablenotemark{f} &  $-$17, $-$16, $-$15, $-$14, $-$13, $-$12, $-11, -10, -9, -4, -1$, 3, 4, 5, 6, 7, 9, 10, 11, 13,
15 & A11, E14, Ma14\\ [1ex]
&
19, 23, 27, 28, 31, 37, 46, 50, 61, 68
& \\[1ex]
SN 2011ei &  $-$14, $-$10, $-$6, $-$4, $-$3, 3, 8, 13, 17, 38, 48, 66, 
90+(3) & M13\\ [1ex]
SN 2011fu &  $-$14, $-$13, $-$11, 2, 17, 19, 42, 71
 & K13
\enddata
\tablenotetext{a}{Phases are in the rest-frame with respect to maximum light and rounded to the nearest whole day. The number in a bracket is the number of spectra with phases larger than 90 days after the date of maximum light, which we include for completeness, but do not analyze here. All dates are indicated with respect to maximum light in the (rest frame) $V$-band, either directly measured, or transformed (see below). The references for the date of $V$-band maximum light are the same as references in the third column or otherwise noted.}
\tablenotetext{b}{References: SNID = in SNID release version 5.0 via templates-2.0 by \citet{Blondin2007}, with the original references in parentheses; J94 = \citet{jeffery94}; B95 = \citet{barbon95};  M00 = \citet{matheson00_93j}; F05 = \citet{fransson05}; M01 = \citet{matheson01}; M14 = \citet{modjaz14}; B02 = \citet{branch02}; H09 = \citet{hamuy09}; S12 = \citet{Silverman12}; M10 = \citet{milisavljevic10}; O12 = \citet{oates12}; A11 = \citet{arcavi11}; E14 = \citet{ergon14}; Ma14 = \citet{marion14}; M13 = \citet{milisavljevic13}; K13 = \citet{kumar13}.}
\tablenotetext{c}{For this SN, we converted the date of maximum light in the $R$-band to the date in the $V$-band using the relationship found by \citet{bianco14}.}
\tablenotetext{d}{\citet{hamuy09} regarded this SN as a broad-line SN IIb because its early spectra are broad and its inferred kinetic energy of explosion is large \citep{mazzali09}.}
\tablenotetext{e}{For this SN, the date of $V$-band maximum light is $2453313.69\pm0.90$, which was determined from the $V$-band photometry in \citet{drout11} through the polynomial fitting in \citet{bianco14}.}
\tablenotetext{f}{The reference for the date of $V$-band maximum light for this SN is \citet{Tsvetkov2012}.}
\label{table_IIb}
\end{deluxetable*}

\begin{deluxetable*}{lcl}
\tablecolumns{3}
\singlespace
\tablecaption{Spectral sample of SNe Ib}
\tablehead{
\colhead{SN Name} &
\colhead{Phases\tablenotemark{a} of Spectra}  &
\colhead{References\tablenotemark{b}}
}
\startdata
SN 1983N &  4, 12,
90+(1) & SNID (Wheeler)\\ [1ex]
SN 1984L &  8, 9, 12, 13, 28, 32, 37, 38, 39, 57, 70, 71 
 & SNID (Wheeler)\\ [1ex]
SN 1990I &  11, 19, 40, 41, 51, 65, 88,
90+(1) & SNID (E04) \\ [1ex]
SN 1990U &  11, 12, 34, 41, 70,
90+(1) & SNID (M01)\\ [1ex]
SN 1998dt\tablenotemark{c} &  2, 3, 6, 9, 10, 13, 14, 19, 34
& SNID (M01), M14\\ [1ex]
SN 1999dn &  $-$7, $-$1, 0, 3, 6, 12, 13, 17, 20, 41, 67
& SNID (M01), B11\\ [1ex]
SN 1999ex\tablenotemark{d} &  $-$5, 0,
9 & SNID (H02)\\ [1ex]
SN 2004dk\tablenotemark{e} &  14, 17, 46
 & M14\\ [1ex]
SN 2004gq\tablenotemark{c} &  $-$9, $-$8, $-$7, $-$6, $-$5, $-$2, $-$1, 0, 16, 17, 18, 21, 21, 25, 42, 55, 70, 73, 78, 84, 89,
90+(3) & M14 \\ [1ex]
SN 2004gv &  13, 15, 19, 48
  & M14 \\ [1ex]
SN 2005bf\tablenotemark{f} &  $-$20$-$(7), $-$9, $-$7, $-$6, $-$5, $-$4, $-$3, $-$2, $-$1, 0, 1, 2, 2, 4, 5, 7, 22, 22, 24, 25, 28, 29, 32
 & SNID (T05, F06), M14\\ [1ex]
SN 2005hg &  $-$14, $-$13, $-$11, $-$10, $-$9, $-$8, $-$7, $-$6, $-$5, $-$4, $-$3, $-$2, $-$1, 12, 16, 22, 26, 87
 & M14 \\ [1ex]
SN 2006ep &  $-$8, 8, 10, 12, 44
 & M14 \\ [1ex]
SN 2007ag &  $-$1, 3, 
9  & S12, M14 \\ [1ex]
SN 2007C &  $-$6, $-$5, $-$1, $-$1, 1, 7, 12, 14, 27, 37, 42, 55, 61, 63, 64,
71  & S12, M14 \\ [1ex]
SN 2007kj &  $-$1,
4  & M14 \\ [1ex]
SN 2007Y &  $-$15, $-$9, $-$2, 5, 8, 13, 20, 38,
90+(4) & SNID (S09)\\ [1ex]
SN 2008D\tablenotemark{g} &  $-$19, $-$18, $-$16, $-$16, $-$15, $-$14, $-$13, $-$11, $-$9, $-$8, $-$4, $-$4, 2, 3, 4, 5, 9, 11, 13, 14, 18, & M09, Mo09, M10, M14\\ [1ex]
&   19, 27, 30, 33, 41, 49, 63
& \\ [1ex]
SN 2009iz &  $-$14, $-$10, 5, 11, 12, 13, 21, 42, 47
 & M14 \\ [1ex]
SN 2009jf &  $-$18, $-$17, $-$16, $-$15, $-$13, $-$11, $-$10, $-10$, $-$9, $-$7, $-$5 to $-$1, 0, 1, 1, 3, 3, 7, 9, 11, 18, & S11, V11, M14\\ [1ex]
 & 24, 25, 27, 30, 31, 33, 34, 47, 49, 55, 59, 61, 67, 70, 80, 81, 82, 
90+(7)  & \\ [1ex]
iPTF13bvn\tablenotemark{h} &  $-16, -15, -14, -13, -12, -11, -9, -7, -6, -6, -2,
-1$, 1, 8, 20, 34 & C13, F14
\enddata
\tablenotetext{a}{Phases are in rest-frame and rounded to the nearest whole day. The number in a bracket is the number of spectra with phases larger than 90 days after or smaller than 20 days before the date of maximum light. All dates of maximum light were measured in the $V$-band. The references for the date of $V$-band max are the same as references in the third column unless otherwise noted.}
\tablenotetext{b}{References: SNID = in SNID release version 5.0 via templates-2.0 by \citet{Blondin2007}, with the original references in parentheses; Wheeler = University of Texas spectral library, which is no longer in use; E04 = \citet{elmhamdi04}; M01 = \citet{matheson01}; M14 = \citet{modjaz14}; B11 = \citet{benetti11}; H02 = \citet{hamuy02}; T05 = \citet{tominaga05}; F06 = \citet{folatelli06}; S12 = \citet{Silverman12}; S09 = \citet{stritzinger09}; M09=\citet{malesani09}; Mo09 = \citet{modjaz09}; M10=\citet{Moskvitin2010}; S11 = \citet{sahu11}; V11 = \citet{valenti11}; C13 = \citet{cao13}; F14 = \citet{Fremling2014}.}
\tablenotetext{c}{For this SN, we converted the date of maximum light in the $R$-band to the date in the $V$-band using the relationship found in \citet{bianco14}.}
\tablenotetext{d}{For this SN, the initial classification was an SN Ib/c event by \citet{hamuy02}, whereas \citet{parrent07} support the Ib classification and SNID also regards it as a Ib SN. Thus, we classify it as a SN Ib in this study.}
\tablenotetext{e}{For this SN, the date of $V$-band maximum light is $2453241.18\pm0.98$, which was determined from the $V$-band photometry in \citet{drout11} through the polynomial fitting in \citet{bianco14}.}
\tablenotetext{f}{The $V$-band light curve of this SN shows two distinct maxima. In this study, the phases are expressed in days from the date of the second maximum light according to \citet{modjaz14}.}
\tablenotetext{g}{This SN had spectra that resemble SN Ic-bl spectra after explosion, i.e., 15 to 10 days before the date of maximum light, but developed narrow-line spectra with helium by the date of maximum light \citep{mazzali08,modjaz09,mazzali09}.}
\tablenotetext{h}{For this SN, the date of $V$-band maximum light is $2456476.27\pm0.02$, which was determined from the $V$-band photometry in \citet{Fremling2014} through the polynomial fitting in \citet{bianco14}.}
\label{table_Ib-norm}
\end{deluxetable*}

\section{Spectral Analysis Methods}
\label{method}

We used two spectral analysis methods that are the same as those in \citet{modjaz15}.  One is constructing average spectra using the whole spectra, as explained in detail in Section \ref{mean_spec}. We used them to directly compare spectra in Sections \ref{sec_H_Ib} and \ref{sec_He_Ic}, and to quantify the spectral diversity of different subtypes of SNe in Section \ref{sec_meanspec}. The other method is to quantify spectral features based on individual lines by measuring their absorption velocities and strengths. We analyzed these measurements in a statistical way, i.e., constructing error bars and average values of the measurements, to conduct a spectroscopic comparison between SNe IIb and SNe Ib in Sections \ref{sec_H_IbIIb}--\ref{Ib_two} and to constrain progenitors of SNe Ic in Section \ref{sec_helium_problem}. 

Before analyzing the spectra, we pre-process them, as discussed in detail in Appendix \ref{pre-process}. In particular, we would like to remove the effect of different dispersions and wavelength ranges due to the use of different telescopes, as well as reddening of the spectra. The spectra in our sample are flattened via the same procedure used to prepare the spectra for SNID, thus the resultant spectra (i.e., SNID templates) have the same format as those in the SNID database templates-2.0. Then we used the SNID templates we constructed above, as well as the SNID templates of stripped SNe that are already in the SNID database templates-2.0, to construct mean spectra (Section \ref{mean_spec}). In order to quantify SN features, we measure their absorption velocities and strengths (Section \ref{vel_pEW_measure}). The corresponding error bars were produced using the uncertainty arrays of spectra  (Appendix \ref{smooth}) and Monte Carlo (MC) simulations (Section \ref{error_bar}). In order to explore bulk properties as a function of different Stripped SN subtypes, we calculated a rolling weighted average of the measurements (Appendix \ref{mean_KS}).

All of the mean spectra, the new SNID templates, and the code that produces uncertainty arrays of spectra are published under DOI 10.5281/zenodo.58766 and DOI 10.5281/zenodo.58767. They are available on our SNYU github website, \footnote{https://github.com/nyusngroup} and also linked from our SNYU webpage.\footnote{http://cosmo.nyu.edu/SNYU/}

\subsection{Constructing Mean Spectra from SN Spectra} 
\label{mean_spec}

Mean spectra, together with the standard deviations from the mean spectra, can characterize the spectral diversity of each SN subtype, thus they can be used to determine if a newly discovered SN belongs to a normal type or a novel type \citep[e.g., Ib-Ca;][]{kasliwal12}. Since our spectra are in relative flux and we care more about SN features than SN continua, here we construct mean spectra, as well as their standard deviations, using the flattened version of our spectral sample (i.e., the corresponding SNID templates) and following the same procedure outlined in \citet{Blondin2007}. Here, we apply the procedure to SNe IIb, SNe Ib and SNe Ic; in \citet{modjaz15}, we also apply it to SNe Ic-bl. Our mean spectra are constructed every five days from $t_{\mathrm{Vmax}}=-20$ to  $t_{\mathrm{Vmax}}=70$ days. Here, we only show them at several specific phases (e.g., $-10$, 0, $10$, and $20$ days). Each phase includes no more than one spectrum per input SN that is within $\pm2$ days of the corresponding phase.\footnote{This phase range is for the phases rounded to the closest integer, which is the same as the phase range of $\pm2.5$ days (i.e., five-day bins) for the phases rounded to tenths. Hereafter, we use $\pm2$ days for convenience.} Compared with multiple spectra per SN, one spectrum per SN in a mean spectrum will avoid bias from a SN if it contributes several spectra to the mean spectrum. If two or more spectra from the same SN are available within $\pm2$ days of the target phase, we choose the one that is nearest to the target phase. If there is still more than one spectrum satisfying the above condition, we use the one that covers the largest optical wavelength range, or the one with the highest signal-to-noise (S/N) ratio. 

In a future paper, we will construct mean spectra that contain the continua, which can be used in photometric classification \citep[e.g.,][]{poznanski02} to generate model magnitudes that represent the expected photometry of different classes of SNe at a range of redshifts and with various filters. Thus, a SN can be classified by comparing the observed photometry to the expected photometry of different types of SNe at the same redshift and with the same filter.

\subsection{Absorption Velocity and Strength Measurements}
\label{vel_pEW_measure}

SN spectra are characterized by the presence or absence of specific lines. In order to quantify the properties of various subtypes of SNe, we shall measure the velocity and strength of specific absorption lines in each spectrum. We did so by following the same procedure used for SN Ia spectra in \citet{blondin06}, \citet[see their figure 15]{blondin11_spec_distance_Ia}, and \citet[see their figure 2]{silverman12_analysis}, as outlined below.

We used the following steps to measure the velocity of absorption features in spectra. First, a quadratic polynomial was fitted around the minimum of the concerned absorption feature to find the exact wavelength position that corresponds to the minimum flux. Second, the relativistic Doppler formula was applied to the wavelength found above. We used the equivalent width (EW) to quantify the strength of the absorption features. First, a median filter was applied to both sides of the absorption feature to find the local maxima. A straight line connecting one local maximum on the blue side and one local maximum on the red side was regarded as the local continuum or pseudo continuum.\footnote{``Local" or ``pseudo" is due to the fact that there is no one source for the continuum.} If there were more than one local maximum on either or both sides, the pair of local maxima that ensures the pseudo continuum slope does not cross the spectrum within the boundaries of the feature was picked. If no pair of local maxima satisfied this condition, the pair with the highest local maxima was picked. If more than one pair of local maxima satisfied this condition, the pair that maximized the wavelength range was picked. The resultant pseudo continuum was used to calculate the EW, which will be called ``pseudo EW" \citep[pEW;][]{blondin11_spec_distance_Ia, silverman12_analysis}. The pEW has the following definition,
\begin{equation}
\mathrm{pEW}=\sum_{i=0}^{N_d-1}\Delta \lambda _{i} \left (\frac{f_{c}(\lambda _{i})-f(\lambda _{i})}{f_{c}(\lambda _{i})} \right ),
\end{equation}
where $N_d$ is the number of data points between two local maxima, $\lambda_{i}$ is the wavelength of the $i$th data point, $\Delta \lambda _{i}$ is the bin size between $\lambda_{i}$ and $\lambda_{i+1}$,  $f_{c}(\lambda _{i})$ is the flux of local continuum at $\lambda_{i}$, and $f(\lambda _{i})$ is the flux at $\lambda_{i}$. One advantage of using the pEW to quantify the strength of absorption lines is that no assumption about the line profile is made. 

We adopt the line identifications for stripped SNe of \citet{branch02, branch06}, \citet{elmhamdi06}, and \citet{dessart12}, as well as other works in the literature included in this study that identified lines in the spectra of a specific stripped SN (see Sections \ref{classify} and \ref{He_pro}). Since stripped envelope SNe have been observed to start transitioning from the photospheric phase to the nebular phase at generally 60 days after the date of maximum light \citep[e.g.,][]{barbon90, filippenko95, iwamoto00, matheson01}, in this work, we only show our measurements at phases before $t_{\mathrm{Vmax}}\simeq60$ days. We note that our line identifications are not definitive; furthermore, we discuss below some of the features for which contradictory identifications exist in the literature. We also assume that the pEW of an absorption feature can reflect the amount of atoms of the corresponding element at the corresponding state in the ejecta. 

%For relative depth, we found the local minimum in the same way as in velocity measurements. The relative depth was calculated as the fraction of the difference between the minimum flux and the corresponding flux on the local continuum with respect to the latter flux at the same wavelength. Thus its value is between 0 and 1. 

\subsection{Error Bars of the Measurements}
\label{error_bar}

\begin{deluxetable*}{llllllllllllll}
\tabletypesize{\scriptsize}
\tablecaption{Measured absorption velocities\label{vels_table}}
\tablehead{
\colhead{Phase\tablenotemark{a}} &
\colhead{He I $\lambda$5876\tablenotemark{b}} &
\colhead{He I $\lambda$6678\tablenotemark{b}} &
\colhead{He I $\lambda$7065\tablenotemark{b}} &
\colhead{H$\alpha$\tablenotemark{b}} &
\colhead{Fe II $\lambda$5169\tablenotemark{b}} & 
\colhead{O I $\lambda$7774\tablenotemark{b}} &\\
\colhead{(days)} &
\colhead{(km s$^{-1}$)} &
\colhead{(km s$^{-1}$)} &
\colhead{(km s$^{-1}$)} &
\colhead{(km s$^{-1}$)} &
\colhead{(km s$^{-1}$)} &
\colhead{(km s$^{-1}$)} &
}
\startdata
\multicolumn{6}{c}{\bf SN 1983N} \\
$4$ &    $-$9600 $\pm$ 600 &       $-$8900 $\pm$ 600 &              \nodata &      $-$14300 $\pm$ 500 &      $-$8900 $\pm$ 1500 & \nodata \\
$12$ &   $-$10200 $\pm$ 700 &       $-$9900 $\pm$ 700 &       $-$7500 $\pm$ 300 &      $-$12500 $\pm$ 500 &       $-$6200 $\pm$ 700 & \nodata
\enddata
\tablecomments{This table is available in its entirety in a machine-readable form in the online journal. A portion is shown here for guidance regarding its form and content.}
\tablenotetext{a}{Rest-frame age of spectrum in days relative to $V$-band maximum. See text for details.} 
\tablenotetext{b}{For multiple velocity measurements for the same SN of different spectra taken on the same night, the weighted average is reported, and the velocity error reported here is the standard deviation of the weighted average.}
\end{deluxetable*}

\begin{deluxetable*}{cccccccccccc}
\tabletypesize{\scriptsize}
\tablecaption{Measured pEW values\label{pew_table}}
\tablehead{
\colhead{Phase\tablenotemark{a}} &
\colhead{He I $\lambda$5876} &
\colhead{He I $\lambda$6678} &
\colhead{He I $\lambda$7065} &
\colhead{H$\alpha$} &
\colhead{O I $\lambda$7774} &\\
\colhead{(days)} &
\colhead{(\AA)} &
\colhead{(\AA)} &
\colhead{(\AA)} &
\colhead{(\AA)} &
\colhead{(\AA)}
}
\startdata
\multicolumn{6}{c}{\bf SN 1983N} \\
$4$ &          114 $\pm$ 4 &           31 $\pm$ 3 &           \nodata &           28 $\pm$ 3 &              \nodata \\
$12$ &          122 $\pm$ 4 &           43 $\pm$ 4 &          105 $\pm$ 8 &          30 $\pm$ 10 &              \nodata 
\enddata
\tablecomments{This table is the same as Tabel \ref{vels_table}, but for pEW values.}
\end{deluxetable*}

We used the uncertainty arrays derived in Appendix \ref{smooth} and a Monte Carlo (MC) sampling method to estimate errors of the above measurements. Assuming the noise at a wavelength bin in a spectrum obeys a Gaussian distribution and equals one standard deviation of the distribution, a synthetic spectrum was generated by drawing a data point for each wavelength bin within a Gaussian distribution centered on the original spectral datum, and with a standard deviation equal to the value for the same wavelength bin in the uncertainty array of the original spectrum. For each synthetic spectrum, we repeated the steps in the velocity and pEW measurements in Section \ref{vel_pEW_measure}. The estimated measurement and error bar were calculated as the mean and standard deviation of the corresponding values across 3000 realizations. The number of realizations should be larger than $N_d$(log$N_d)^2$ \citep{babu83}, where $N_d$ is the number of data points in the portion of the resampled spectrum. The 3000 realizations are sufficient in our case since we only have a few tens of data points for each SN feature. 

For the velocity and pEW measurements, the typical MC uncertainties are $\sim\pm1000$ km s$^{-1}$ and $\sim\pm10$ \AA, respectively. The sources of error include the spectral quality (i.e., S/N ratio) and line blending that gives rise to broad features. 

A subset of our measurements is shown in Tables \ref{vels_table} and \ref{pew_table}. A full version of this table is available in a machine-readable form in the online journal.

%\section{Discussions}
%\label{discuss}

%We checked the presence or absence of H$\alpha$ in spectra of SNe Ib by comparing mean spectra of SNe Ib with mean spectra of SNe IIb in \ref{sec_H_Ib}. We did similar things to check if He I lines present in spectra of SNe Ic in \ref{sec_He_Ic}. In section \ref{classify}, we used the velocity and pEW measurements of H$\alpha$ and He I lines to spectroscopically compare SNe Ib and SNe IIb in a statistical way. In section  \ref{sec_helium_problem}, we discussed the He problem in the progenitors of SNe Ic using the pEW of O I $\lambda$7774 and the velocity of Fe II $\lambda$5169 as the indicators to test two competing models. We also explored the spectral diversity within each SN subtype using mean spectra in section \ref{sec_meanspec}.

\section{Is there a continuum between progenitors of SNe IIb and SNe Ib?}
\label{classify}

Our ultimate goal is to map different progenitor models to different SN subtypes. Thus, it is important to correctly classify SNe as the first step. In general, SNe are classified based on spectra and light curves. It is not always easy to classify SNe because SN spectra are time dependent and even the spectra at the same phase are diverse within a SN subtype. The time-dependent nature of spectra is claimed to affect most classifications of SNe IIb and Ib \citep[e.g.,][]{milisavljevic13}. That is because by definition, SNe IIb are similar to SNe II at early phases, but over time they evolve to appear more analogous to SNe Ib. Thus, there are concerns that a SN IIb may be misclassified as a SN Ib if it is not discovered early enough \citep[e.g.,][]{milisavljevic13}. In this section, we show that this concern is not well-founded by investigating the presence of H$\alpha$ in SNe Ib (Section \ref{sec_H_Ib}), exploring the behavior of important common lines in SNe IIb and Ib, and concluding that we can use the pEW of H$\alpha$ (Section \ref{sec_H_IbIIb}) to distinguish between these two SN subtypes. Moreover, there are observational continua between the spectral properties of SNe IIb and those of SNe Ib, based on measurements of H$\alpha$, He I lines, and Fe II $\lambda$5169 (Sections  \ref{sec_He_IbIIb_vel}--\ref{Fe_vabs_IIbIb}, respectively). We also explore the possibility of the existence of subtypes within SNe Ib (Section \ref{Ib_two}).

\subsection{Is There H$\alpha$ in the Spectra of SNe Ib?}
\label{sec_H_Ib}

\begin{figure}[t]
\includegraphics[scale=0.5,angle=0,trim = 0mm 0mm 0mm 0mm, clip]{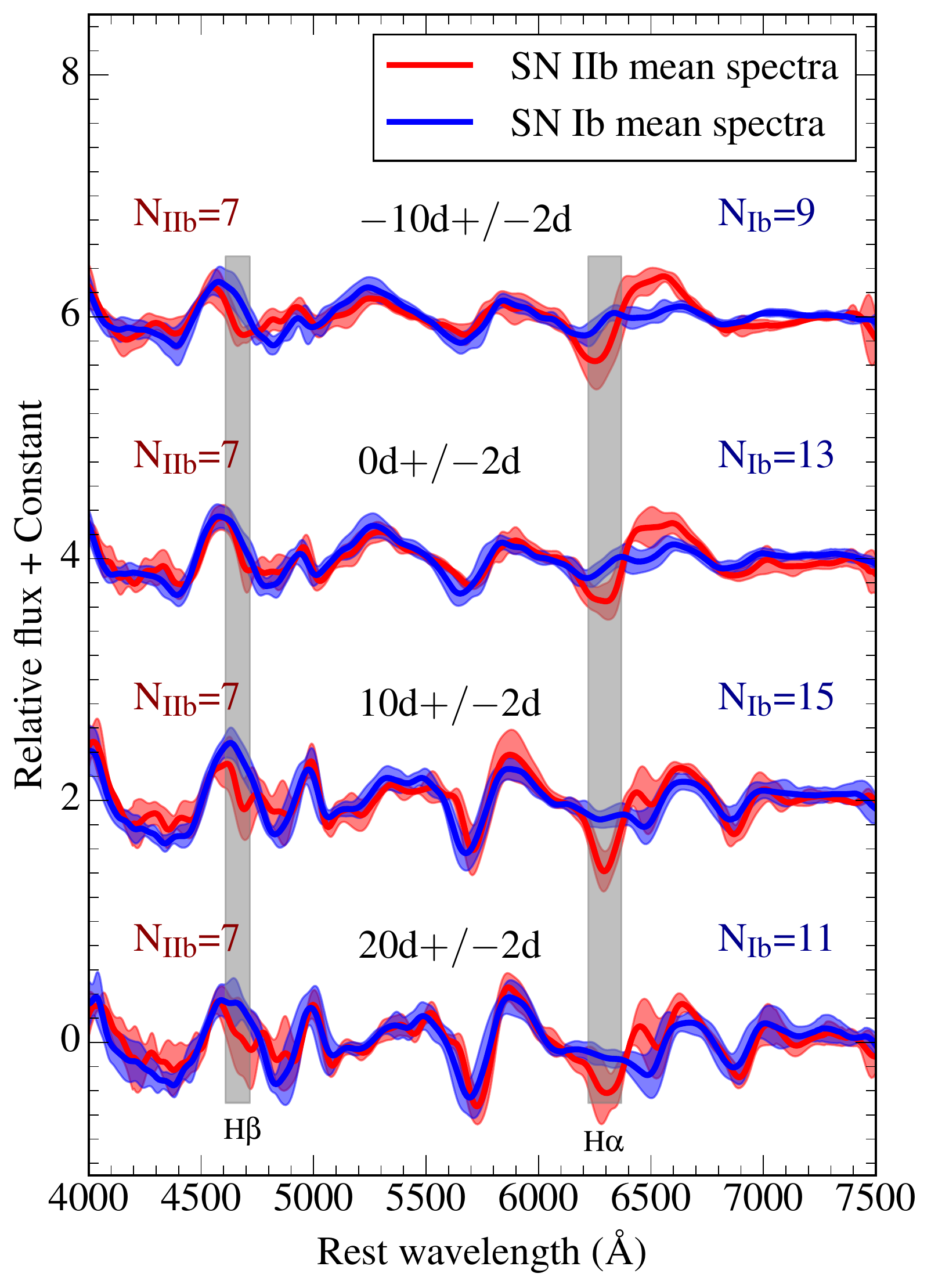}
\caption{Mean spectra and their corresponding standard deviations of SNe IIb (red) and SNe Ib (blue) at four different phase ranges: $t_{\mathrm{Vmax}}=-12$ to $-8$, $-2$ to 2, 8 to 12, and 18 to 22 days. Each mean spectrum only includes one spectrum per SN even if multiple spectra have been taken within the phase range. N$_{\mathrm{IIb}}$ represents the number of spectra (which is also the number of SNe) included in the mean spectrum of SNe IIb at each phase, and N$_{\mathrm{Ib}}$ represents the number of spectra for SNe Ib. The gray vertical bands indicate the expected positions of H$\alpha$ and H$\beta$ at velocities of $-$9000 km s$^{-1}$ to $-$16000 km s$^{-1}$.}
\label{fig_mean_IIbIb_h}
\end{figure}

While the absence of H$\alpha$ is supposed to be the hallmark of SNe Ib, there are good reasons to assume that there may be some H$\alpha$ in SNe Ib \citep[e.g.,][]{deng00, branch02, elmhamdi06, parrent07, James10,yoon10,parrent15}. Since the progenitor stars may have different amounts of hydrogen present before explosion (since the various mechanisms may not have removed all of the hydrogen), we may expect different amounts, including small amounts, of hydrogen still present at the time of explosion.  However, it is not easy to identify H$\alpha$ because the absorption feature at the expected position of H$\alpha$---the ``6300 \AA~absorption line"---can be due to C II $\lambda$6580, Ne I $\lambda$6402 or Si II $\lambda$6355 as well \citep[e.g.,][]{branch02, tominaga05, elmhamdi06,parrent07}. 

Previous investigations include modeling progenitors of SNe Ib to investigate whether there is a hydrogen layer before explosion, as well as comparing the synthetic spectrum from radiative transfer calculations to the observed spectrum. Applying stellar evolutionary models that include the effects of rotation to binary systems, \citet{yoon10} found that the progenitor of SNe Ib should have a thin hydrogen layer before explosion. Thus, it is likely that the spectra of SNe Ib show H$\alpha$ in absorption. Moreover, many works \citep[e.g.,][]{deng00, branch02, elmhamdi06, parrent07, James10,parrent15} identified H$\alpha$ in the spectra of SNe Ib using various spectral synthesis codes such as SYNOW \citep[a parameterized SN spectrum-synthesis code;][]{parrent10}, PHOENIX \citep[a generalized non-local thermodynamic equilibrium stellar atmospheres code;][]{hauschildt99,hauschildt04}, or non-local thermodynamics equilibrium time-dependent radiative-transfer calculations \citep{dessart12}. 

Here, we use a statistical and data-driven approach to investigate whether there is H$\alpha$ present in SN Ib spectra. We use the fact that SNe IIb, by definition, show unambiguous lines of H$\alpha$ in their spectra and compare the mean spectra of SNe IIb with those of SNe Ib. First, we identified H$\alpha$ and H$\beta$ in the mean spectra of SNe IIb. Then, we searched for H$\alpha$ and H$\beta$ at similar velocities (i.e., blueshift) in the mean spectra of SNe Ib. In Figure \ref{fig_mean_IIbIb_h}, we present the mean spectra of SNe IIb and SNe Ib, which  were constructed using the flattened version of our SN sample, at several specific phase ranges: $t_{\mathrm{Vmax}}=-10\pm2$, $0\pm2$, $10\pm2$, and $20\pm2$ days. The mean spectra of SNe IIb show a strong H$\alpha$ P-Cygni profile at $t_{\mathrm{Vmax}}=-10$ and 0 days, a pronounced H$\alpha$ absorption feature (assuming the absorption feature around 6300\AA~is due to H$\alpha$) at $t_{\mathrm{Vmax}}=10$ and 20 days, and a weak H$\beta$ absorption feature (assuming the absorption feature around 4700\AA~is due to H$\beta$) at all phases. In the mean spectra of SNe Ib, there are absorption features at the expected positions of H$\alpha$ at $t_{\mathrm{Vmax}}=-10$, 0, and 10 days. Thus, it is reasonable to identify these absorption features as H$\alpha$. At $t_{\mathrm{Vmax}}\simeq-10$ days and $t_{\mathrm{Vmax}}\simeq0$ day, the H$\alpha$ absorption features in the SN Ib mean spectra are weaker and at a higher velocity than those in the SN IIb mean spectra. At $t_{\mathrm{Vmax}}\simeq10$ days, the H$\alpha$ absorption feature in the SN Ib mean spectrum is much weaker than that in the SN IIb mean spectrum but the H$\alpha$ absorption features are at a similar velocity in both mean spectra. After $t_{\mathrm{Vmax}}\simeq20$ days, it is very hard to detect the H$\alpha$ absorption feature in the spectra of SNe Ib. 

While we cannot conclusively verify the detection of H$\alpha$ in all SNe Ib, in the following sections we assume that the weak absorption feature at $\sim$ 6250 \AA ~around the date of maximum light in SN Ib spectra is due to H$\alpha$ and report its pEW values and absorption velocities. We caution that the same absorption feature may be due to different elements at different phases.\footnote{For example, as claimed in \citet{tominaga05}, the line near 6300 \AA~in spectra of SN 2005bf is reproduced as a blend of H$\alpha$ and Si II $\lambda$6355 at the time of the first peak, while it is reproduced as Si II $\lambda$6355 alone at the time of the second peak.}

%\begin{figure*}[!ht]
%\subfigure[Ib]{%
%\includegraphics[scale=0.4,angle=0]{Ib_H.pdf}
%\label{fig_Halpha_Ib}}
%\quad
%\subfigure[IIb]{%
%\includegraphics[scale=0.4,angle=0]{IIb_H.pdf}
%\label{fig_Halpha_IIb}}
%
%\caption{Spectra of Ib/IIb around maximum. Dotted lines represent expected positions of H lines with a velocity between -12000 km s$^{-1}$ and -8000 km s$^{-1}$. Dashed lines represent expected positions of H lines with a velocity between -3000 km s$^{-1}$ and -7000 km s$^{-1}$.}
%\label{fig_Halpha_IbIIb}
%\end{figure*}

%Figure~\ref{fig_Halpha_Ib} and figure~\ref{fig_Halpha_IIb} show ten spectra of Ib and six spectra of IIb around maximum respectively. Dotted lines represent expected positions of H lines with a velocity between -12000 km s$^{-1}$ and -8000 km s$^{-1}$. Dashed lines represent expected positions of H lines with a velocity between -3000 km s$^{-1}$ and -7000 km s$^{-1}$. All Ib spectra have obvious H features. All of them have similar velocity of H-$\beta$. Eight of them have quite similar velocity of H$\alpha$The other two spectra of sn2007C and sn2005bf have high velocity of H$\alpha$Five IIb spectra have two components H$\alpha$and they are not fully developed. SN 2011ei has one component and it's fully developed. Thus it's reasonable that Ib have H$\alpha$

\subsection{The H$\alpha$ Line Behaves Differently in SNe IIb and SNe Ib}
\label{sec_H_IbIIb}

\begin{figure*}[t]
\subfigure{%
\includegraphics[scale=0.4,angle=90,trim = 0mm 5mm 0mm 10mm, clip]{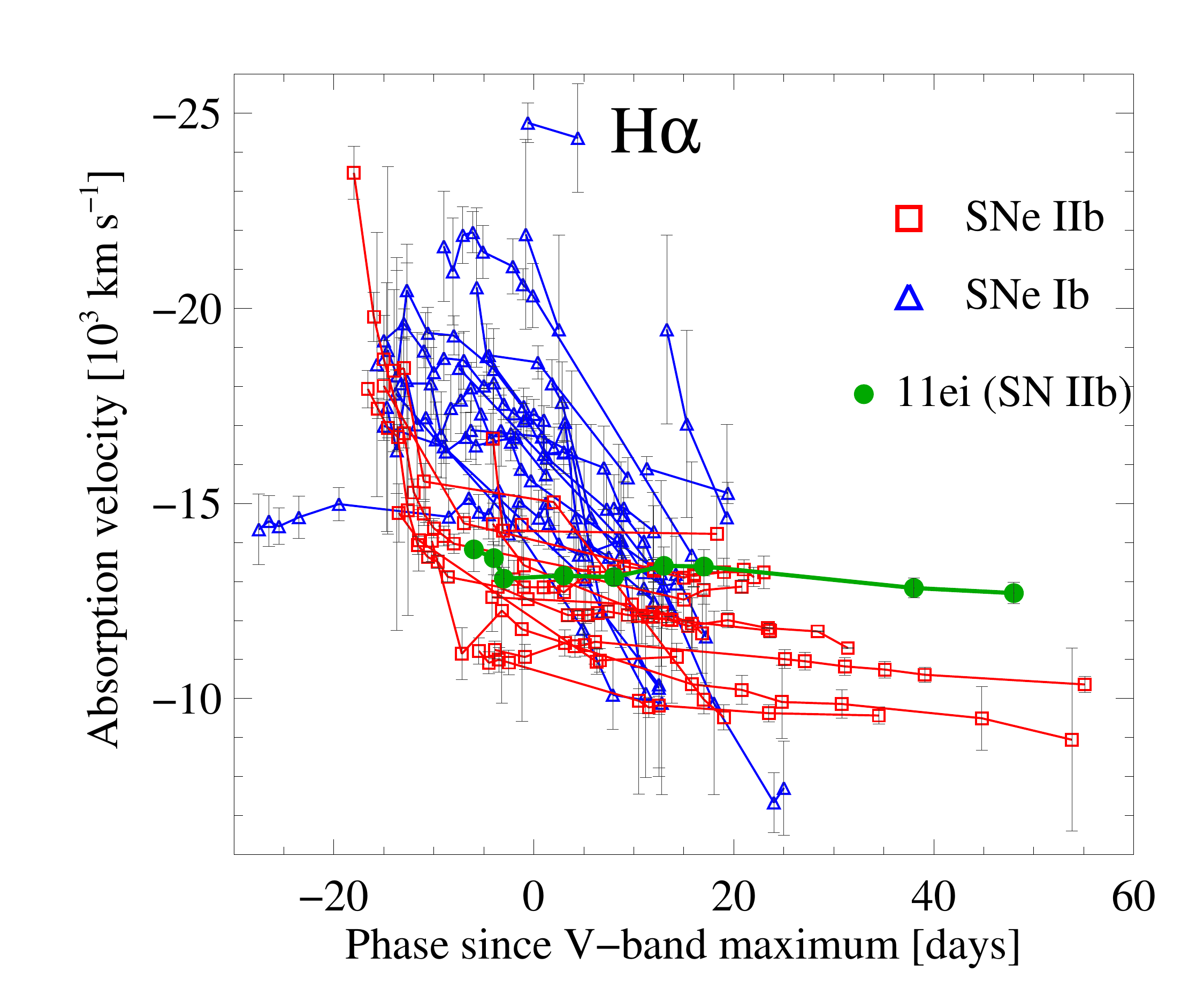}
}
\quad
\subfigure{%
\includegraphics[scale=0.4,angle=90]{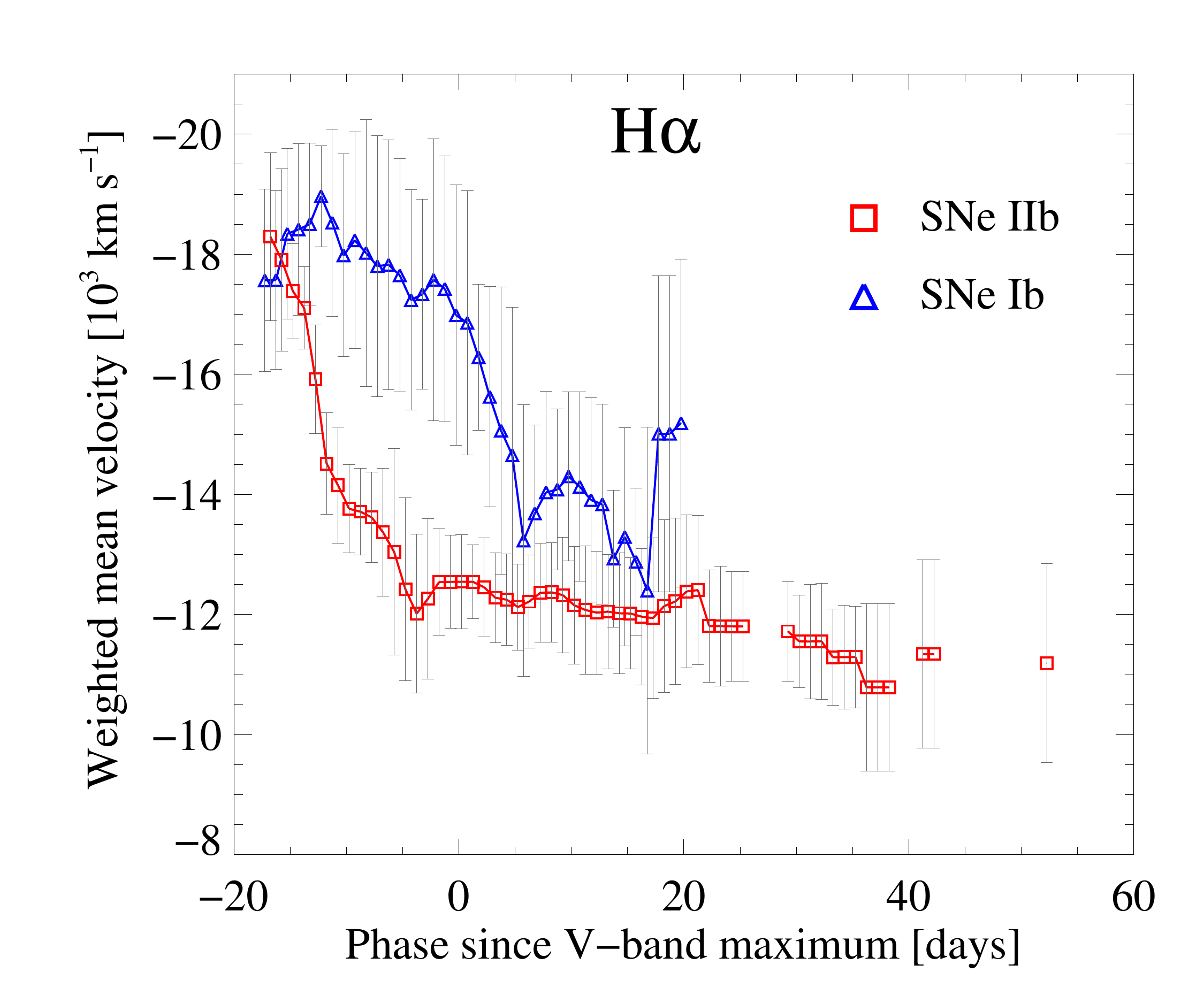}
}
\quad
\subfigure{%
\includegraphics[scale=0.4,angle=90,trim = 0mm 5mm 0mm 10mm, clip]{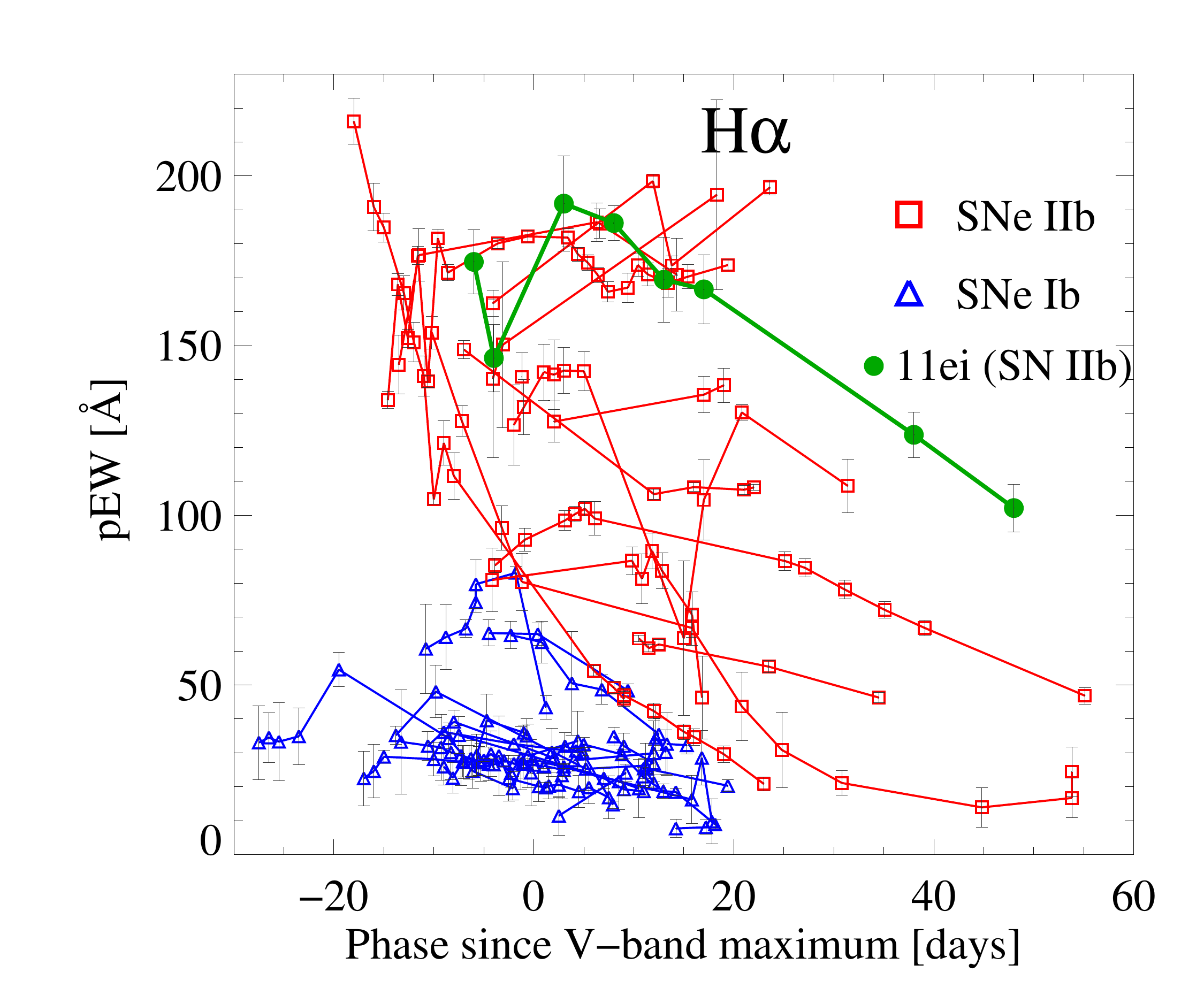}
}
\quad
\subfigure{%
\includegraphics[scale=0.4,angle=0]{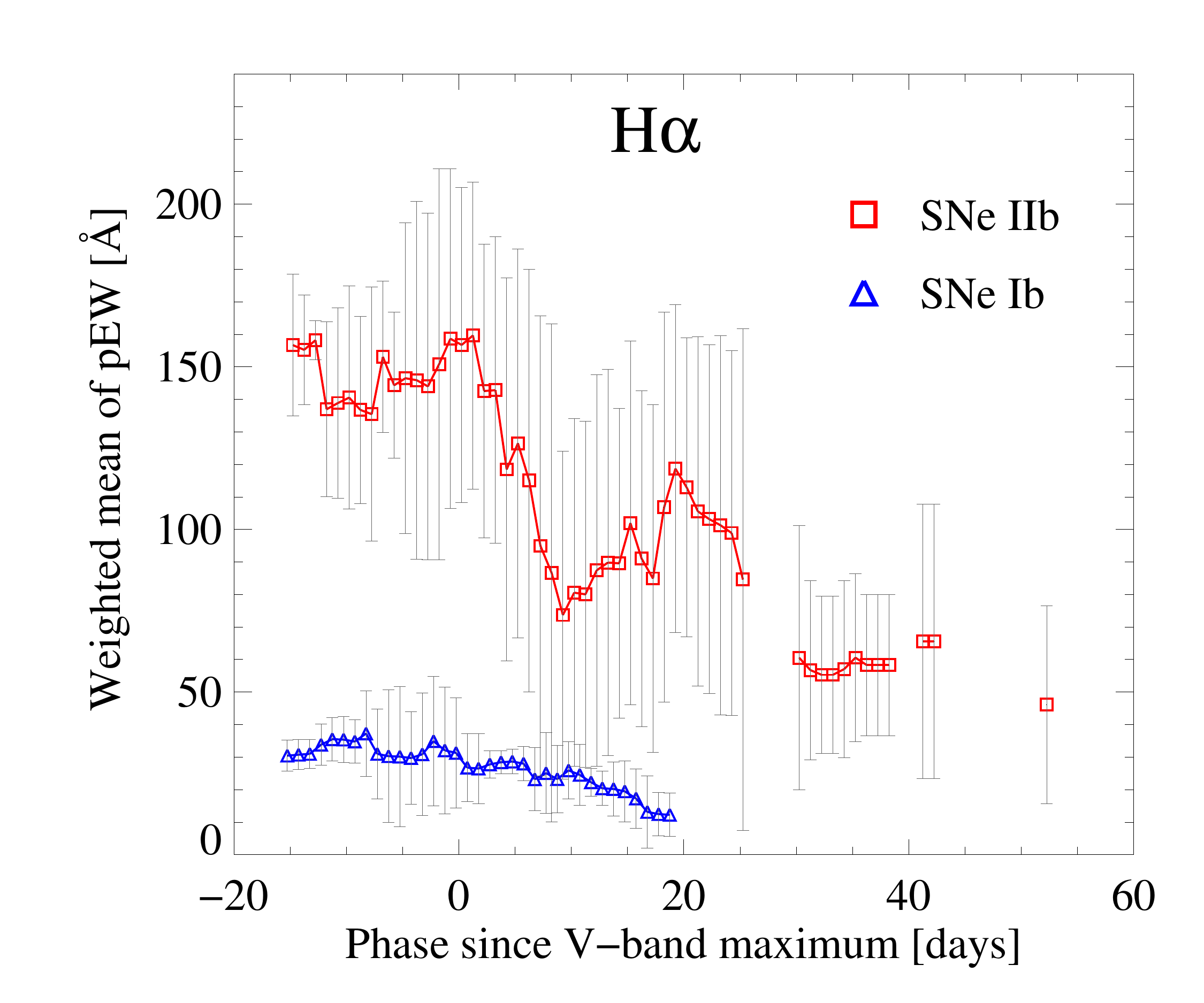}
}

\caption{Measured H$\alpha$ velocities (\textit{Top}) and pEW values (\textit{Bottom}) for SNe IIb (red squares) and SNe Ib (blue triangles). \textit{Left}: velocities and pEW values for individual SNe. Data of the same SN are connected by a line. The SN Ib that has data earlier than $t_{\mathrm{Vmax}}=-20$ days is SN 2005bf. The evolution of SN 2011ei is also shown (see text for more discussion). In the lower left panel, the three SNe Ib and one SN IIb that have nearly overlapping values are SNe 1999dn, 1999ex, iPTF13bvn, and 2008ax. \textit{Right}: Rolling weighted average values for SNe IIb and SNe Ib, with a bin size of five days for phases before $t_{\mathrm{Vmax}}=30$ days and a bin size of 10 days for phases thereafter. In order to show the distribution of the data, the error bars on the mean values represent the standard deviation of the contributing data points. For the standard deviation to be meaningful, only weighted average values constructed from more than three SNe are shown. There are no data for SNe Ib at $t_{\mathrm{Vmax}}>25$ days because it is very difficult to identify H$\alpha$ in the spectra of SNe Ib at those phases. Note that there is almost no overlap between pEW values of H$\alpha$ in SNe IIb and SNe Ib, thus the H$\alpha$ pEW defines SNe IIb and Ib, i.e.,  the H$\alpha$ pEW values can be used to differentiate SNe IIb from SNe Ib at all epochs.}
\label{fig_vabs_Halpha_f3}
\end{figure*}

\begin{figure*}[t]
\subfigure{
\includegraphics[scale=0.4,angle=90,trim = 0mm 5mm 0mm 10mm, clip]{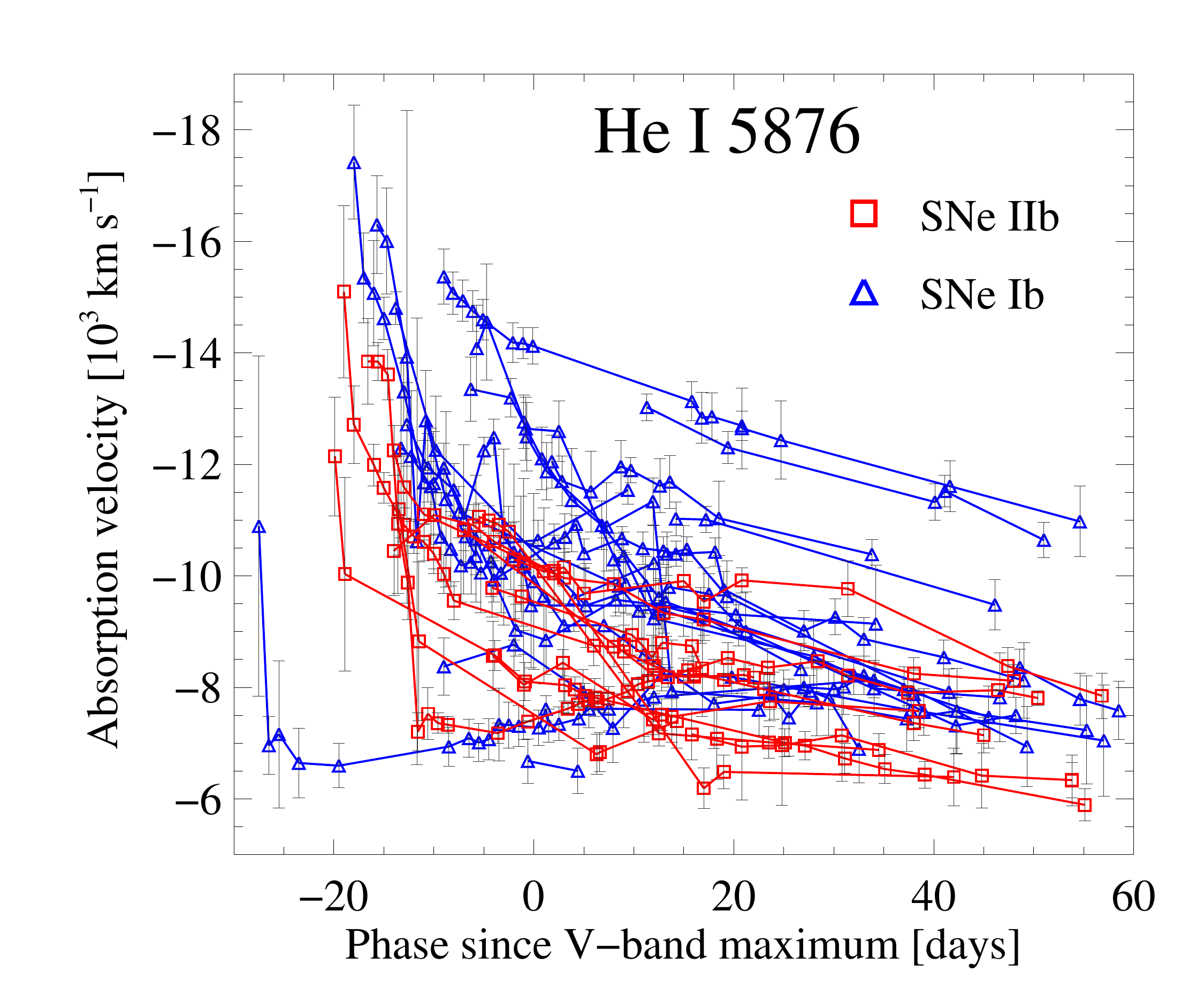}
}
\quad
\subfigure{
\includegraphics[scale=0.4,angle=90]{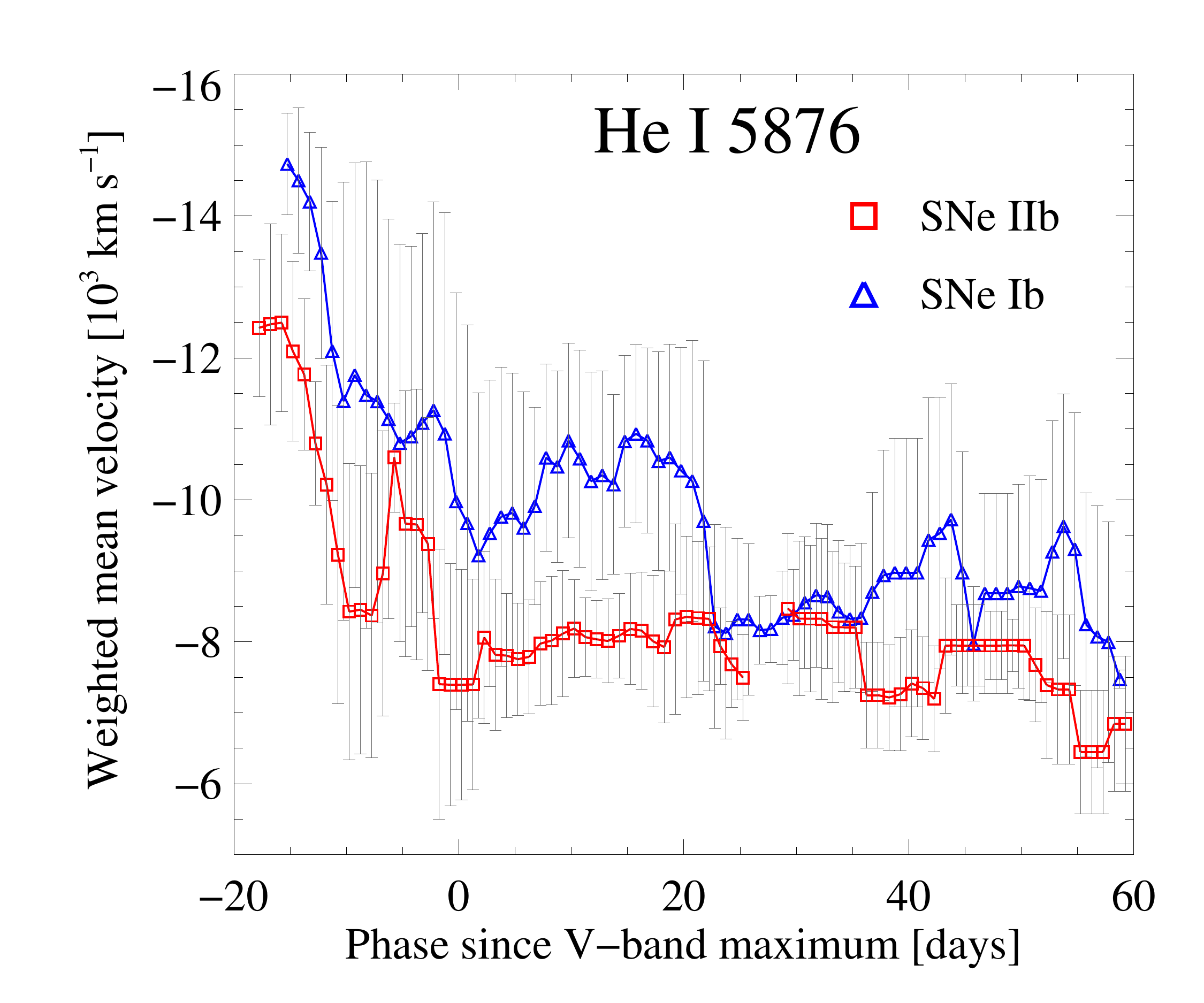}
}
\quad
\subfigure{
\includegraphics[scale=0.4,angle=90,trim = 0mm 5mm 0mm 10mm, clip]{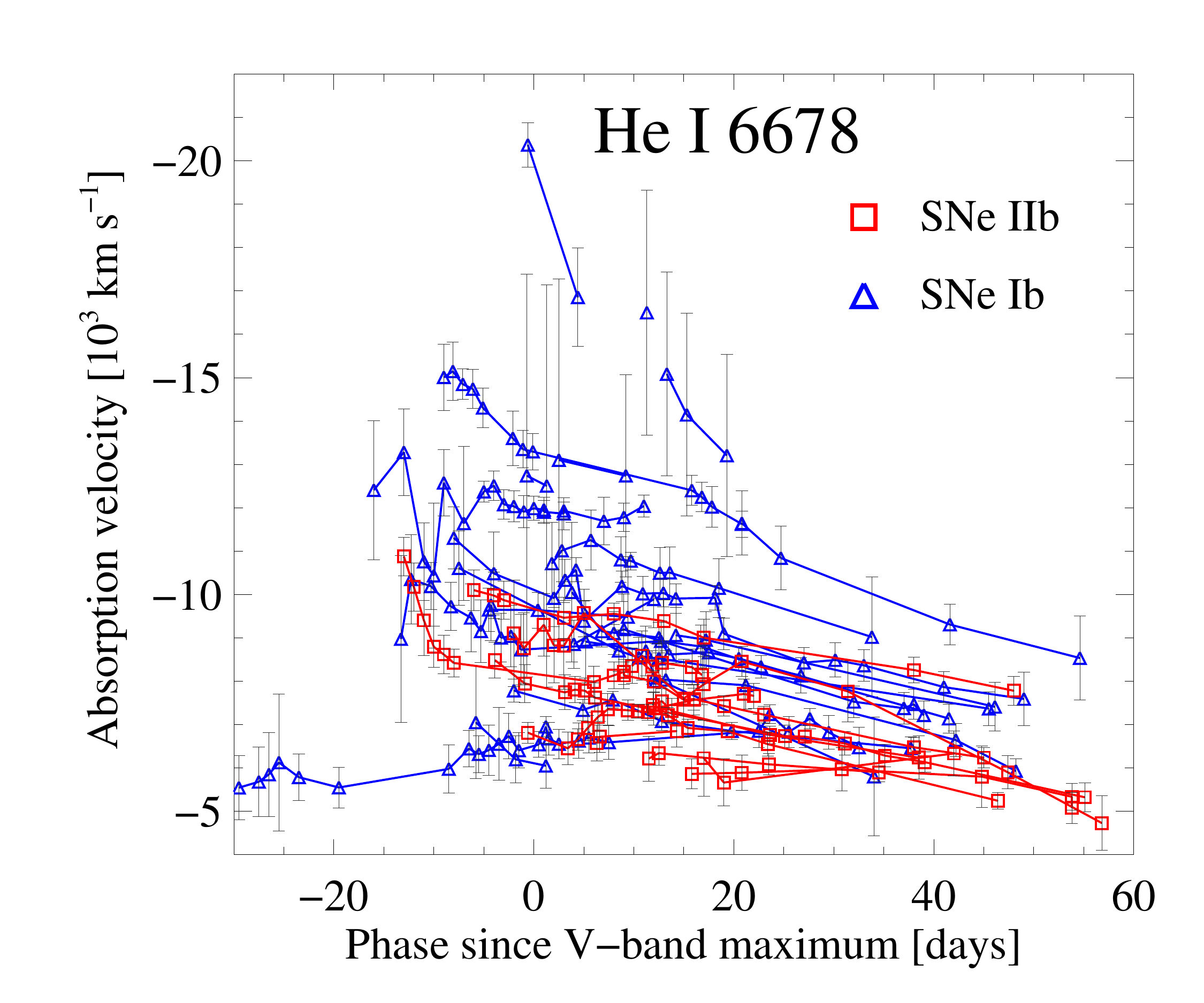}
}
\quad
\subfigure{
\includegraphics[scale=0.4,angle=90,trim = 0mm 0mm 0mm 5mm, clip]{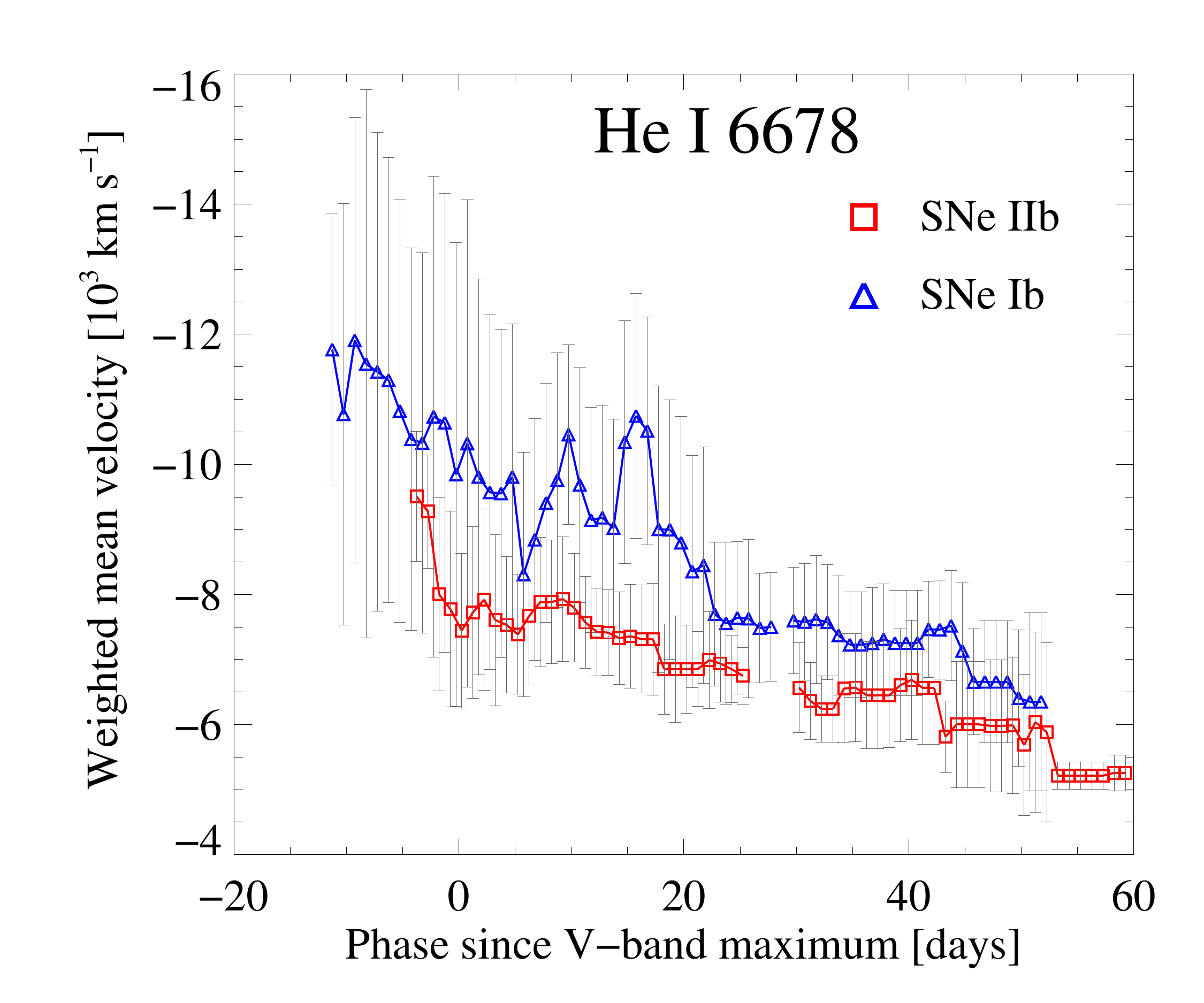}
}
\quad
\subfigure{
\includegraphics[scale=0.4,angle=90,trim = 0mm 5mm 0mm 10mm, clip]{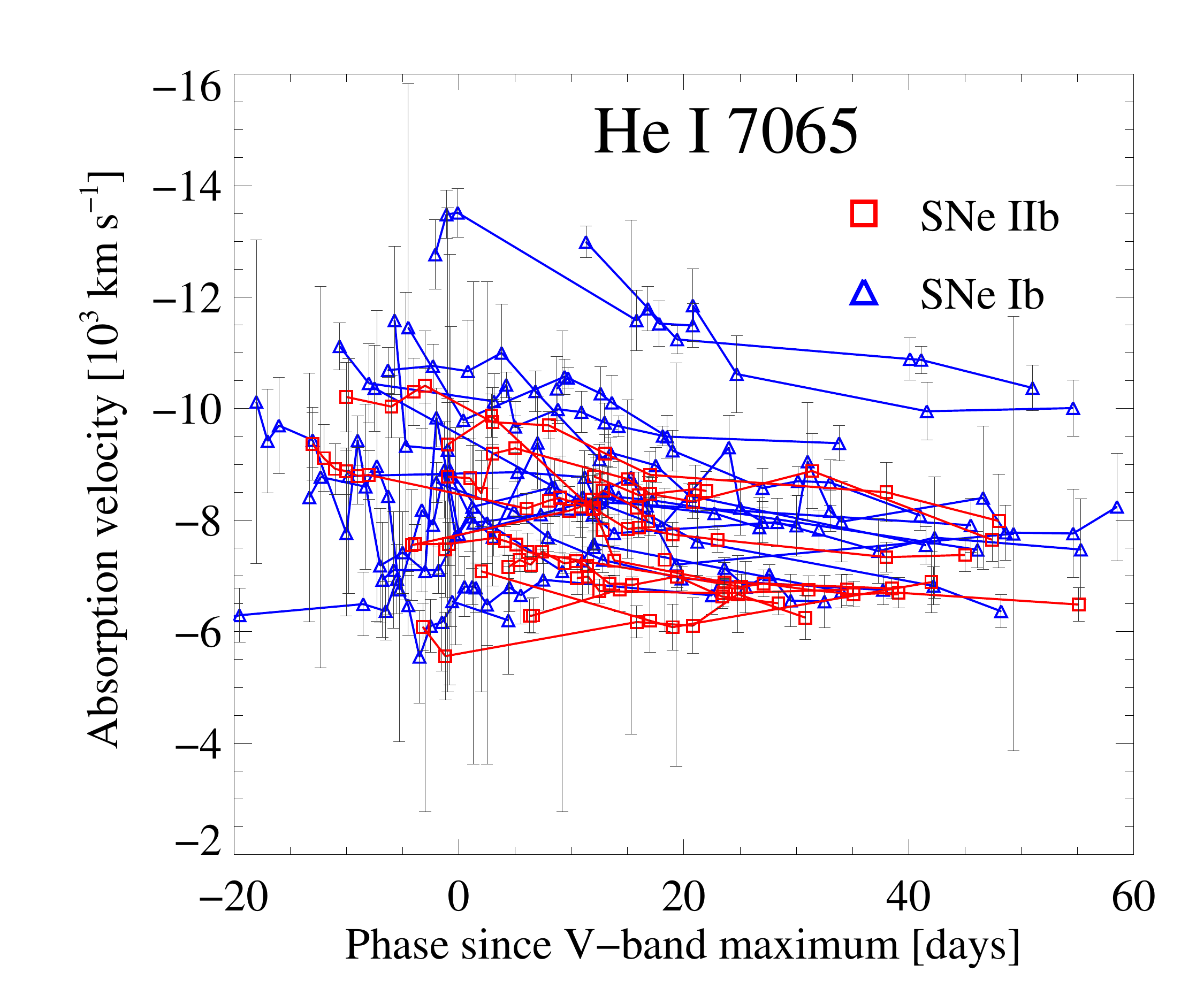}
}
\quad
\subfigure{
\includegraphics[scale=0.4,angle=90]{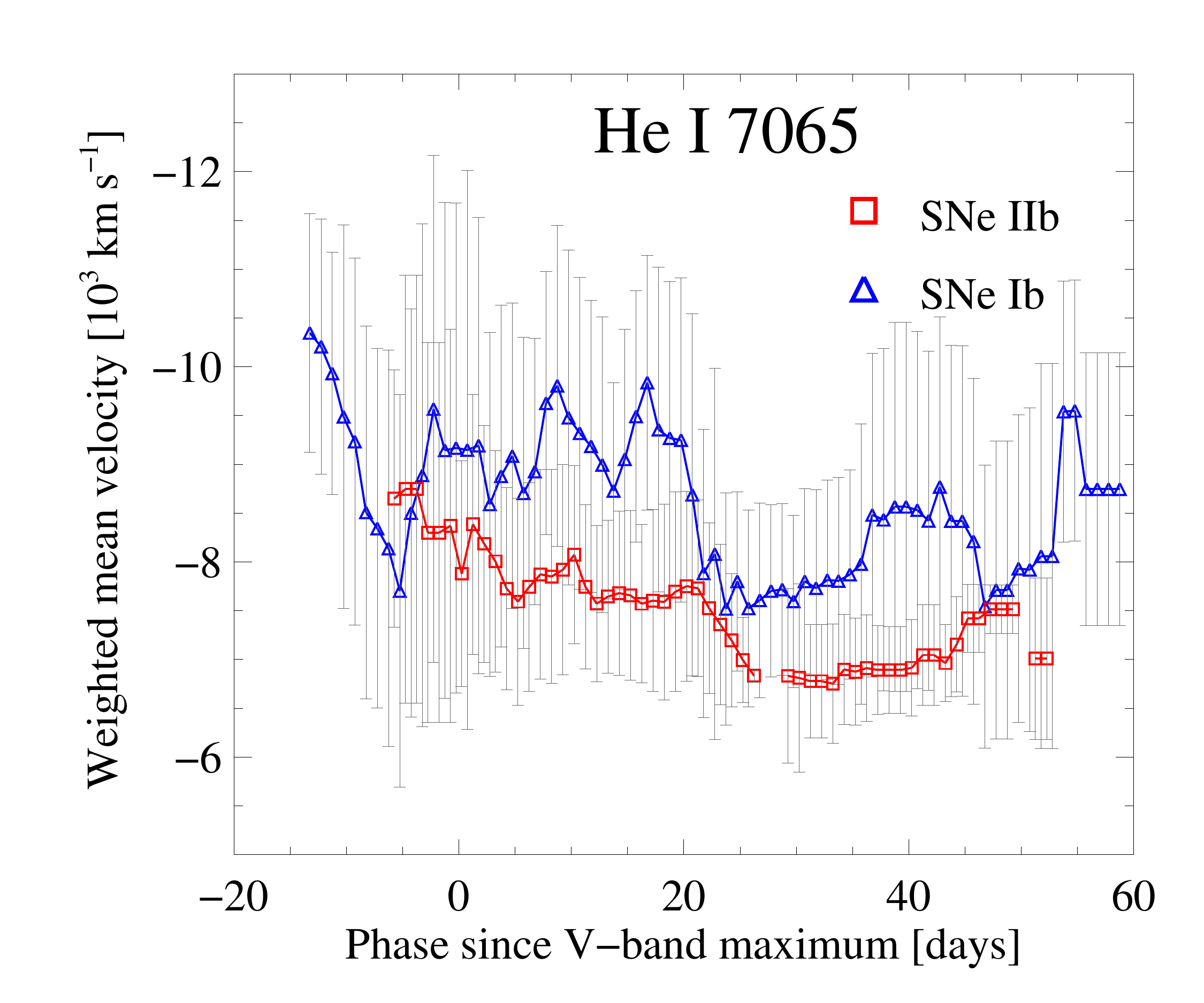}
}
\caption{The same as Figure \ref{fig_vabs_Halpha_f3}, but for measurements of He I $\lambda$5876 (\textit{Top}), He I $\lambda$6678 (\textit{Middle}), and He I $\lambda$7065 (\textit{Bottom}) velocities. In the bottom left panel, the two SNe Ib with the highest He I $\lambda$7065 velocities are SNe 1990I and 2004gq.}
\label{fig_vabs_HeI5876}
\end{figure*}

%\begin{figure*}[t]
%\subfigure{%
%\includegraphics[scale=0.4,angle=0,trim = 10mm 0mm 10mm 0mm, clip]{IbIIb_Halpha_pEW.pdf}
%}
%\quad
%\subfigure{%
%\includegraphics[scale=0.4,angle=0]{mean_plot_Halpha_pEW.pdf}
%}
%%
%\caption{The same as Figure \ref{fig_vabs_Halpha_f3}, but for measurements of H$\alpha$ pEW. In the left panel, the three SNe Ib on the lower border of the SN IIb sample are SNe 1999dn, 1999ex, and iPTF13bvn. Since there is almost no overlap between SNe IIb and SNe Ib, the H$\alpha$ pEW defines SNe IIb \& Ib, i.e.,  the H$\alpha$ pEW values can be used to differentiate SNe IIb from SNe Ib at all epochs.}
%\label{fig_pEW_Halpha_f3}
%\end{figure*}

As shown in Section \ref{sec_H_Ib}, we assume that some H$\alpha$ is present in SNe Ib in our sample, and thus we can use it as a diagnostic to compare SNe IIb with SNe Ib. The H$\alpha$ line in the spectra of SNe IIb becomes weaker and even disappears with time. Hence, there are concerns that a SN IIb may be misclassified as a SN Ib if it is not discovered early enough \citep[e.g.,][]{dessart11, milisavljevic13}. In this section, we show that these concerns may not apply since the pEW values of H$\alpha$ evolve differently in SNe IIb compared to those in SNe Ib. Thus, the observed spectra of SNe IIb are different from those of SNe Ib, even at later phases, indicating that SNe IIb are distinguishable from SNe Ib. 

Assuming the absorption feature around 6300\AA~ is due to H$\alpha$, the temporal evolution of the H$\alpha$ velocity and pEW values of individual SNe IIb and Ib, as well as the rolling weighted averages (see Appendix \ref{mean_KS}) for the two SN subtypes, are displayed in Figure \ref{fig_vabs_Halpha_f3}.\footnote{Due to the difficulty of detecting H$\alpha$ in the spectra of SNe Ib at phases later than $t_{\mathrm{Vmax}}\simeq25$ days, no data points of SNe Ib are shown at these phases.} In particular, the weighted averages at $t_{\mathrm{Vmax}}\simeq0$ day are listed in Table \ref{table_mean}. As mentioned before, five and seven SNe in our sample, respectively, were also included in \citet{branch02} and \citet{elmhamdi06}. Both papers used SYNOW to determine H$\alpha$ velocities; these are consistent with H$\alpha$ velocities via line identifications in this work. We observe that the average H$\alpha$ velocities in SNe Ib are systematically higher (around 40\% at $t_{\mathrm{Vmax}}\simeq0$ day) than in SNe IIb. If SNe IIb and Ib have almost the same explosion energies, then SNe Ib have a thinner hydrogen envelope or a lower hydrogen mass than SNe IIb. The H$\alpha$ velocities of individual SNe Ib show that they drop rapidly over phase intervals during which they are detected. However, the rolling weighted average velocities stay approximately the same after $t_{\mathrm{Vmax}}\simeq10$ days due to one SN Ib (SN 2004gv) with high H$\alpha$ velocities starting at $t_{\mathrm{Vmax}}\simeq15$ days and ending at $t_{\mathrm{Vmax}}\simeq20$ days. The H$\alpha$ velocities in SNe IIb decrease rapidly from $t_{\mathrm{Vmax}}\simeq-10$ days to $t_{\mathrm{Vmax}}\simeq0$ day and have a relatively flat evolution after that. There is much overlap between the SN IIb and SN Ib velocities around $t_{\mathrm{Vmax}}=15$ days.

The average H$\alpha$ pEW values in SNe IIb are systematically higher ($\sim$ 4 times at $t_{\mathrm{Vmax}}\simeq0$ day) than in SNe Ib. This confirms that, compared to SNe Ib, SNe IIb have more hydrogen in their progenitors. There is a larger scatter in the pEW values of SNe IIb than in SNe Ib. The pEW values of H$\alpha$ form two distinct groups for SNe IIb and SNe Ib, except for three SNe Ib (SNe 1999dn, 1999ex, and iPTF13bvn) on the lower bound of the SN IIb sample. Since there is almost no overlap between SNe IIb and SNe Ib, the H$\alpha$ pEW can be used to differentiate SNe IIb from SNe Ib at all epochs. The concern of \citet{milisavljevic13} that if a SN IIb is discovered several days after the date of maximum light, the SN IIb may be misclassified as a Ib, is not well-founded since even at later phases, a SN IIb will always have a larger H$\alpha$ pEW than a SN Ib. 

%Note that the weighted mean values in SNe Ib are not shown at phases $\sim30$ and 40 due to the difficulty of detecting H$\alpha$ in SNe Ib spectra at these phases. This applies to other H$\alpha$ related plot for SNe Ib as well. Using a significance level of 0.05, the K-S probability shows that the velocity of H$\alpha$ can be used to differentiate SN Ib from SNe IIb at phases $\sim-10$ and 0 days whereas the pEW of H$\alpha$ can be used at phases $\sim-10$, 0, 10, and 20 days. If we can identify H$\alpha$ at a spectrum later than phase $\sim25$ days and measure its pEW to be larger than $\sim40$ \AA, then we can classify it as a SN IIb. Thus, we conclude that the pEW of H$\alpha$ can be used to do classification at all phases.

\citet{milisavljevic13} suggested using the ratios of He I $\lambda$5876 pEW values to H$\alpha$ pEW values to differentiate SNe IIb from SNe Ib. We found that this criterion is driven by the pEW values of H$\alpha$, since as shown in Section \ref{sec_He_IbIIb_pEW}, the pEW values of He I $\lambda$5876 in SNe IIb and SNe Ib are similar. Thus, we suggest to use the pEW values of H$\alpha$ to classify SNe IIb and Ib. In particular, we plot SN 2011ei, the SN in \citet{milisavljevic13}, in Figure \ref{fig_vabs_Halpha_f3}, which indeed shows that even if SN 2011ei was discovered after $t_{\mathrm{Vmax}}\simeq10$ days, it would not have been classified as a SN Ib.\footnote{\citet{milisavljevic13} identified a two-component H$\alpha$ absorption in SN 2011ei. For SN 2011ei and other SNe that have two-component H$\alpha$ absorptions, we measure absorption velocity using the dominant component and pEW using both components.} To mimic the situation where SN 2011ei was discovered one week after $t_{\mathrm{Vmax}}\simeq0$ day or later, we ran SNID on its spectra at $t_{\mathrm{Vmax}}=8$, 13, and 17 days. The SNID code identifies these spectra as SN IIb spectra as well. In particular, using the default definition of matched SN spectra in SNID, the ratio between the number of matched SNe IIb and that of SNe Ib ranges from 1.3 to 2.3.

%\begin{table}[t]
%\caption{K-S probability of H$\alpha$ velocity and pEW in SNe Ib and SNe IIb at different phases}
%\centering
%\begin{tabular}{c || c c | c c}
%\hline\hline
%Phase range (days) & No. of Ib & No. of IIb & prob\_vel\footnote{K-S probability for H$\alpha$ velocity.} & prob\_pEW\footnote{K-S probability for H$\alpha$ pEW.}\\ [0.5ex]
%\hline
%    $-$15 to    $-$5 &     10 &      7 &   0.001 &  0.003 \\ [1.0ex]
%     $-$5 to     5 &     14 &     10 &    0.002 & 0.000\\ [1.0ex]
%      5 to    15 &     14 &      9 &   0.097 &  0.000\\ [1.0ex]
%     15 to    25 &      4 &     11 &    0.862 & 0.035\\ [1.0ex]
%     25 to    35 &      0 &      4 &   N/A & N/A\\ [1.0ex]
%     35 to    45 &      0 &      3 &   N/A & N/A\\ [1.0ex]
%\hline
%\end{tabular}
%\label{table_num_ks_vabs_pEW_Halpha_t3}
%\end{table}

\subsection{Absorption Velocities of He I Lines in SNe IIb and SNe Ib}
\label{sec_He_IbIIb_vel}

%\begin{figure*}[t]
%\subfigure{
%\includegraphics[scale=0.4,angle=90,trim = 0mm 5mm 0mm 10mm, clip]{IbIIb_HeI6678_vabs.pdf}
%}
%\quad
%\subfigure{
%\includegraphics[scale=0.4,angle=90]{mean_plot_HeI6678_vabs.pdf}
%}
%\caption{The same as Figure \ref{fig_vabs_Halpha_f3}, but for measurements of He I $\lambda$6678 velocities.}
%\label{fig_vabs_HeI6678}
%\end{figure*}
%
%\begin{figure*}[t]
%\subfigure{
%\includegraphics[scale=0.4,angle=90,trim = 0mm 10mm 0mm 10mm, clip]{IbIIb_HeI7065_vabs.pdf}
%}
%\quad
%\subfigure{
%\includegraphics[scale=0.4,angle=90]{mean_plot_HeI7065_vabs.pdf}
%}
%\caption{The same as Figure \ref{fig_vabs_Halpha_f3}, but for measurements of He I $\lambda$7065 velocities. In the left panel, the two SNe Ib with the highest He I $\lambda$7065 velocities are SNe 1990I and 2004gq.}
%\label{fig_vabs_HeI7065}
%\end{figure*}

In the optical spectra of SNe IIb and Ib, the three strongest He I lines are He I $\lambda\lambda\lambda$5876, 6678, and 7065. In the following, we will first present the temporal evolution of the velocities of these He I lines. Then we will explore the so called ``flat-velocity" SNe IIb \citep{folatelli14} in our sample as well. 

Figure \ref{fig_vabs_HeI5876} presents the temporal velocity evolution of He I $\lambda\lambda\lambda$5876, 6678, and 7065 for SNe IIb and Ib, respectively. The left panels show the velocity evolutions of individual SNe IIb and Ib, while the right panels show the corresponding rolling weighted averages for the two SN subtypes. In particular, the weighted averages at $t_{\mathrm{Vmax}}\simeq0$ day are listed in Table \ref{table_mean}. As mentioned before, five and seven SNe in our above sample were in \citet{branch02} and \citet{elmhamdi06}, respectively. Both papers determined He I velocities using SYNOW. For the SNe we have in common, their He I $\lambda\lambda\lambda$5876, 6678, 7065 velocities via line identifications in this work are consistent with the values reported by \citet{branch02} and \citet{elmhamdi06}.

There is a wide overlap in the He I velocities for SNe IIb and SNe Ib, indicating a continuum of some physical parameters such as the positions of the helium layer in the progenitors, the conditions for non-thermal excitation, or both. The average velocities of He I $\lambda\lambda\lambda$5876, 6678, 7065 in SNe Ib are slightly higher than those in SNe IIb. Assuming SNe IIb and Ib have similar explosion energies, SNe Ib may have lost some of the helium layer, which results in a smaller ejecta mass and a higher velocity in the remaining helium layer. 

\begin{figure}[t]
\includegraphics[scale=0.4,angle=90,trim = 0mm 0mm 0mm 10mm, clip]{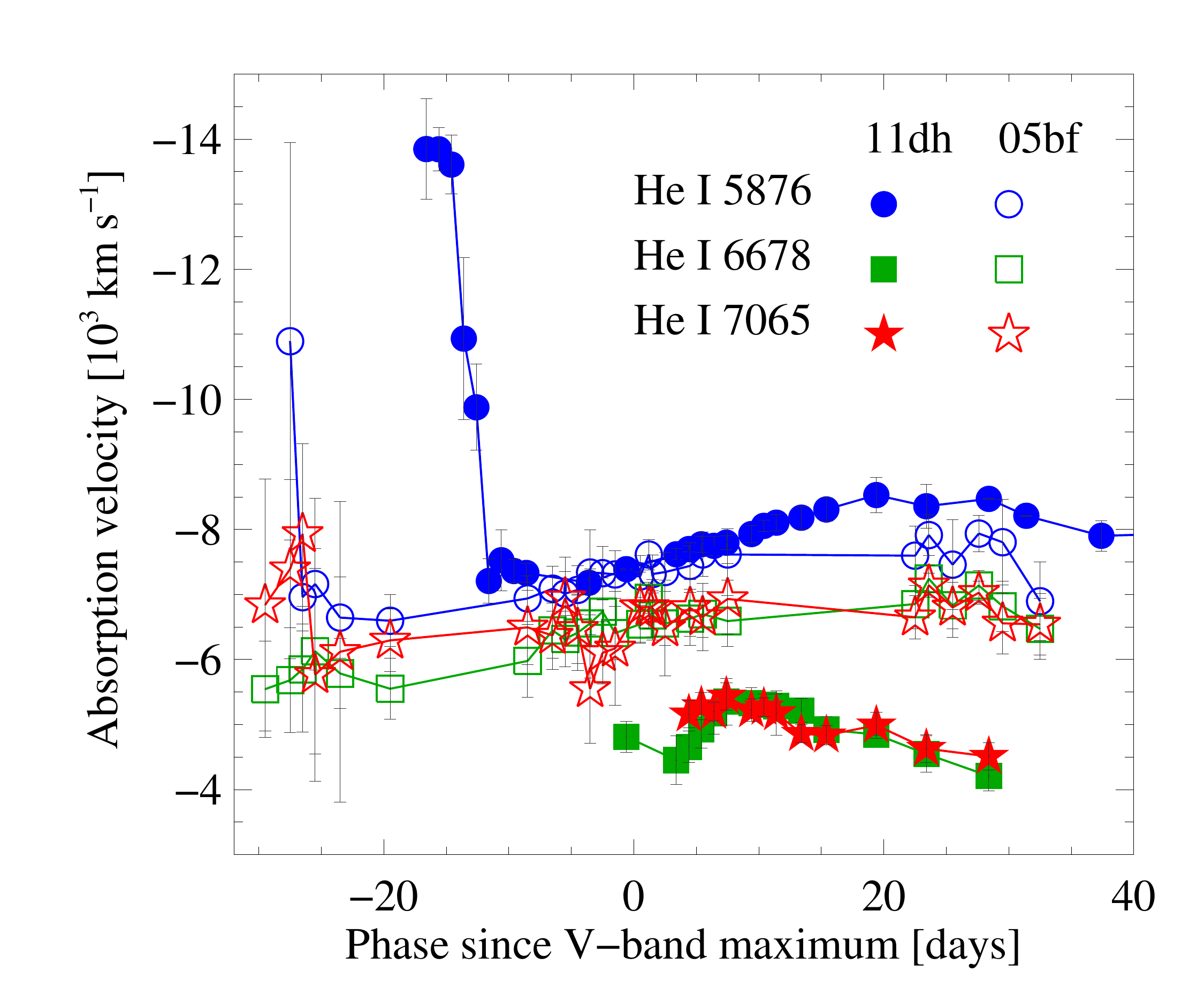}
\caption{Measurements of He I $\lambda\lambda\lambda$5876, 6678, and 7065 absorption velocities for SNe 2011dh (a SN IIb) and 2005bf (a SN Ib). For clarity, the velocities for SN 2011dh are shifted downwards by 2000 km s$^{-1}$. We note that the velocity values do not decrease monotonically with time, which is surprising.}
\label{fig_vabs_He_05bf}
\end{figure}

In general, the He I velocities evolve to lower values over time, with the He I $\lambda\lambda$5876, 6678 velocities in SNe IIb and Ib decreasing more rapidly than the He I $\lambda$7065 velocities by a factor of 70\% to 100\% at $t_{\mathrm{Vmax}}\simeq0$ day. However, as shown in Figure \ref{fig_vabs_He_05bf}, the He I velocities in SNe 2005bf and 2011dh increase with time, if the He I $\lambda$5876 velocities at the earliest phase are ignored. As explained in \citet{tominaga05}, the increasing He I velocities may be the result of increasing non-thermal excitation.  

\begin{deluxetable*}{cccccc}
\tabletypesize{\scriptsize}
\tablecaption{Classifications of three flat-velocity SNe \label{table_classify}}
\tablehead{
\colhead{SN name} &
\colhead{Classification in \citet{folatelli14}} &
\colhead{Classification in some works and this work} 
}
\startdata
SN 1999ex &    IIb  &    Ib \citep{parrent07} \tablenotemark{a} \\
SN 2005bf &  IIb   &  Ib \citep{modjaz14}  \tablenotemark{b} \\
SN 2007Y &  IIb   &  Ib \citep{stritzinger09} \tablenotemark{c}
\enddata
\tablenotetext{a}{Although the H$\alpha$ pEW value in SN 1999ex is within the nearly overlapping region for SN IIb and SN Ib samples, the H$\alpha$ velocity and behavior of He I lines in its spectra are consistent with SNe Ib in our sample.}
\tablenotetext{b}{SN 2005bf does not have strong H-alpha line (as shown in the bottom panel of Figure 2), which by definition is the case for SNe Ib instead of SNe IIb.}
\tablenotetext{c}{Although the time evolution of He I $\lambda$7065 pEW values in SN 2000Y is comparable to that of SNe IIb, the pEW values and velocities of H$\alpha$ and other He I lines in this SN are more consistent with those of SNe Ib in our sample. }
\end{deluxetable*}

\citet{folatelli14} introduced a family of flat-velocity SNe IIb (SNe fvIIb) whose velocities stay roughly the same---between 6000 and 8000 km s$^{-1}$---during the photospheric phase. They argue that this may indicate a dense shell in the ejecta, though its formation lacks a physical mechanism. In particular, \citet{folatelli14} showed the evolution of the He I $\lambda$5876 velocity in their SN sample and claimed five SNe fvIIb: SNe 1999ex, 2005bf, 2007Y, 2010as, and 2011dh. We agree that the velocities of He I lines for these SNe evolve slowly compared to other SNe. However, we argue that the first three SNe are flat-velocity SNe Ib (SNe fvIb), based on classifications in other literature and measurements in this work, as summarized in Table \ref{table_classify}. Except for SN 2010as, which is not included in our sample (since we only include SN spectra available before September of 2014 and spectra of SN 2010as were not accessible before that date), we confirm that the remaining four flat-velocity SNe have relatively flat velocities of He I $\lambda\lambda$6678 and 7065 as well. 

\begin{figure}[t]
\includegraphics[scale=0.4,angle=90,trim = 0mm 0mm 0mm 10mm, clip]{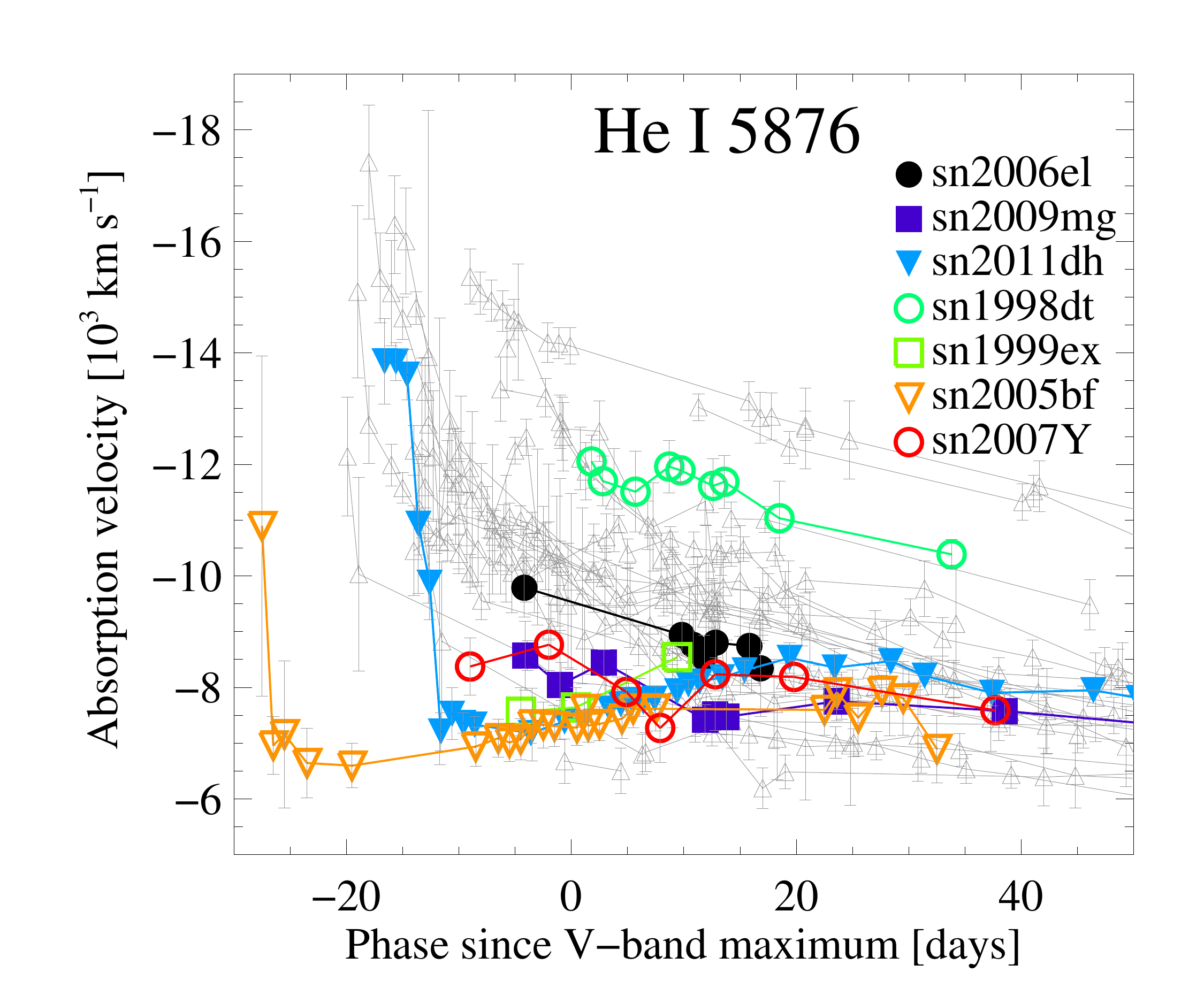}
\caption{Measurements of He I $\lambda$5876 absorption velocities for seven flat-velocity SNe IIb and Ib in our sample. The three SNe (SNe 2006el, 2009mg, and 2011dh) with filled symbols are SNe IIb and the four SNe (SNe 1998dt, 1999ex, 2005bf, and 2007Y) with open symbols are SNe Ib. For comparison, the gray triangles are the remaining SNe IIb and Ib in our sample. Data points of the same SN are connected by a line.}
\label{fig_vabs_HeI5875_IIbIb}
\end{figure}

Figure \ref{fig_vabs_HeI5875_IIbIb} shows seven SNe fvIIb and fvIb in our sample that satisfy the following conditions: each SN has spectra around the date of maximum light, i.e., $-5<t_{\mathrm{Vmax}}<5$ days; spectra cover more than 10 days; ignoring early phases ($t_{\mathrm{Vmax}}<-10$ days), the velocities for He I $\lambda\lambda\lambda$ 5876, 6678, and 7065 stays roughly the same, i.e., the change in velocity is smaller than 2000 km s$^{-1}$. In addition to the SNe mentioned in \citet{folatelli14}, we identify SNe 1998dt, 2006el and 2009mg as new flat-velocity SNe. Thus, using the He I velocities, we suggest there are four SNe fvIIb (SNe 2006el, 2009mg, 2010as, and 2011dh) and four SNe fvIb (SNe 1998dt, 1999ex, 2005bf, and 2007Y) based on our work. Since none of them show flat-velocity Fe II lines and only two of them (SNe 2006el and 2005bf) have flat-velocity H$\alpha$ lines, the flat-velocity He I lines formed in these SNe could rather be due to e.g., excitation effects since the He I lines are due to non-thermal excitation \citep{lucy91}, instead of a dense shell as proposed by \citet{folatelli14}, since otherwise, other lines would have also exhibited the same flat velocities.

\subsection{Strengths of He I Lines in SNe IIb and Ib}
\label{sec_He_IbIIb_pEW}

\begin{figure*}[t]
\subfigure{%
\includegraphics[scale=0.4,angle=90,trim = 0mm 5mm 0mm 10mm, clip]{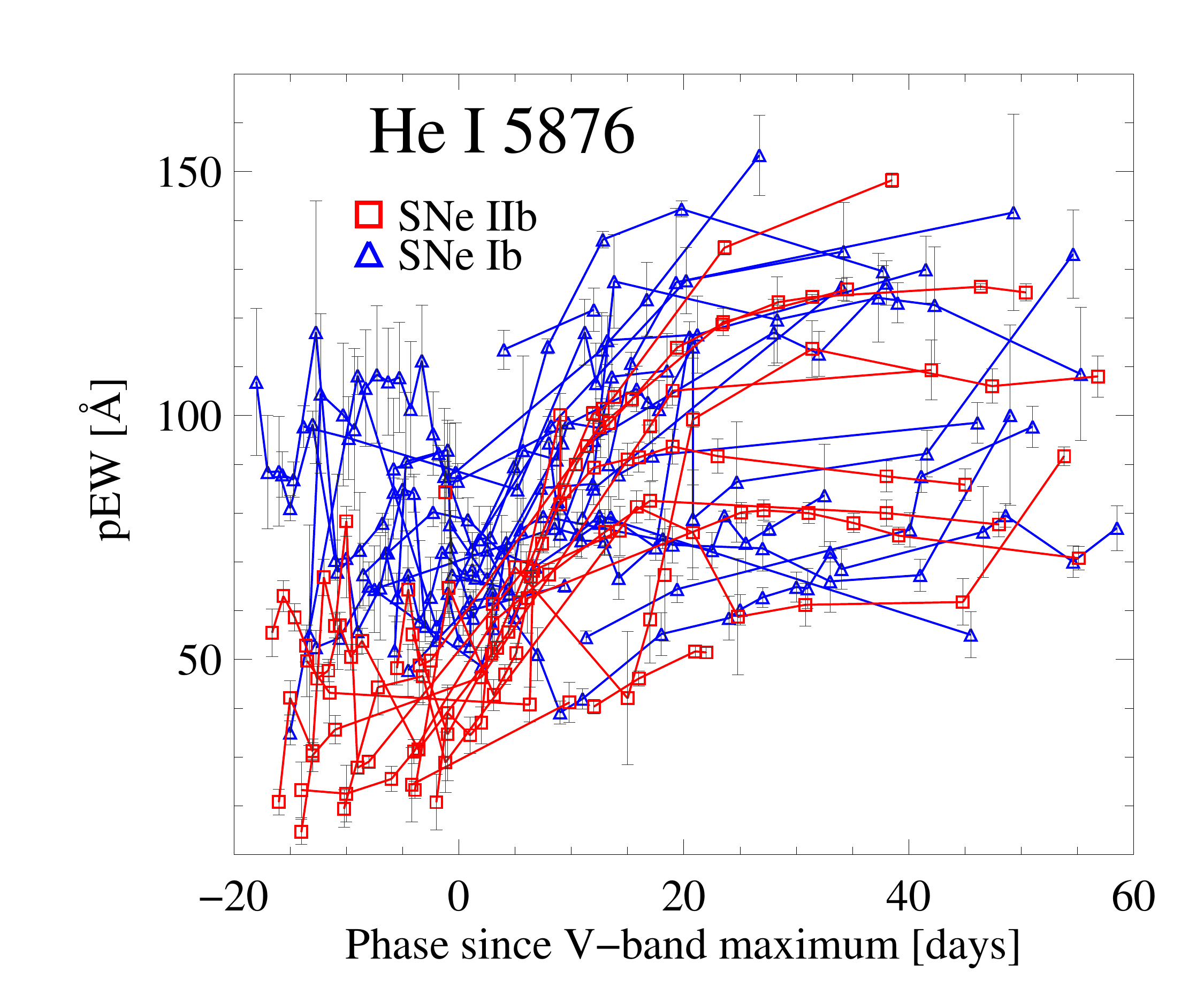}
}
\quad
\subfigure{%
\includegraphics[scale=0.4,angle=0,trim = 5mm 0mm 0mm 0mm, clip]{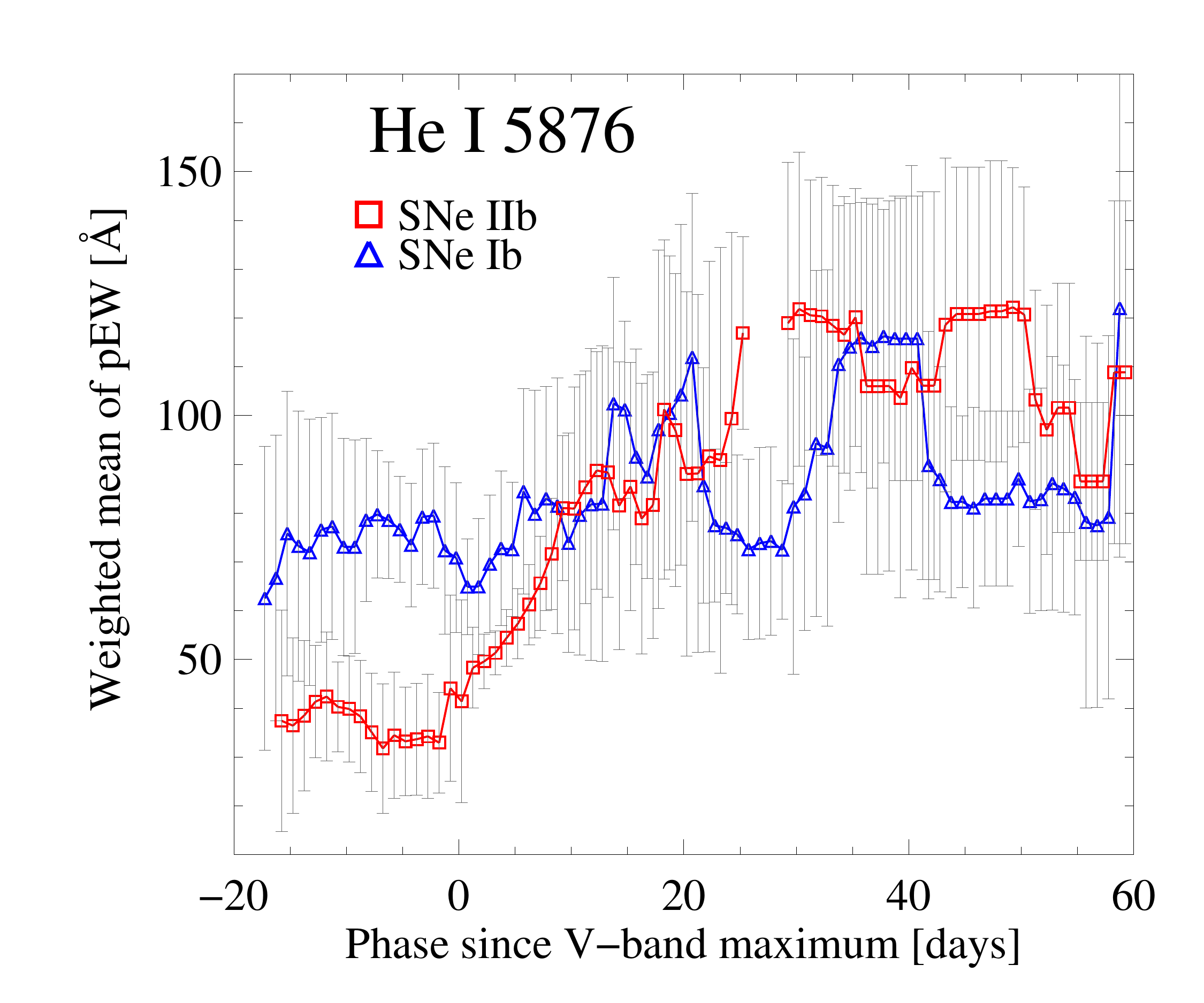}
}
\subfigure{%
\includegraphics[scale=0.4,angle=90,trim = 0mm 5mm 10mm 10mm, clip]{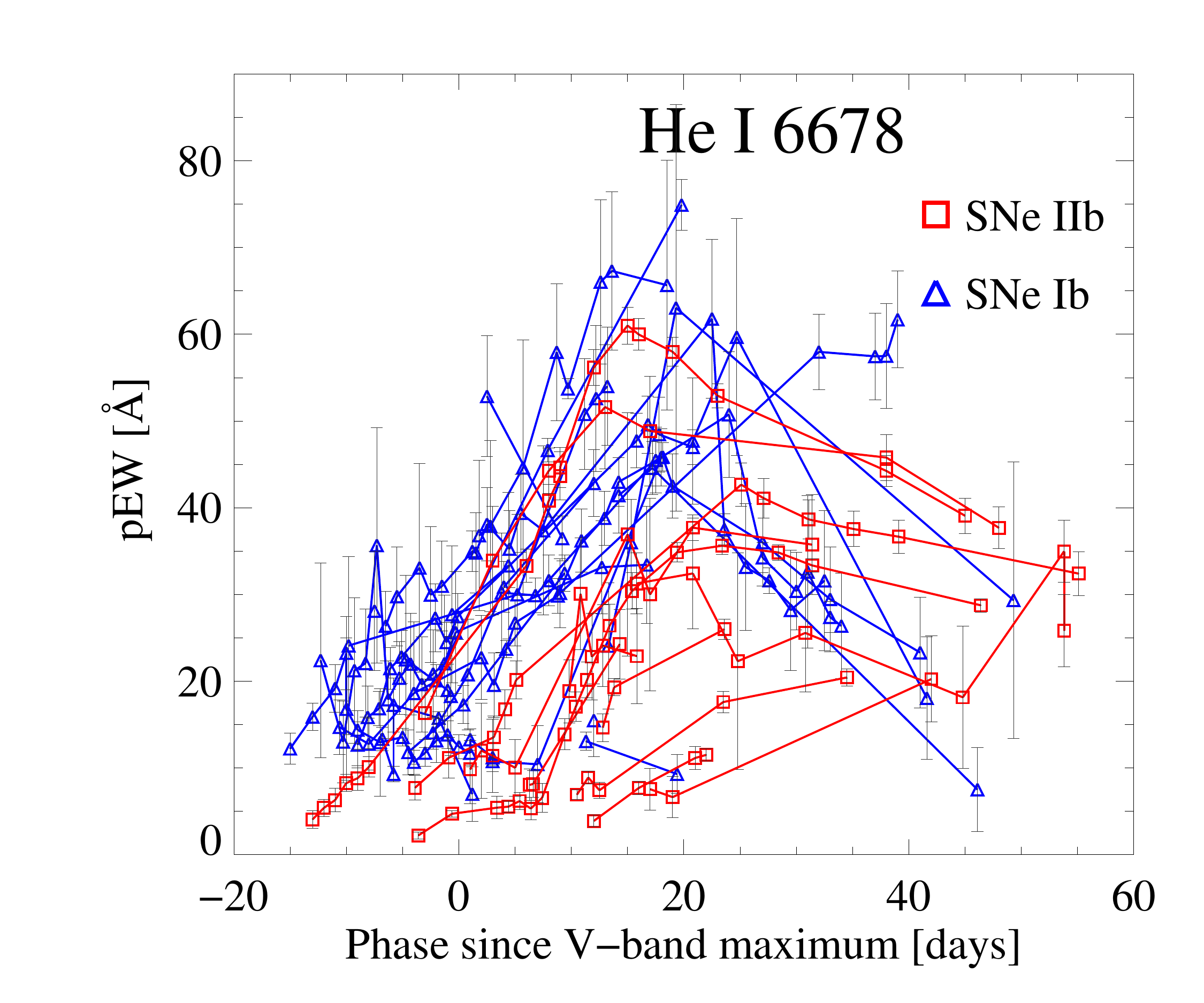}
}
\quad
\subfigure{%
\includegraphics[scale=0.4,angle=0,trim = 5mm 0mm 0mm 10mm, clip]{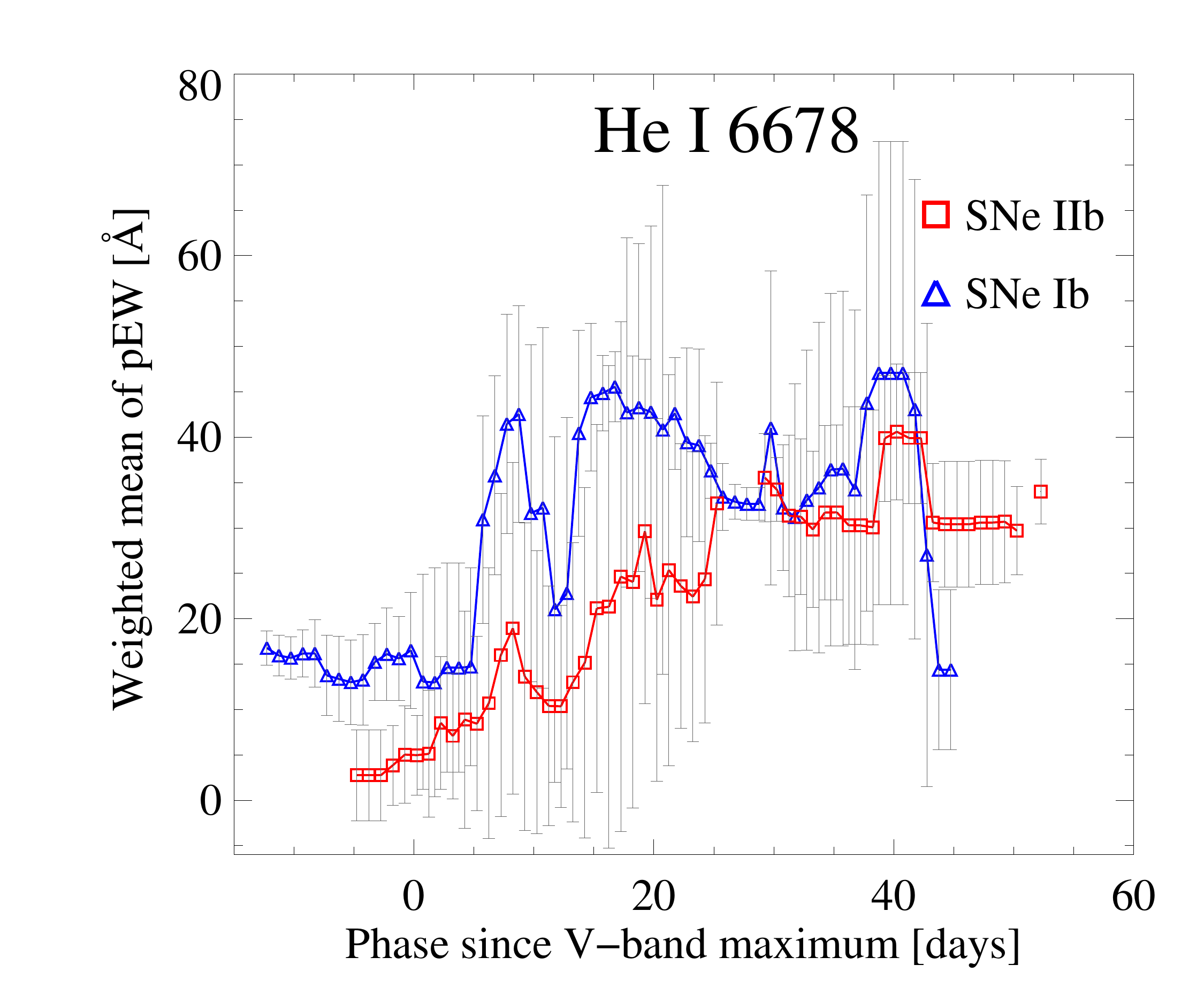}
}
\subfigure{%
\includegraphics[scale=0.4,angle=90,trim = 0mm 5mm 10mm 10mm, clip]{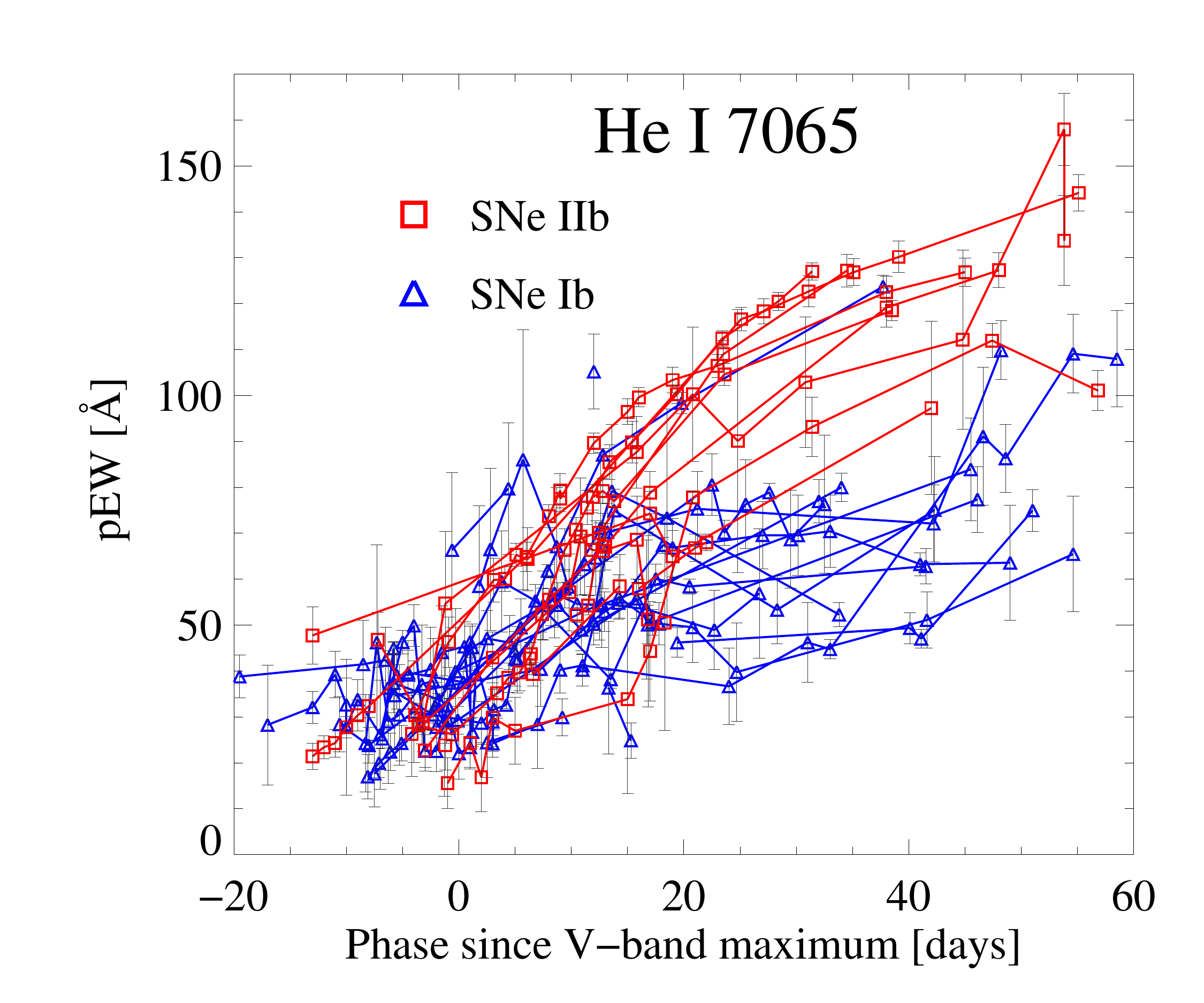}
}
\quad
\subfigure{%
\includegraphics[scale=0.4,angle=0,trim = 5mm 0mm 0mm 10mm, clip]{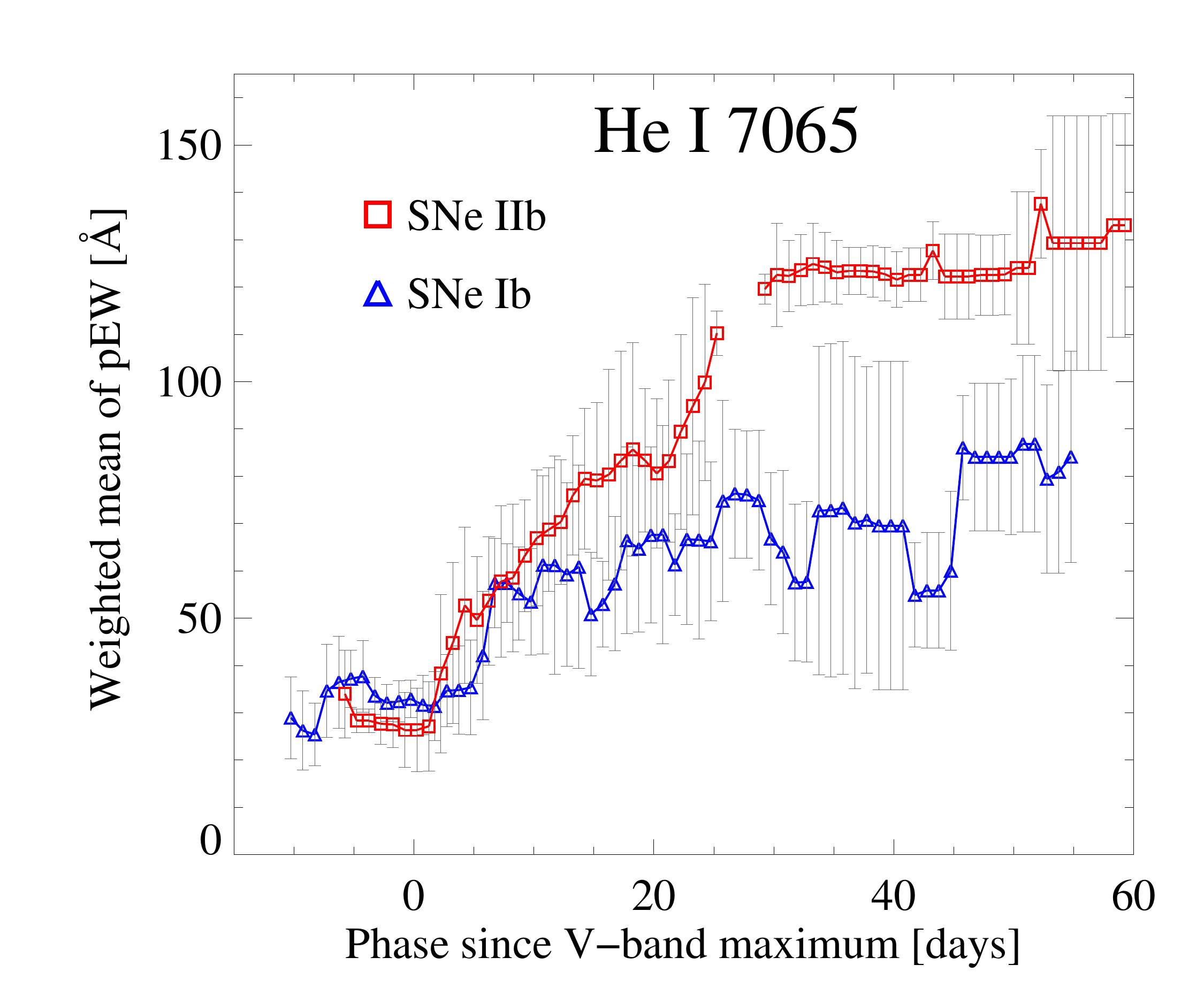}
}
\caption{The same as Figure \ref{fig_vabs_Halpha_f3}, but for pEW measurements of He I $\lambda$5876 (\textit{Top}), He I $\lambda$6678 (\textit{Middle}), and He I $\lambda$7065 (\textit{Bottom}). In the top left panel, the SN IIb that lies in the SN Ib sample at $t_{\mathrm{Vmax}}\simeq0$ day is SN 2004ff. In the middle left panel, the two SNe IIb with the strongest He I $\lambda$6678 pEW values are SNe 2008ax and 2011ei. In the bottom left panel, the SN Ib that lies in the SN IIb sample after $t_{\mathrm{Vmax}}\simeq20$ days is SN 2007Y.}
\label{fig_HeI5876_pEW_IbIIb}
\end{figure*}

%\begin{figure*}[t]
%\subfigure{
%\includegraphics[scale=0.4,angle=90,trim = 0mm 5mm 0mm 10mm, clip]{IbIIb_HeI6678_pEW.pdf}
%}
%\quad
%\subfigure{
%\includegraphics[scale=0.4,angle=0]{mean_plot_HeI6678_pEW.pdf}
%}
%\caption{The same as Figure \ref{fig_vabs_Halpha_f3}, but for measurements of He I $\lambda$6678 pEW. In the left panel, the two SNe IIb with the strongest He I $\lambda$6678 pEWs are SNe 2008ax and 2011ei.}
%\label{fig_HeI6678_pEW_IbIIb}
%\end{figure*}
%
%\begin{figure*}[t]
%\subfigure{
%\includegraphics[scale=0.4,angle=90,trim = 0mm 5mm 0mm 10mm, clip]{IbIIb_HeI7065_pEW.pdf}
%}
%\quad
%\subfigure{
%\includegraphics[scale=0.4,angle=0]{mean_plot_HeI7065_pEW.pdf}
%}
%\caption{The same as Figure \ref{fig_vabs_Halpha_f3}, but for measurements of He I $\lambda$7065 pEW. In the left panel, the SN Ib that lies in the SN IIb sample after $t_{\mathrm{Vmax}}\simeq20$ days is SN 2007Y.}
%\label{fig_HeI7065_pEW_IbIIb}
%\end{figure*}
%

In this section, we first show the temporal evolution of pEW values of He I  $\lambda\lambda\lambda$5876, 6678, 7065 in SNe IIb and in SNe Ib. Then, we compare the trends we found with the observations in \citet{matheson01}. 

The temporal pEW evolution of He I $\lambda\lambda\lambda$5876, 6678, and 7065 for SNe IIb and Ib are displayed in Figure \ref{fig_HeI5876_pEW_IbIIb}. The pEW evolution of individual SNe IIb and Ib are shown in the left panels, and the corresponding rolling weighted averages are shown in the right panels. In particular, the weighted averages at $t_{\mathrm{Vmax}}\simeq0$ day are listed in Table \ref{table_mean}. There is a wide overlap in the pEW values of He I lines between SNe IIb and SNe Ib, which indicates a continuum of physical parameters, e.g., the thickness of the helium layer, in the progenitors. The He I $\lambda$5876 pEW values are indistinguishable between SNe Ib and SNe IIb at most epochs while before $t_{\mathrm{Vmax}}\simeq0$ day, the average He I $\lambda$5876 pEW values in SNe Ib are systematically higher than those in SNe IIb although a SN IIb data point at $t_{\mathrm{Vmax}}\simeq0$ day (SN 2004ff) is among values of our SN Ib sample. For He I $\lambda$6678 pEW values, the average values are higher in SNe Ib than in SNe IIb at most phases. The two SNe IIb that have the highest He I $\lambda$6678 pEW values are SNe 2008ax and 2011ei. The He I $\lambda$7065 pEW values behave similarly in SNe Ib and SNe IIb before $t_{\mathrm{Vmax}}\simeq10$ days. However, SNe IIb show much stronger He I $\lambda$7065 than SNe Ib afterwards, although a SN Ib (SN 2007Y) lies in the SN IIb sample.

We did not find the same trends as those reported in \citet{matheson01} concerning the pEW ratios of He I lines. They claimed that for SNe Ib, He I $\lambda$5876 and $\lambda$7065 grow in strength relative to He I $\lambda$6678 over time. They chose He I $\lambda$6678 as the reference feature for two reasons. First, He I $\lambda$5876 is likely to be contaminated by Na I D and He I  $\lambda$7065 is generally weaker than He I $\lambda$6678. Thus, He I $\lambda$6678 would be the cleanest and easiest one to measure in SNe IIb and Ib. Second, He I  $\lambda$6678 arises from the singlet state whereas He I $\lambda\lambda$5876, 7065 arise from the triplet state. Thus, He I $\lambda\lambda$5876, 7065 should be tracked together and may be different from He I $\lambda$6678. Compared with the sample in \citet{matheson01}, we have a larger sample and more spectra at earlier phases. Their sample is composed of three SNe Ib (SNe 1998dt, 1999di, and 1999dn), and their data begin at $t_{\mathrm{Vmax}}\simeq10$ days; We have 21 SNe Ib (including SNe 1998dt and 1999dn), and our data begin at $t_{\mathrm{Vmax}}\simeq-10$ days. We find that for individual SNe IIb and Ib, the pEW ratios mentioned above decrease at the beginning and then stay relatively unchanged or increase slightly.

\subsection{Absorption Velocity of Fe II $\lambda$5169 in SNe IIb and Ib}
\label{Fe_vabs_IIbIb}

\begin{figure}[t]
\includegraphics[scale=0.4,angle=0,trim = 0mm 0mm 0mm 0mm, clip]{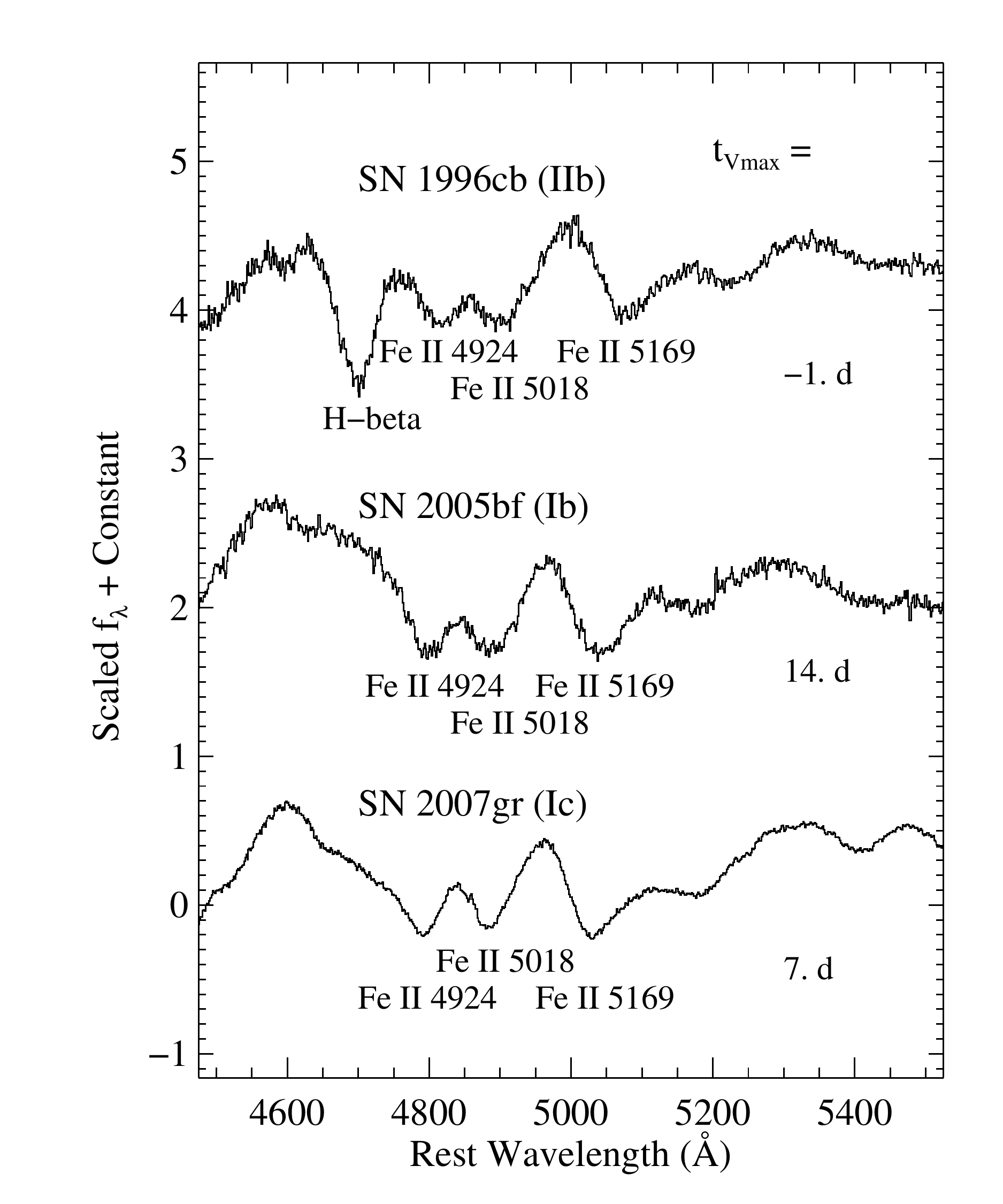}
\caption{The Fe II triplet region for spectra of three different SNe at various phases.}
\label{fig_FeII}
\end{figure}

Figure \ref{fig_FeII} shows our phenomenological way to identify Fe II $\lambda\lambda\lambda$4924, 5018, and 5169, which are the main Fe II features between 4800 \AA~and 5100 \AA. If the three lines appear blended into a ``w" feature, which is more common than the lines appearing discernibly separate, the bluer absorption component is regarded as a blend of Fe II $\lambda$4924 and 5018 and the redder absorption component as Fe II $\lambda$5169. Given that Fe II $\lambda\lambda$4924 and 5018 are more difficult to identify than Fe II $\lambda$5169, here we report the velocities of Fe II $\lambda$5169 in SNe IIb and Ib, as has been done previously in the literature \citep[e.g.,][]{branch02}.

\begin{figure*}[t]
\subfigure{
\includegraphics[scale=0.4,angle=90,trim = 0mm 5mm 0mm 10mm, clip]{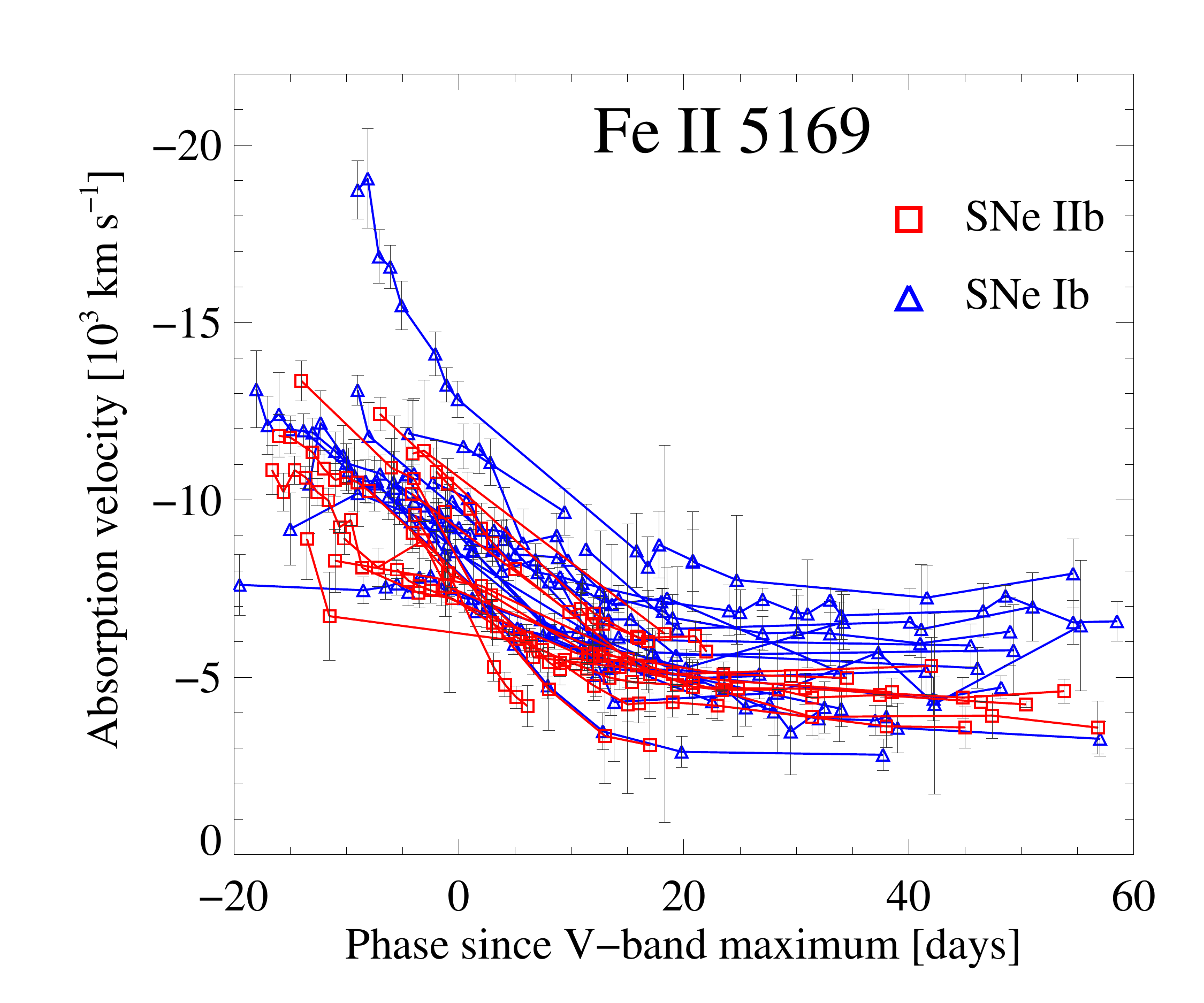}
}
\quad
\subfigure{
\includegraphics[scale=0.4,angle=90]{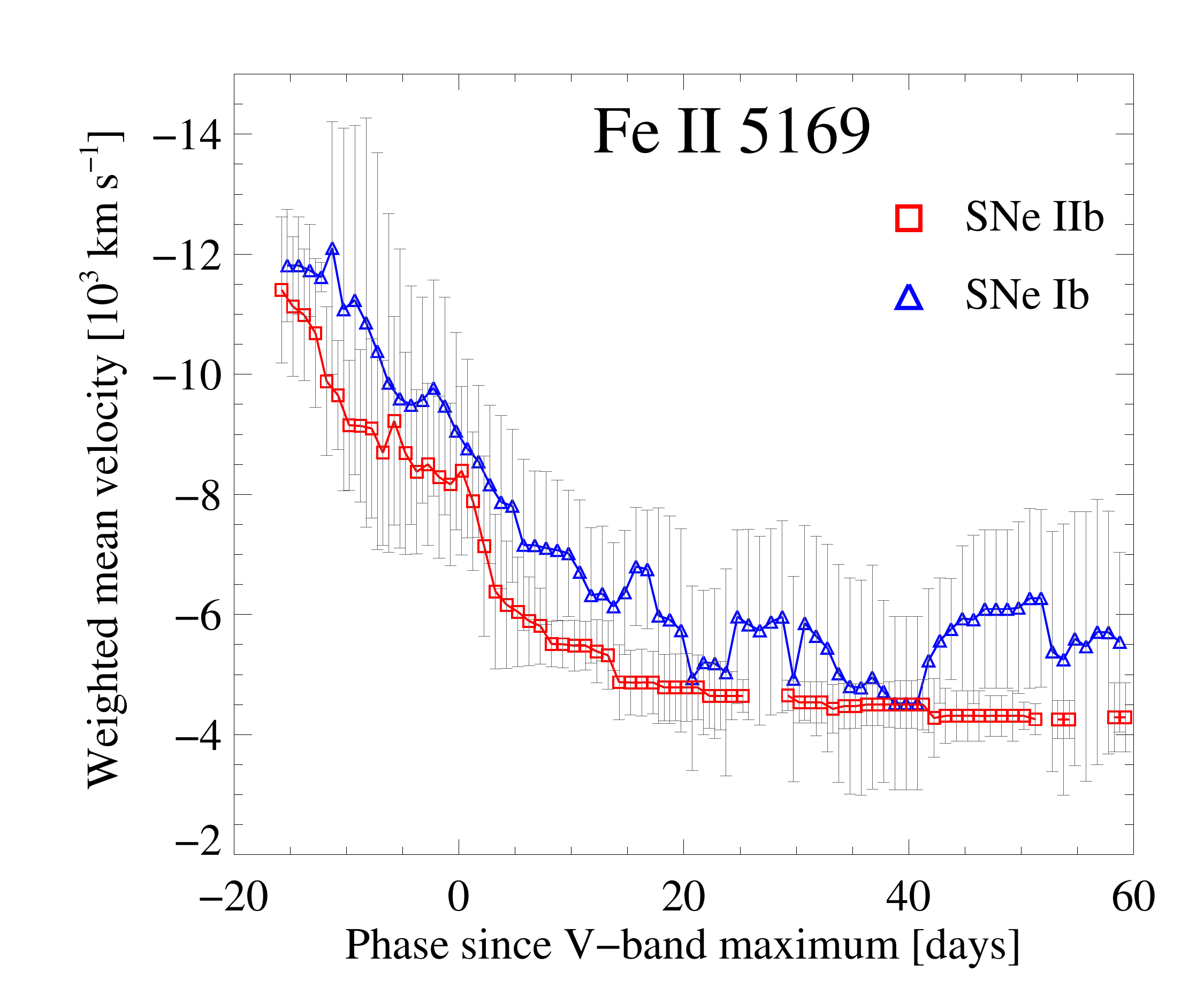}
}
\caption{The same as Figure \ref{fig_vabs_Halpha_f3}, but for measurements of Fe II $\lambda$5169 velocities.}
\label{fig_vel_FeII5169_IbIIb}
\end{figure*}

As shown in Figure \ref{fig_vel_FeII5169_IbIIb}, the weighted average velocities of Fe II $\lambda$5169 in SNe Ib are slightly higher than those in SNe IIb, which is consistent with the trends of He I velocities observed in SNe IIb and Ib. The two possible explanations are the same as those in Section \ref{sec_He_IbIIb_vel}. We also notice that the relationship between the Fe II $\lambda$5169 velocities and phase is not as tight as that claimed in \citet{branch02}. For example, at $t_{\mathrm{Vmax}}\simeq0$ day, the spread of Fe II $\lambda$5169 velocities in our SN Ib sample is $\sim$ 4000 km s$^{-1}$, while in the SN Ib sample of \citet{branch02}, it is $\sim$ 1000 km s$^{-1}$ (their figure 22).

Table \ref{table_mean} summarizes the weighted absorption velocities and pEW values for different lines in SNe IIb and Ib at $t_{\mathrm{Vmax}}\simeq0$ day.

\begin{deluxetable*}{cccccccccccccccccccc}
\tabletypesize{\scriptsize}
\tablecaption{Summary of weighted absorption velocity and pEW measurements for SNe IIb and Ib at $t_{\mathrm{Vmax}}\simeq0$ day\label{table_mean}}
\tablehead{
\colhead{SN type} &
\colhead{V$_{\mathrm{H}\alpha}$} &
\colhead{pEW$_{\mathrm{H}\alpha}$} &
\colhead{V$_{\mathrm{HeI5876}}$} &
\colhead{V$_{\mathrm{HeI6678}}$} &
\colhead{V$_{\mathrm{HeI7065}}$} &
\colhead{V$_{\mathrm{FeII5169}}$} & 
\colhead{pEW$_{\mathrm{HeI5876}}$} &
\colhead{pEW$_{\mathrm{HeI6678}}$} &
\colhead{pEW$_{\mathrm{HeI7065}}$} &\\
\colhead{} &
\colhead{(10$^3$kms$^{-1}$)} &
\colhead{(\AA)} &
\colhead{(10$^3$kms$^{-1}$)} &
\colhead{(10$^3$kms$^{-1}$)} &
\colhead{(10$^3$kms$^{-1}$)} &
\colhead{(10$^3$kms$^{-1}$)} &
\colhead{(\AA)} &
\colhead{(\AA)} &
\colhead{(\AA)} &
}
\startdata
SNe IIb &    $-12.5$ $\pm$ 0.8&       157 $\pm$ 48 &              $-$7.4$\pm1.6$  &      $-$7.4 $\pm$ 1.2 &      $-$7.9 $\pm$ 1.2  &      $-$8.4 $\pm$ 1.4 & 41$\pm$21 & 5$\pm$4 & 26$\pm$9\\
SNe Ib &   $-$17.0 $\pm$ 2.2 &       31 $\pm$ 17 &       $-$10.0 $\pm$ 3.0 &      $-$9.9 $\pm$ 3.6 &       $-$8.9 $\pm$ 2.5  &      $-$9.1 $\pm$ 1.6 & 71$\pm$15 & 17$\pm$6 & 33$\pm$4
\enddata
\tablecomments{The error is the standard deviation of data that contribute the weighted average value, which is consistent with the errors in the figures that show weighted average values.}
\end{deluxetable*}

\subsection{Are there two classes of SNe Ib?}
\label{Ib_two}

\begin{figure}[t]
\includegraphics[scale=0.4,angle=0,trim = 10mm 0mm 5mm 0mm, clip]{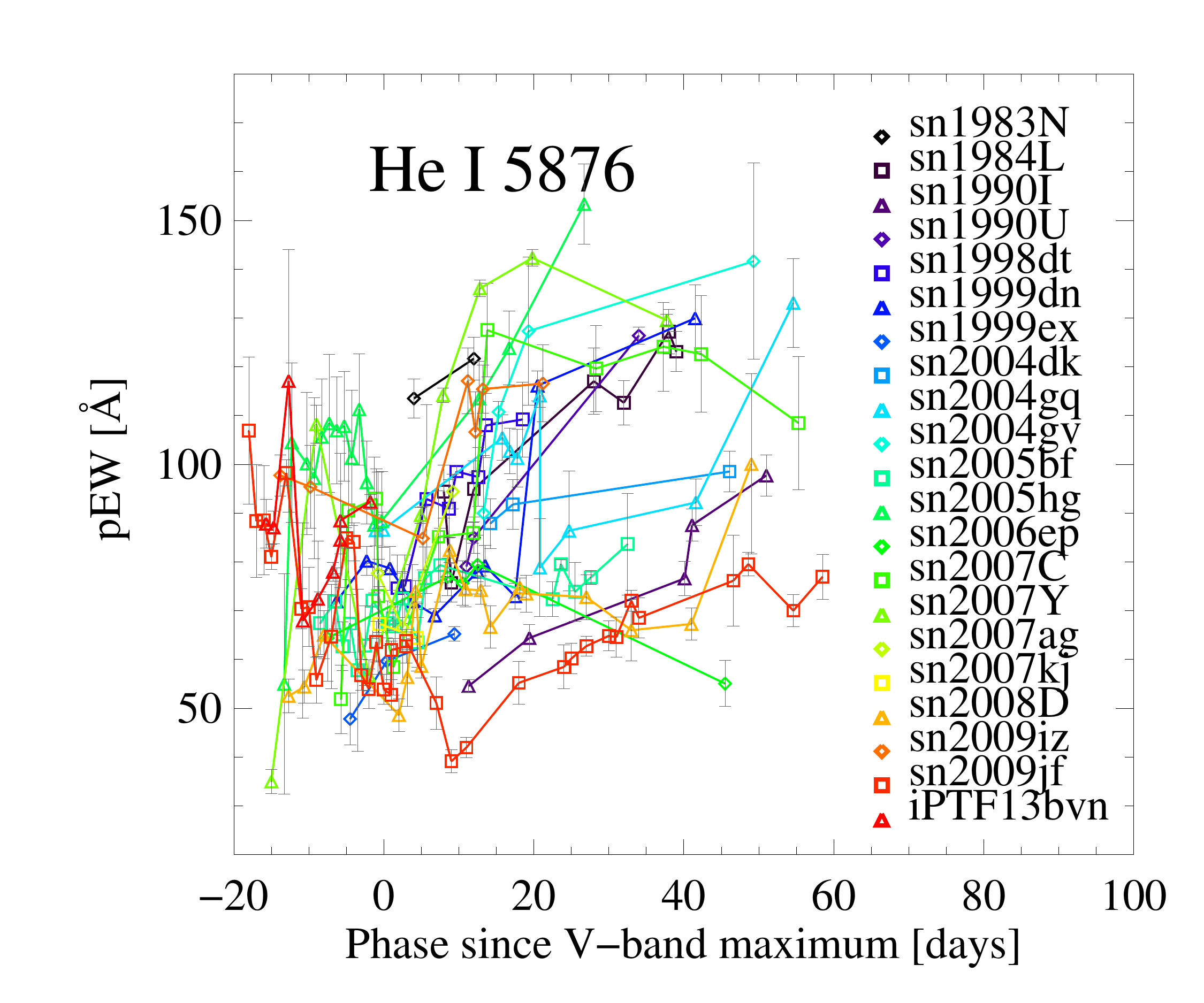}
\caption{Measurements of He I $\lambda$5875 pEW values for each SN Ib in our sample. Data points of the same SN are connected by a line.}
\label{fig_pEW_HeI5875_Ib}
\end{figure}

\citet{valenti11} was the first to observe that some SNe Ib (e.g., their SN 2009jf) show weak He I $\lambda$5876 whose strength stays roughly the same after the date of maximum while others show strong He I $\lambda$5876 that increases in strength over time. We do not observe this trend using our large dataset. As shown in Figure \ref{fig_pEW_HeI5875_Ib}, although SNe 2009jf and 2005hg seem to represent the two kinds of SNe Ib mentioned in \citet{valenti11}, other SNe Ib significantly overlap in the He I $\lambda$5876 pEW measurements around the date of maximum light and after $t_{\mathrm{Vmax}}\simeq50$ days. Moreover, no sign of two SN Ib subclasses is found based on the pEW values of He I $\lambda$$\lambda$6678 and 7065 (see the middle left and bottom left panels of Figure \ref{fig_HeI5876_pEW_IbIIb}). Thus, we do not divide SNe Ib into two classes. 

%We cannot use these pEW ratios of He I lines to differentiate SNe Ib from SNe IIb. Although mean pEW ratio of He I $\lambda$5876 to $\lambda$6678 and pEW ratio of He I $\lambda$7065 to $\lambda$6678 in SNe IIb is systematically higher than that in SNe Ib, they are heavily overlapped at all phases.  

%\subsection{Summary of spectral behaviour of SNe IIb and Ib}
%\label{sec_summary_IbIIb}

%We summarized ways to differentiate SNe IIb from SNe Ib in table \ref{table_ks_method}. The letter ``Y" means that we can use a given item to differentiate SNe Ib from SNe IIb at a given phase whereas the letter ``N" means that we cannot. The letter ``N/A" means that we cannot tell due to a lack of data points. Different ways can be used to cross check. If a SN IIb/Ib is not discovered before phase $\sim20$ days, we can use pEW of He I $\lambda$7065 or H$\alpha$ to correctly classify it. Although pEW ratio of He I lines to H$\alpha$ can be used to do classification as well, we did not list them to table \ref{table_ks_method} because it is largely driven by pEW of H$\alpha$. It seems like that pEW of H$\alpha$ is a good indicator to differentiate SNe IIb from SNe Ib.

\section{The Helium problem for SNe Ic}
\label{He_pro}

While SNe Ic are defined by the apparent lack of He I lines in their spectra, there is a long-standing debate whether there is weak He or hidden He in SN Ic spectra. For the former, as discussed below in Section \ref{sec_He_Ic}, some authors have claimed the detection of weak He I $\lambda$5876 or weak He I $\lambda$10830, while others disagree about these identifications. For the latter, as discussed in Section \ref{sec_helium_problem}, some models suggest that helium may be present in SN Ic progenitors, but that they may not be excited via non-thermal processes due to insufficient mixing of $^{56}$Ni \citep[e.g.,][]{dessart12}, while others argue that the SN Ic progenitors are helium-free \citep[e.g.,][]{hachinger12, frey13, cano14a}. Resolving whether or not the progenitor stars of SNe Ic are truly helium-free has large implications for a number of fields, including stellar evolution of massive stars \citep[e.g.,][]{yoon05, yoon10, langer12, yoon12} and the SN-GRB connection \citep[e.g.,][]{modjaz15}.

\subsection{Are there weak He I lines in spectra of SNe Ic?}
\label{sec_He_Ic}

\begin{figure}[t]
\includegraphics[scale=0.5,angle=0]{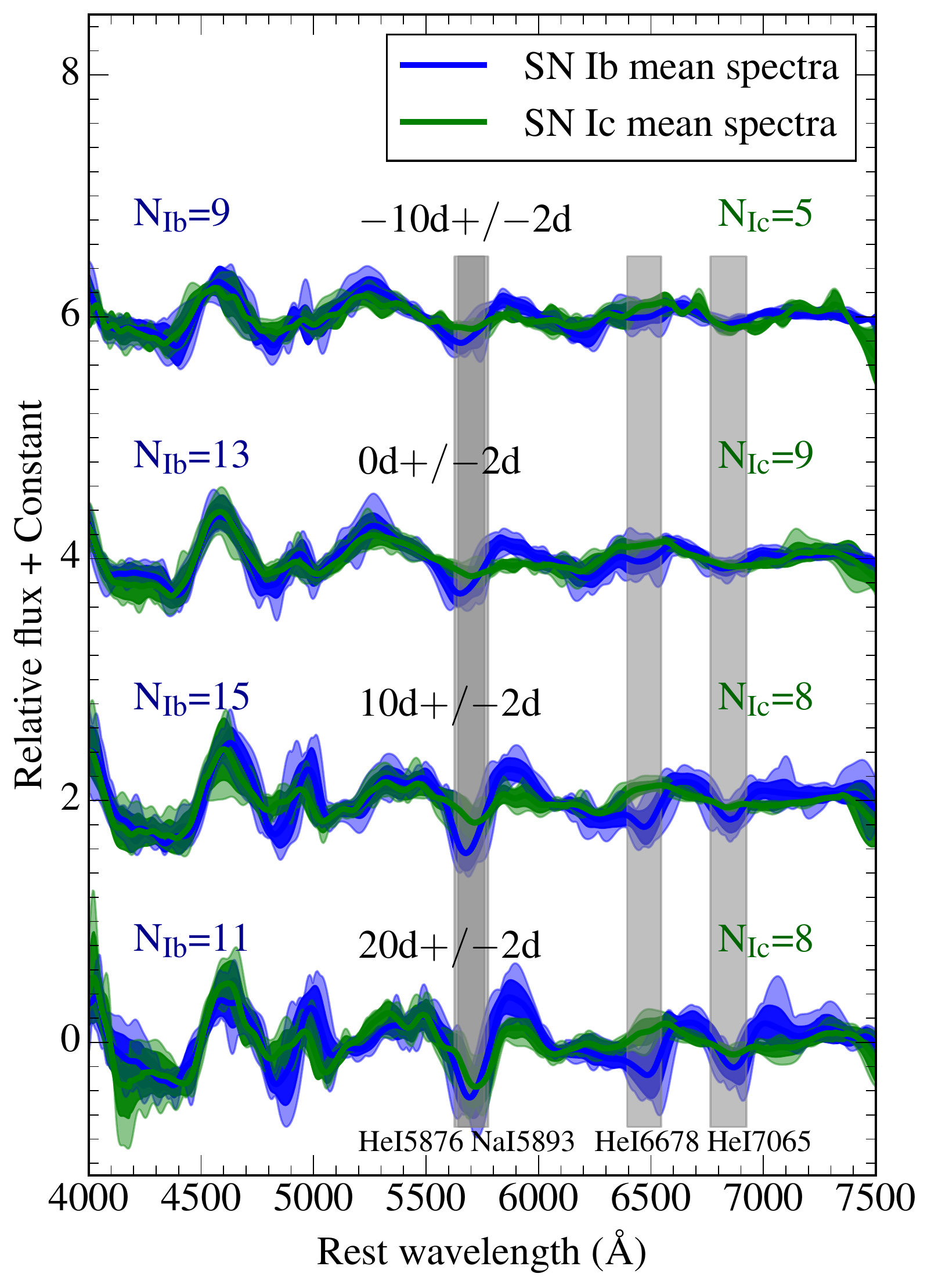}
\caption{The same as Figure \ref{fig_mean_IIbIb_h}, but for SNe Ib (blue) and SNe Ic (green). N$_{\mathrm{Ib}}$ represents the number of spectra or SNe that go into each mean spectrum of SNe Ib, and N$_{\mathrm{Ic}}$ represents the number for SNe Ic. The gray vertical bands indicate the expected positions of He I $\lambda$5876, Na I D, He I $\lambda\lambda$ 6678, and 7065 at velocities of $-$6000 km s$^{-1}$ to $-$13000 km s$^{-1}$.}
\label{fig_mean_IcIb_he}
\end{figure}

By definition, there are no conspicuous He I lines in the spectra of SNe Ic. However, it is instructive to attempt to detect weak He I lines in spectra of SNe Ic to see if a spectroscopic link with SNe Ib could be established. This link would be a constraint on progenitor models. 

Prior works investigated the presence of He I lines in SN Ic spectra by comparing the velocity evolution (as defined from line identification) of one potential He I line to that of another potential He I line \citep{clocchiatti96, matheson01}, as well as comparing synthetic spectra---calculated via SYNOW or a Monte Carlo transport spectral synthesis code \citep{abbott85, mazzali93, lucy99, mazzali00}---to observed spectra \citep{elmhamdi06, sauer06}. However, no agreement on whether there are He I lines in the spectra of SNe Ic has been reached. In particular, the identifications of  He I $\lambda$5876 and He I $\lambda$10830 in SNe Ic are highly debated \citep{clocchiatti96, matheson01, elmhamdi06, sauer06, dessart15}

In this work, we use a statistical approach to explore the question of whether the spectra of SNe Ic show weak He I lines, just as we explored the potential presence of weak H lines in SN Ib spectra (Section \ref{sec_H_Ib}). Using the same approach, we first identified He I lines in the mean spectra of SNe Ib. Then we searched for He I lines at comparable velocities (i.e., blueshift) in the mean spectra of SNe Ic. 

In Figure \ref{fig_mean_IcIb_he}, we show the mean spectra and their corresponding standard deviations of SNe Ib and SNe Ic at various phase ranges. In the mean spectra of SNe Ib, He I $\lambda$5876 is visible after $t_{\mathrm{Vmax}}=-10$ and He I $\lambda\lambda$6678, 7065 become visible starting at $t_{\mathrm{Vmax}}=0$. At the expected positions of He I $\lambda$5876, the mean spectra of SNe Ic show a broad feature but no obvious absorption feature at $t_{\mathrm{Vmax}}\simeq-10$ and 0 days, a weak absorption feature at $t_{\mathrm{Vmax}}\simeq10$ days, and a strong absorption feature at $t_{\mathrm{Vmax}}\simeq20$ days. However, this feature, which one might identify as He I $\lambda$5876, could be due to other elements such as Na I D, as claimed by many authors \citep[e.g.,][]{branch02, elmhamdi06, kumar13, marion14} who use spectral synthesis calculations. Most importantly, the mean spectra of SNe Ic show no convincing signs of He I $\lambda\lambda$6678 and 7065 either. Thus, we conclude that no obvious He I lines are detected in the mean spectra of SNe Ic. Another interesting observation is that SN Ib spectra and SN Ic spectra differ at wavelengths other than the expected positions of He I lines, e.g., the Fe II feature at $\sim4900$ \AA~in SNe Ib is deeper than that in SNe Ic.

In summary, there has been no agreement on whether there are He I lines in the spectra of SNe Ic. Here, we find that there is no convincing sign of He I $\lambda\lambda$6678 and 7065 in the mean spectra of SNe Ic, and while there is a trough at the expected position of He I $\lambda$5876, it could be due to Na I D as claimed by many authors \citep[e.g.,][]{branch02, elmhamdi06, kumar13, marion14}. Thus, in this study, we will not identify He I $\lambda$5876 in our SN Ic sample. We suggest the use of non-LTE codes that properly treat non-thermal excitations to explore all our spectra for the presence of He I lines. We also suggest that for a SN Ib/c to be truly identified as a SN Ib, a very strong He I $\lambda$5876 absorption feature needs to be detected before $t_{\mathrm{Vmax}}\simeq0$ or the three optical He I lines need to be detected in the same spectrum at $0<t_{\mathrm{Vmax}}<40$ days.
 
%\begin{figure*}[!ht]
%\subfigure[Ic]{%
%\includegraphics[scale=0.4,angle=0]{Ic_HeI.pdf}
%\label{fig_spec_HeI5876_Ic}}
%\quad
%\subfigure[Ib]{%
%\includegraphics[scale=0.4,angle=0]{Ib_HeI.pdf}
%\label{fig_spec_HeI5876_Ib}}
%%
%\caption{Spectra of different Ib/c SNe around maximum. Dotted lines represent expected positions of He I lines at a velocity between -15000 km s$^{-1}$ and -19000 km s$^{-1}$. Dashed lines represents expected positions of He I lines at velocities of -5000 km s$^{-1}$ to -10000 km s$^{-1}$.}
%\label{fig_spec_HeI5876_IcIb}
%\end{figure*}

%Figure~\ref{fig_spec_HeI5876_Ic} shows spectra of different Ic SNe around maximum. Dotted lines represent expected positions of He I lines with a velocity between -15000 km s$^{-1}$ and -19000 km s$^{-1}$. Dashed lines represents expected positions of He I lines with a velocity between -5000 km s$^{-1}$ and -9000 km s$^{-1}$. sn1994I: low He 5, high He 6. sn2004aw: high He 6. sn2005az: no obvious He features. sn2005ek: no obvious He features. sn2005eo: similar to sn1994I. sn2005mf: similar to sn1994I. To compare, sn2007C is a little weird. sn2009er: high He. other 8: quite normal and consistent low He lines.

%The lack of He I lines in SNe Ic may result from a variety of causes: a genuine helium deficiency; strongly asymmetric mixing; weak mixing; or a more massive, perhaps single, progenitor characterized by a larger oxygen-rich core.

\subsection{Is There Helium in Progenitors of SNe Ic?}
\label{sec_helium_problem}

\begin{figure*}[!ht]
\subfigure{%
\includegraphics[scale=0.4,angle=90,trim = 0mm 5mm 0mm 10mm, clip]{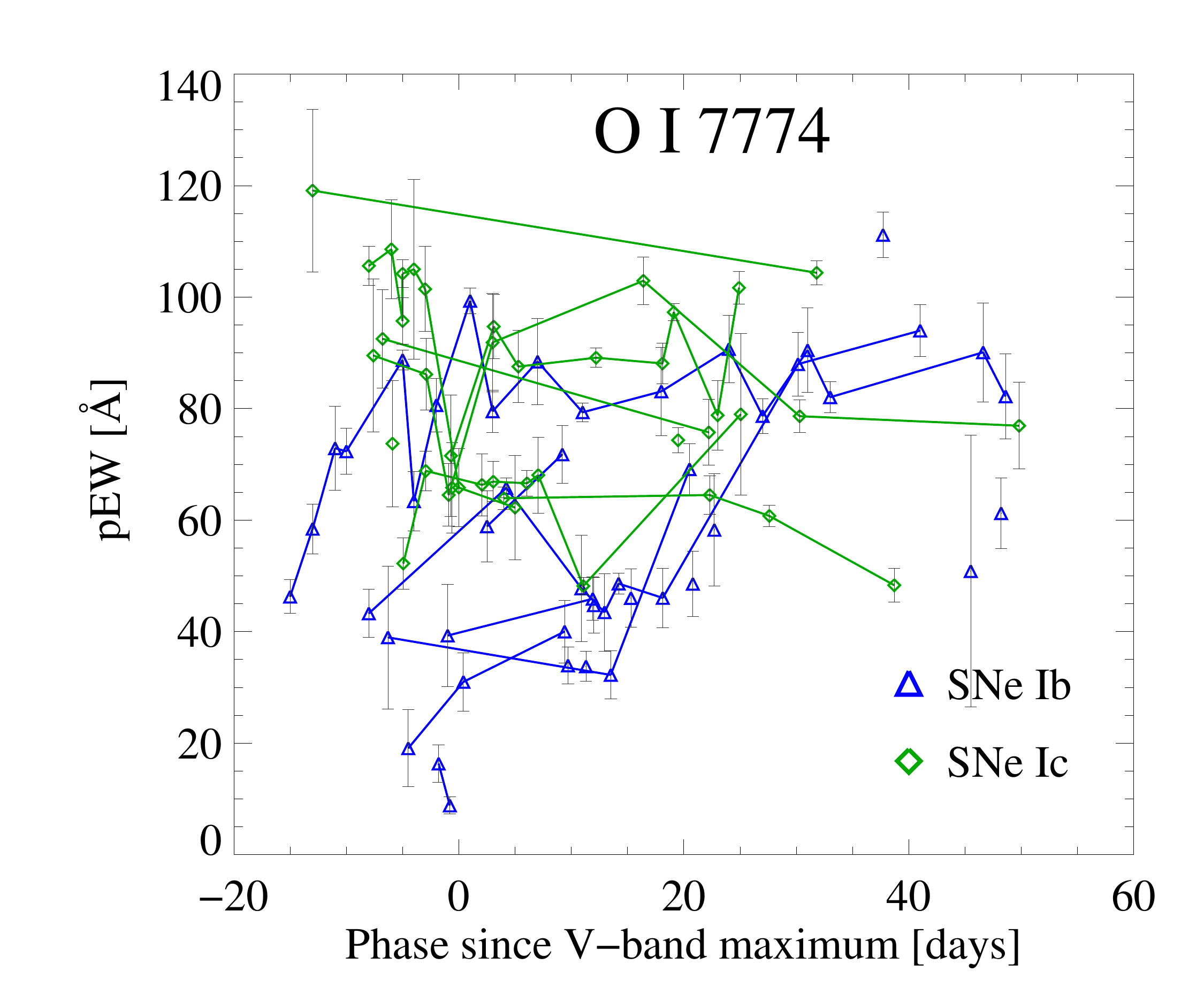}
}
\quad
\subfigure{%
\includegraphics[scale=0.4,angle=0]{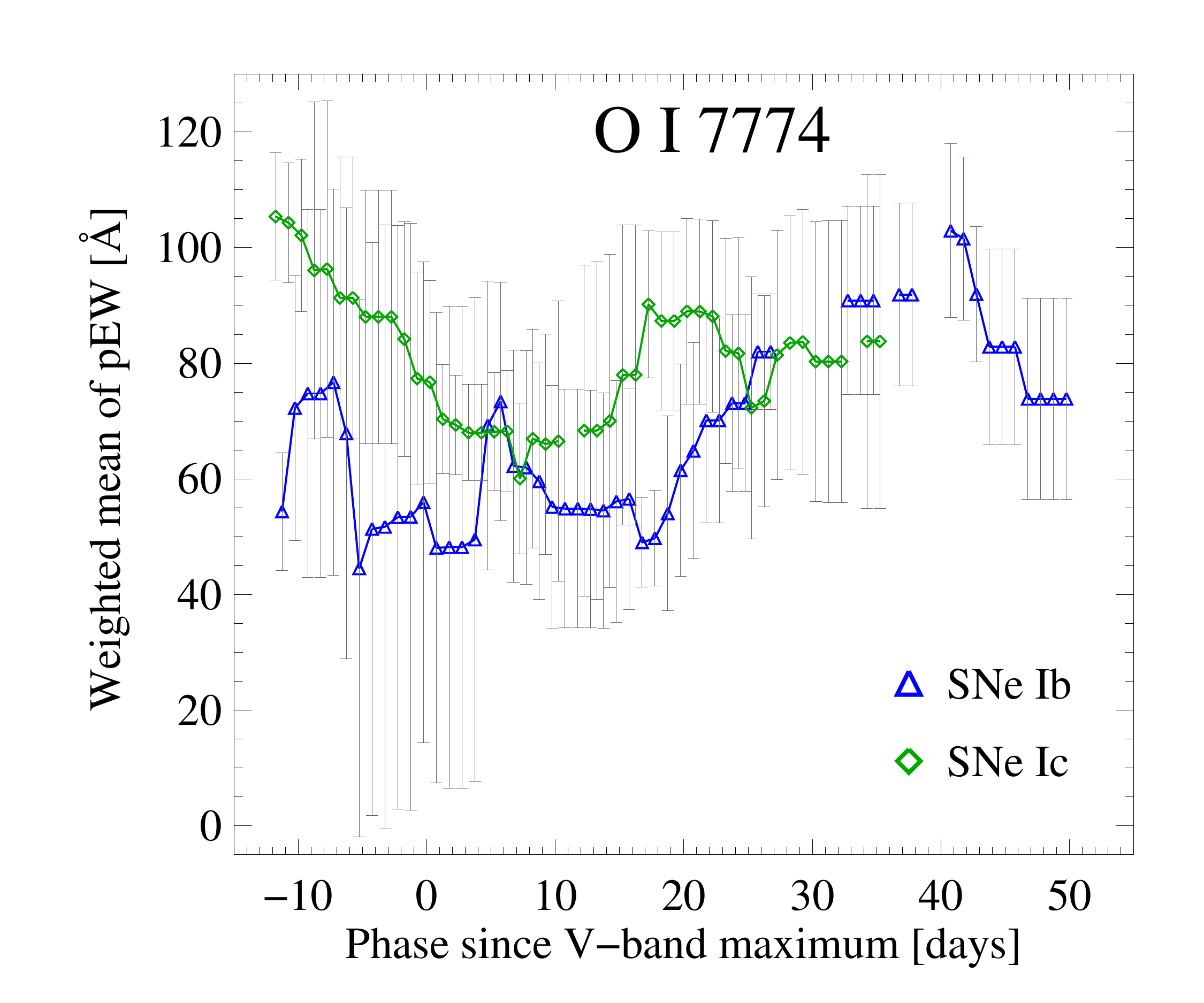}
}
\caption{The same as Figure \ref{fig_vabs_Halpha_f3}, but for measurements of O I $\lambda$7774 pEW for SNe Ib (blue triangles) and SNe Ic (green diamonds). In the left panel, the SN Ib that falls within the SN Ic sample at $t_{\mathrm{Vmax}}\simeq0$ day is SN 2009jf.}
\label{fig_OI7774}
\end{figure*}

\begin{figure*}[t]
\subfigure{%
\includegraphics[scale=0.4,angle=90,trim = 0mm 0mm 0mm 10mm, clip]{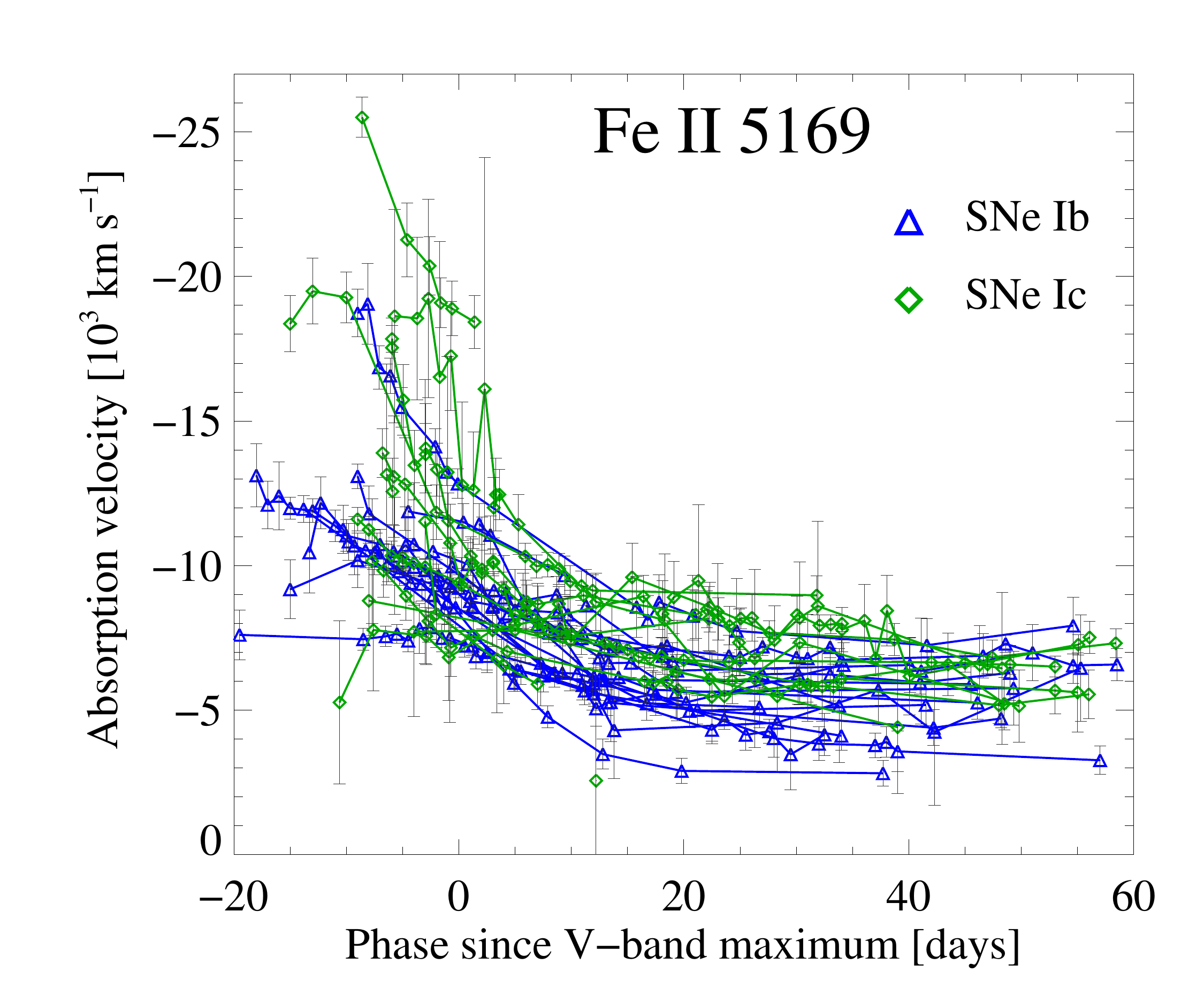}}
\quad
\subfigure{%
\includegraphics[scale=0.4,angle=90]{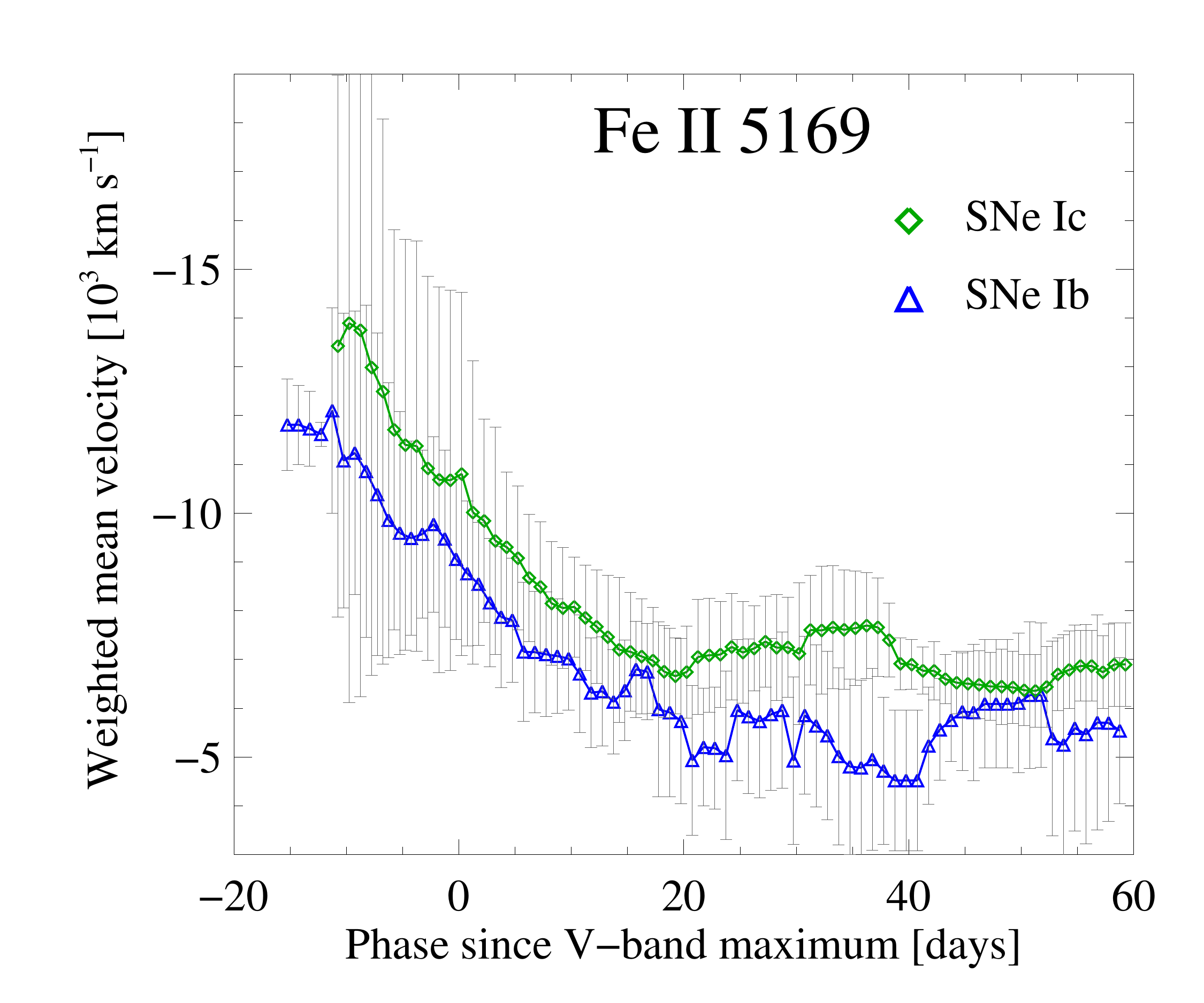}}
\quad
\subfigure{%
\includegraphics[scale=0.4,angle=90,trim = 0mm 0mm 0mm 10mm, clip]{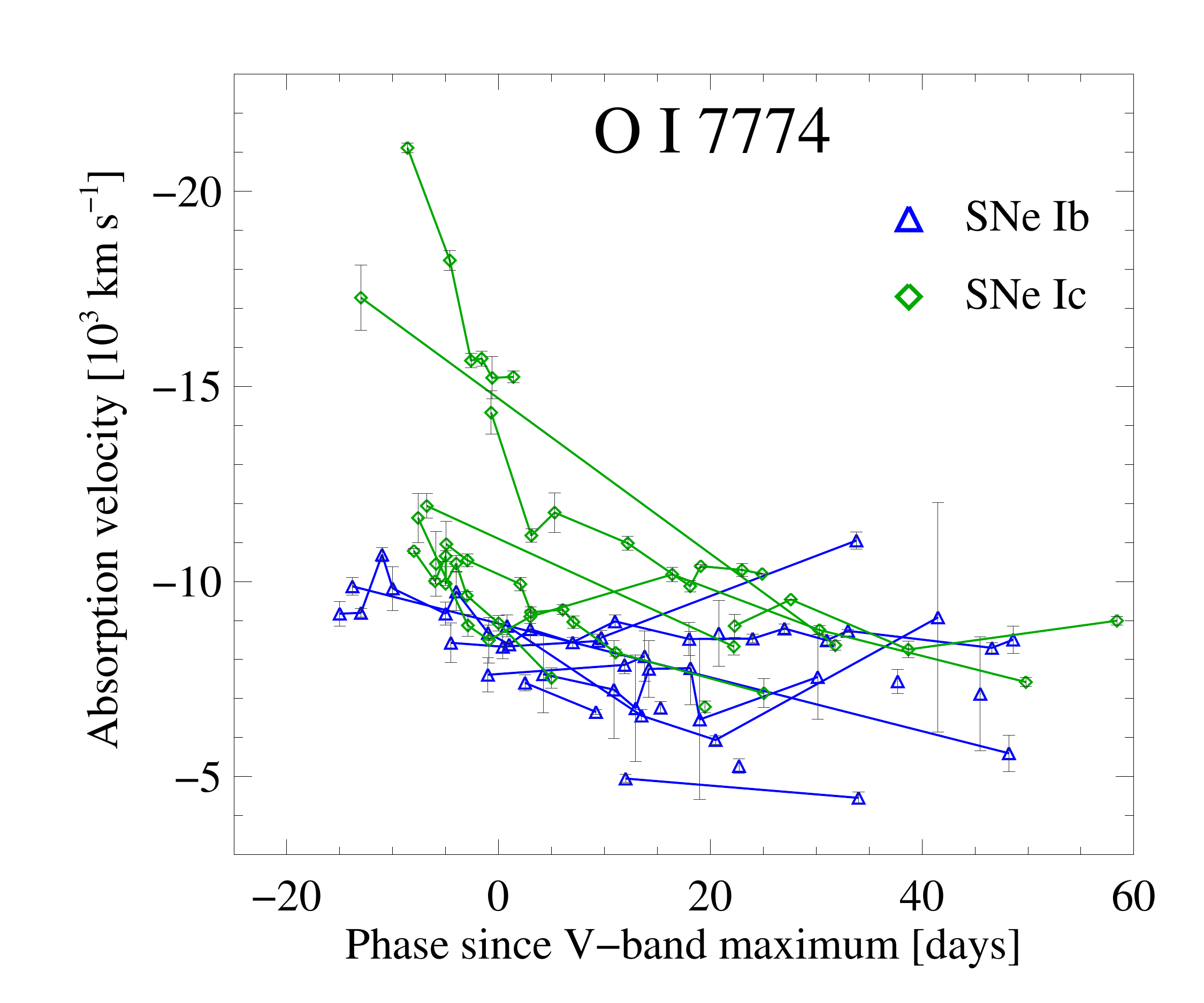}}
\quad
\subfigure{%
\includegraphics[scale=0.4,angle=90]{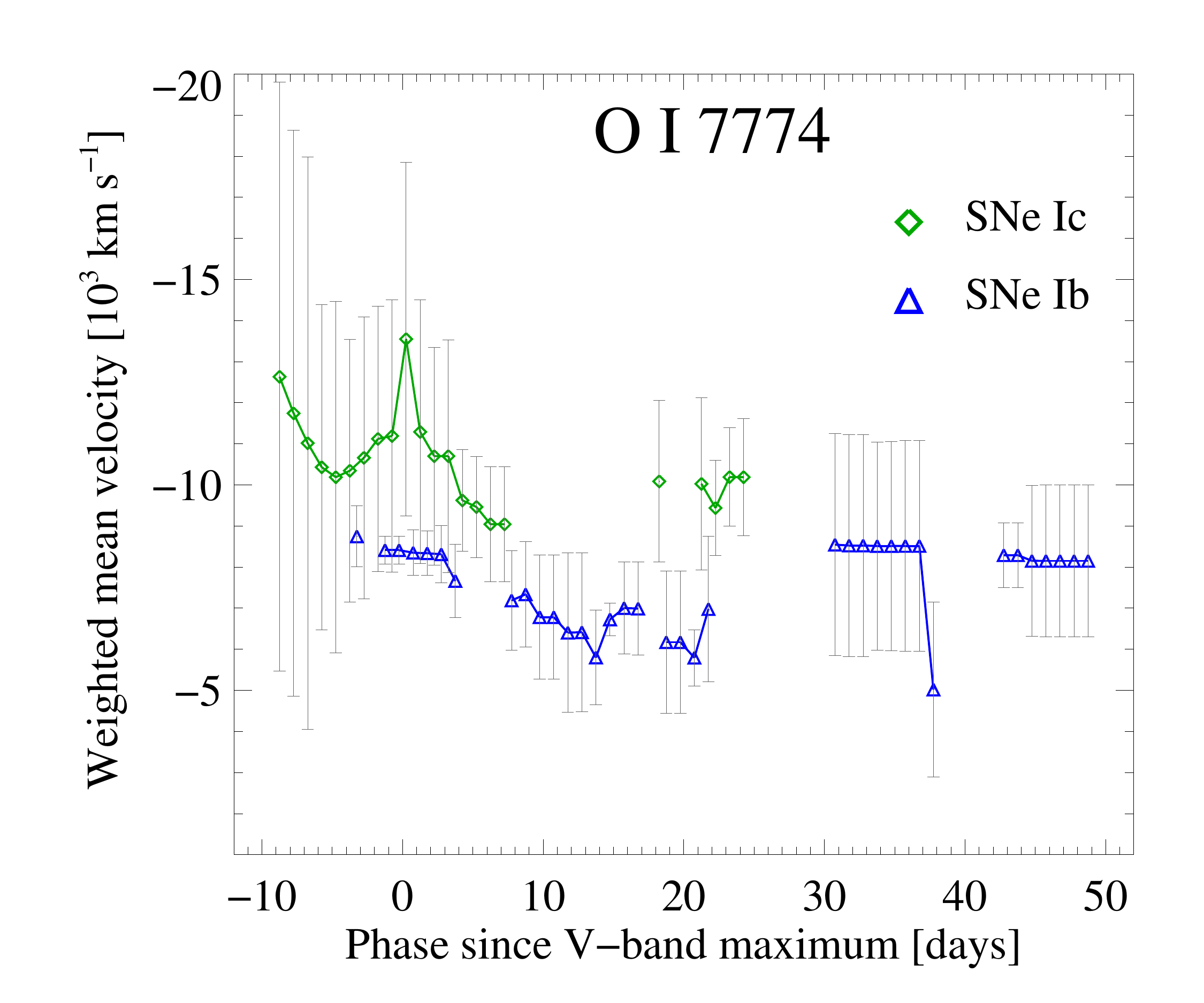}}

\caption{The same as Figure \ref{fig_vabs_Halpha_f3}, but for measurements of Fe II $\lambda$5169 velocities (\textit{Top}) and O I $\lambda$ 7774 velocities (\textit{Bottom}) for SNe Ib (blue triangles) and SNe Ic (green diamonds).}
\label{fig_vel_FeII5169_f2&f3}
\end{figure*}

While in the above section, we found that there is no convincing sign of weak He I lines in SN Ic spectra, it is still debated whether the absence of He I lines in optical spectra necessarily indicates the absence of helium in the ejecta. \citet{dessart12} argued that the progenitors of SNe Ic can have helium since it can be hidden by a low level of mixing. \citet{frey13}, on the other hand, argued that SN Ic progenitors are helium free. 

Using non-LTE radiative transfer calculations, \citet{dessart12} calculated the spectra produced by the explosions of two binary-star models with initial masses of 18 and 25 M$_{\astrosun}$ on the main sequence. They found that the same 18 M$_{\astrosun}$ model can reproduce spectra of SNe IIb, SNe IIc,\footnote{As explained in \citet{dessart12}, SNe IIc refer to SNe that show a strong H$\alpha$ line initially but then show no H I, He I, or Si II in the optical after $t_{\mathrm{Vmax}}\simeq15$ days. This SN subtype has not been observed.} and SNe Ib, depending on the $^{56}$Ni mixing level (with SNe Ib having the highest level). Similarly, they found that the same 25 M$_{\astrosun}$ model can reproduce the spectra of SNe Ic, if there is no $^{56}$Ni mixing or weak mixing, and can reproduce the spectra of SNe Ib, if there is enhanced $^{56}$Ni mixing. Their results indicate that SNe Ic and SNe Ib could have the same progenitors but different amounts of mixing. If that is the case, \citet{dessart12} predicted that SNe Ib should have redder colors, broader line profiles, and higher photospheric velocities than SNe Ic (see their figure 12). 

In contrast, using a new mixing algorithm that is based on a physical analysis of 3D hydrodynamic simulations of convection, \citet{frey13} argued that helium is absent in the progenitors of SNe Ic because most of it is burned to oxygen. They used the TYCHO code that had been updated with this new mixing algorithm in three dimensions to calculate stellar structure of four progenitor models with zero-age main sequence masses of 15, 21, 23, and 27 M$_{\astrosun}$, respectively. They found that the stars had a hydrogen layer preceding their explosions, but much of the helium was brought into deeper and hotter layers due to enhanced mixing, and finally burned to oxygen. Assuming binary interactions can remove the hydrogen layer but not the helium layer, these stars would explode as SNe Ib/c. As shown in their figure 1, only the star with an initial mass of 15 M$_{\astrosun}$ had a normal helium shell (which is also the outermost layer) before the explosion. With the increase of initial mass, the fraction of helium in the outermost layer decreases. For the star with an initial mass of 23 M$_{\astrosun}$, the helium faction was below 15\% in the outermost layer, which had a total mass of 2 M$_{\astrosun}$. In contrast, applying the classical mixing length theory to the same star resulted in 90\% helium in the outermost layer that had the same total mass of 2 M$_{\astrosun}$ \citep{woosley02}.  Using an Eulerian radiation-hydrodynamics code called RAGE (Radiation Adaptive Grid Eulerian) and the SPECTRUM code to calculate spectra and light curves, \citet{frey13} found that their models exhibited stronger oxygen and weaker helium lines than models based on the classical mixing length theory in \citet{woosley02}, mainly because the helium in their models was burned into oxygen. 

In order to test the predictions of \citet{dessart12} and \citet{frey13}, which represent two competing models, we measured the strength and velocity of O I $\lambda$7774 and the velocity of Fe II $\lambda$5169 in SNe Ib and Ic, since each of these models predicts a different behavior of one or all observables.

\subsubsection{Strength of O I $\lambda$7774}
\label{sec_OI7774}

As discussed above, \citet{dessart12} argued that if the progenitors of SNe Ib and SNe Ic have the same composition, but different $^{56}$Ni mixing levels (i.e., they both have the helium layer but SNe Ib have a higher $^{56}$Ni mixing level than SNe Ic), then SNe Ib will have stronger lines than SNe Ic. \citet{frey13} found that their models---which have adapted a new mixing algorithm---show stronger oxygen and weaker helium lines than models based on the classical mixing length theory in \citet{woosley02}. In this section, we choose to measure the strength of the O I $\lambda$7774 line to test predictions made by both papers.\footnote{The strength of the Fe II $\lambda$5169 line is also an ideal indicator to test the predictions of \citet{dessart12}, since iron should have a uniform mass fraction through most of the ejecta during the photospheric phase. However, the Fe II $\lambda$5169 strength is not easy to measure since the local continuum is difficult to define due to the appreciable difference between the height of the two local maxima around the absorption line.} 

Another reason to measure the strength of O I $\lambda$7774 is to compare it with the observation of \citet{matheson01} that the line is stronger in SNe Ic than in SNe Ib. \citet{matheson01} used fractional line depth (i.e., the line depth with respect to the corresponding continuum) to quantify the strength of absorption features. They found that the fractional line depth of O I $\lambda$7774 increases from SNe Ib to SNe Ic. They argued that if the helium envelope dilutes the strength of the oxygen line, this increase in strength of O I $\lambda$7774 could indicate a decreasing envelope mass from SNe Ib to SNe Ic. We revisit this question, since compared to the dataset in \citet{matheson01}, we have in our sample twice and three times more SNe Ib and SNe Ic,\footnote{We only include the subset of SNe Ib and Ic whose spectra cover the O I $\lambda$7774 absorption feature.} respectively. Moreover, we use uncertainty arrays of spectra to calculate the error bars of our measurements, whereas \citet{matheson01} estimated the error bars of their measurements to be 10\% or 20\% of their corresponding measured values. 

As discussed in Section \ref{vel_pEW_measure}, we use pEW to quantify strength of absorption features. The O I $\lambda$7774 pEW values of individual SNe and the corresponding rolling weighted averages are displayed in Figure~\ref{fig_OI7774}. Although there is a wide overlap in the pEW values between the two SN subtypes, on average, SNe Ic have stronger O I $\lambda$7774 than SNe Ib from $t_{\mathrm{Vmax}}\simeq-10$ days to $t_{\mathrm{Vmax}}\simeq25$ days. At $t_{\mathrm{Vmax}}\simeq0$, the pEW values of O I $\lambda$7774 for individual SNe Ic range from 60 \AA~to 110 \AA, while the pEW values range from 10 \AA~to 60 \AA~for individual SNe Ib, except for SN 2009jf, which has pEW values within the range of SNe Ic. We agree with the conclusion in \citet{matheson01} that there is an increase in strength of O I $\lambda$7774 from SNe Ib to SNe Ic in terms of average values, while having used a larger sample than \citet{matheson01} did. Our observations support the predictions of \citet{frey13} that SNe Ic have stronger O I $\lambda$7774 than SNe Ib, though the new mixing algorithm used in \citet{frey13} needs to be verified. However, under certain assumptions, our observations contradict predictions made by \citet{dessart12} that SNe Ib have stronger lines than SNe Ic, if the only difference between SNe Ib and SNe Ic had been the level of mixing. This indicates that progenitors of SNe Ic may not contain helium before the explosion, i.e., the mixing is not the only difference between SNe Ib and SNe Ic, as also supported by our Fe II $\lambda$5169 and O I $\lambda$7774 velocity measurements (see the section below). 

We note that if the progenitors of SNe Ic have larger core masses than those of SNe Ib, SNe Ic could have helium in their envelopes that will not be excited (i.e., they could have ``hidden" helium), and at the same time, show stronger O I $\lambda$7774 than SNe Ib, since even for efficient mixing, a star with a more massive CO core will not be able to produce a SN Ib because too little $^{56}$Ni will be mixed into the outer layers of the ejecta \citep{dessart15}. However, various works have shown that SNe Ib and SNe Ic have similar ejecta masses \citep[e.g.,][Bianco et al. in prep]{drout11, lyman14}, indicating that SNe Ib and SNe Ic may have comparable CO cores before explosion. Thus, our observation that SNe Ic have stronger O I $\lambda$7774 than SNe Ib may not be due to a more massive CO core in SN Ic progenitors. We conclude that progenitors of SNe Ic may not contain helium before the explosion. Moreover, \citet{hachinger12} show that based on their non-LTE radiative transfer models, SNe Ic cannot hide more than 0.06 M$_{\astrosun}$ of helium. In the next section, we will explore this topic further with another observational indicator.

\subsubsection{Absorption Velocity of Fe II $\lambda$5169 and O I $\lambda$7774}
\label{sec_FeII5169}

Using non-LTE radiative transfer calculations, \citet{dessart12} argued that if SNe Ib and SNe Ic have the same progenitors but SNe Ib have a higher mixing level, (i.e., the progenitors of SNe Ic have the helium layers but insufficient $^{56}$Ni mixing), SNe Ib will have higher photospheric velocities than SNe Ic. That is because with more mixing, there will be more opacity due to metals, thus the photosphere, where the optical depth is $\tau=2/3$, will be farther out. We test this prediction by measuring the absorption velocity of Fe II $\lambda$5169 and O I $\lambda$7774 in a statistically significant set of SNe Ib and Ic, since they are suggested to trace the velocity at the photosphere by \citet{branch02} and \citet{dessart15},\footnote{See \citet{dessart15} and \citet{modjaz15} for discussions on what the definition of ``photospheric" velocity is and which line to use.}  respectively.

%The Fe II features around 5000 \AA~are weak lines and are suggested to trace the velocity at the photosphere \citep{branch02}\footnote{See Modjaz et al. (2015) about discussions on what the definition of ``photospheric" velocity is and which line to use.}. Compared with weak lines, strong lines have a larger line formation region with some of the layers above the photosphere. These layers have velocities higher than the velocity at the photosphere in homologous expansion. As a result, the velocity derived from strong absorption lines is higher than the velocity at the photosphere. In contrast, weak lines are formed around the photosphere and should be used to determine the velocity of the photosphere \citep{patat96}. 

Figure \ref{fig_FeII} in Section \ref{Fe_vabs_IIbIb} shows our phenomenological method to identify Fe II $\lambda\lambda\lambda$4924, 5018, and 5169. Given that Fe II $\lambda\lambda$4924 and 5018 are usually blended together, here we report the velocities of Fe II $\lambda$5169 as the photospheric velocities. We emphasize that even if the Fe II $\lambda$5169 feature we identified is not due to Fe, our relative comparisons are still valuable since we measure the same feature in the same way. The Fe II $\lambda$5169 velocities and O I $\lambda$7774 velocities of both individual SNe and the corresponding rolling weighted averages are displayed in Figure \ref{fig_vel_FeII5169_f2&f3}. Although at any given epoch, the velocity measurements for SNe Ib and Ic are consistent with each other, the weighted average velocities of both Fe II $\lambda$5169 and O I $\lambda$7774 velocities in SNe Ic are systematically higher than those in SNe Ib at all epochs. This observation is inconsistent with the predictions of \citet{dessart12}. Therefore, the progenitors of SNe Ib and SNe Ic should be different not only in the mixing level of $^{56}$Ni but also in their composition. Since we have good reasons to believe that the ejecta masses of SNe Ib and SNe Ic are comparable \citep[e.g.,][Bianco et al. in prep]{drout11, lyman14}, the progenitors of SNe Ib and SNe Ic may have identical CO cores. Thus, we conclude that the difference between progenitors of SNe Ib and SNe Ic may lie in the outer layers, i.e., SN Ic progenitors may have a very thin or absent helium layer.

Table \ref{table_IbIc} lists the weighted average of our observations for SNe Ib and SNe Ic at $t_{\mathrm{Vmax}}\simeq0$ day. 

\begin{deluxetable}{cccccccc}
\tabletypesize{\scriptsize}
\tablecaption{Summary of weighted average of the measurements for SNe Ib and Ic at $t_{\mathrm{Vmax}}\simeq0$ day\label{table_IbIc}}
\tablehead{
\colhead{SN type} &
\colhead{pEW (O I $\lambda$7774)} &
\colhead{$v$ (Fe II $\lambda$5169)} & 
\colhead{$v$ (O I $\lambda$7774)} &\\
\colhead{} &
\colhead{(\AA)} &
\colhead{(km s$^{-1}$)} &
\colhead{(km s$^{-1}$)} &
}
\startdata
SNe Ib &    56 $\pm$ 42 &       $-$9100 $\pm$ 1600 &  $-$8400$\pm$ 300\\
SNe Ic &   77 $\pm$ 18 &       $-$11000 $\pm$ 3700 & $-$13600 $\pm$ 4300
\enddata
\tablecomments{The error is the standard deviation of data that contribute the weighted average value, which is consistent with the errors in the figures that show weighted average values.}
\end{deluxetable}

\section{Using Mean Spectra and Their Corresponding Standard Deviations to Characterize Spectral Diversity}
\label{sec_meanspec}

\begin{figure*}[t]
\subfigure{%
\includegraphics[scale=0.5,angle=0,trim =10mm 0mm 5mm 5mm, clip]{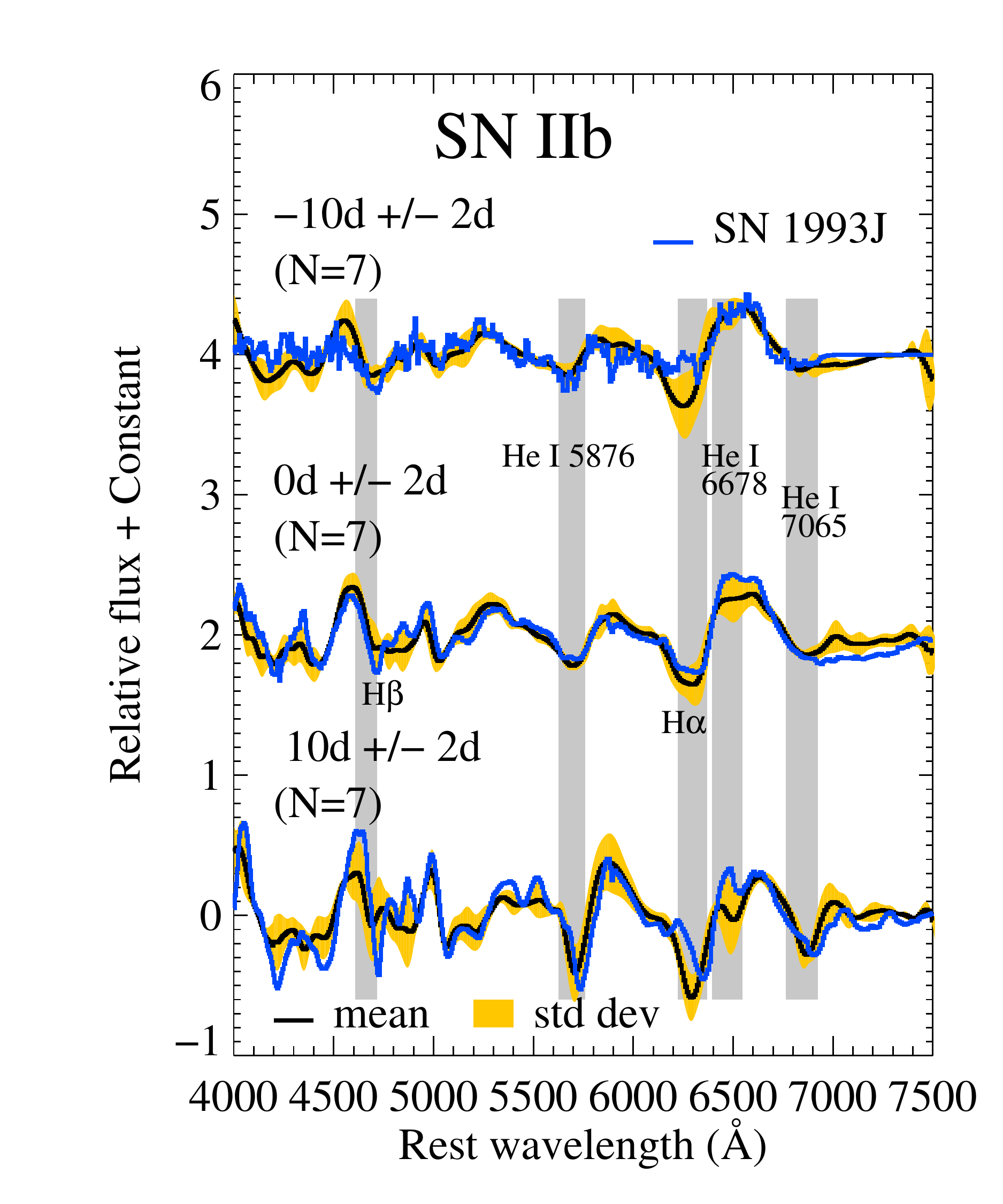}
}
\quad
\subfigure{%
\includegraphics[scale=0.5,angle=0,trim =15mm 0mm 0mm 5mm, clip]{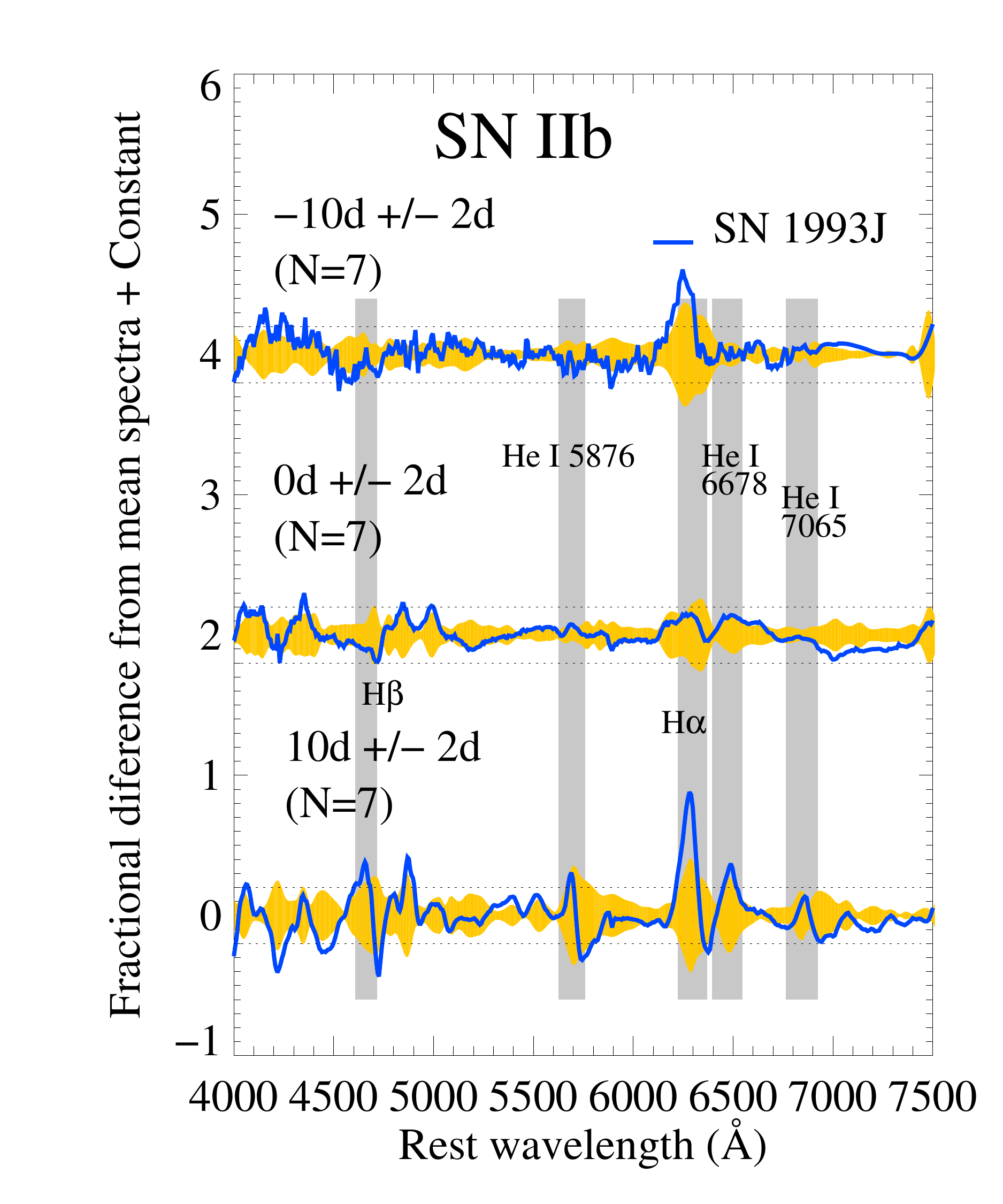}
}
\caption{\textit{Left}: Mean spectra of SNe IIb (in black) with corresponding standard deviations (in yellow) at phase ranges: $-10$ ($+/-2$) days (top), 0 ($+/-2$) days (middle), and 10 ($+/-2$) days (bottom). In blue, we show the flattened (or continuum-divided) spectra of SN 1993J at $t_{\mathrm{Vmax}}\simeq-11$, 1, and 11 days. \textit{Right}: The standard deviation of the mean spectrum divided by the mean spectrum is shown in yellow.  For comparison, the difference between the SN 1993J spectrum and the mean spectrum, also divided by the mean spectrum, is shown in blue. Dotted horizontal lines represent the 20\% of the mean spectrum. The number of spectra used to build each mean spectrum is shown in parentheses. The gray vertical bands indicate the expected positions of He I $\lambda\lambda\lambda$5876, 6678, and 7065 at velocities of $-$6000 km s$^{-1}$ to $-$13000 km s$^{-1}$, or the expected positions of H$\alpha$ and H$\beta$ at velocities of $-$9000 km s$^{-1}$ to $-$16000 km s$^{-1}$.}
\label{fig_mean_residual_spec_IIb}
\end{figure*}

\begin{figure*}[t]
\subfigure{%
\includegraphics[scale=0.5,angle=0,trim =10mm 0mm 5mm 5mm, clip]{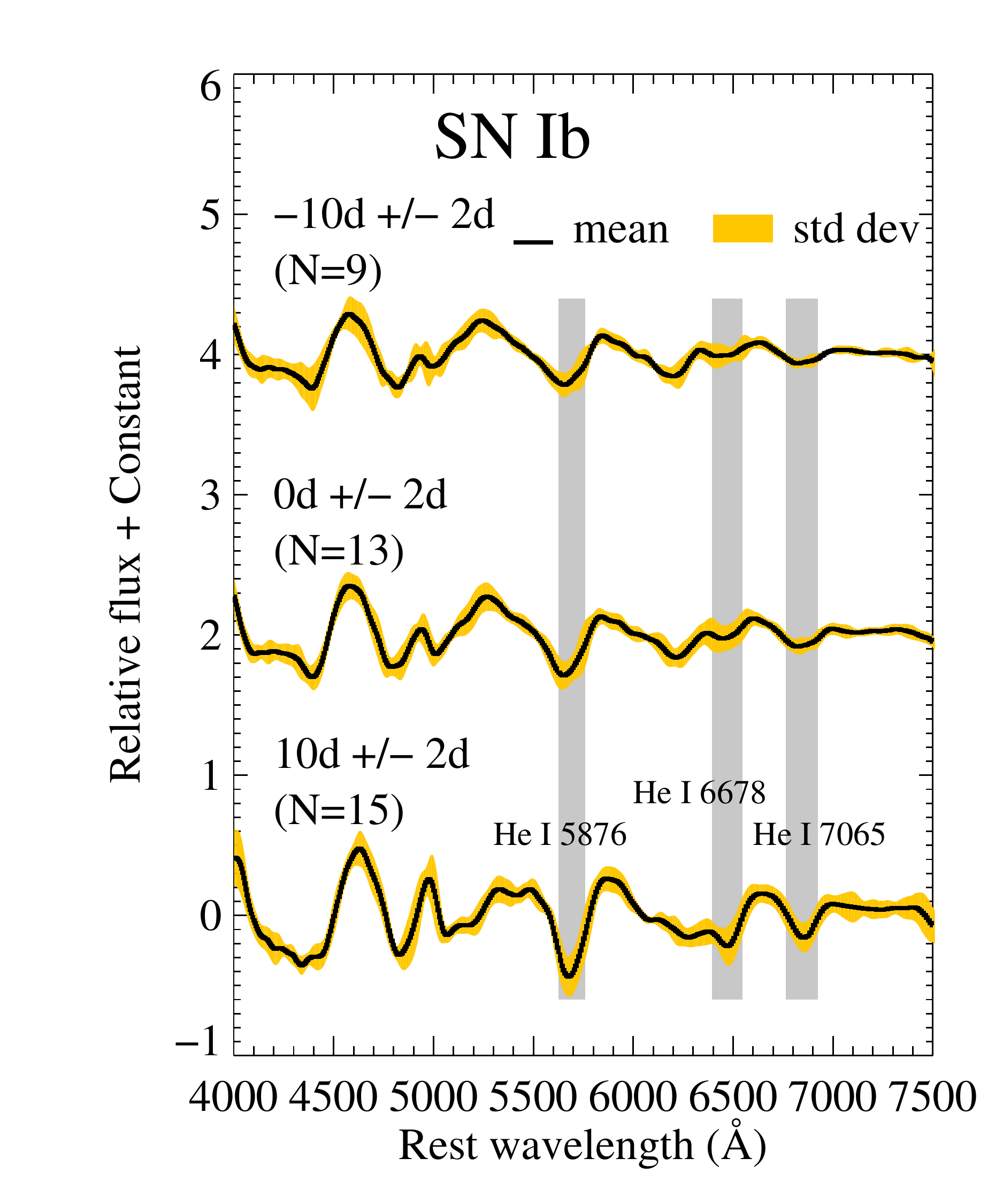}
}
\quad
\subfigure{%
\includegraphics[scale=0.5,angle=0,trim = 15mm 0mm 0mm 5mm, clip]{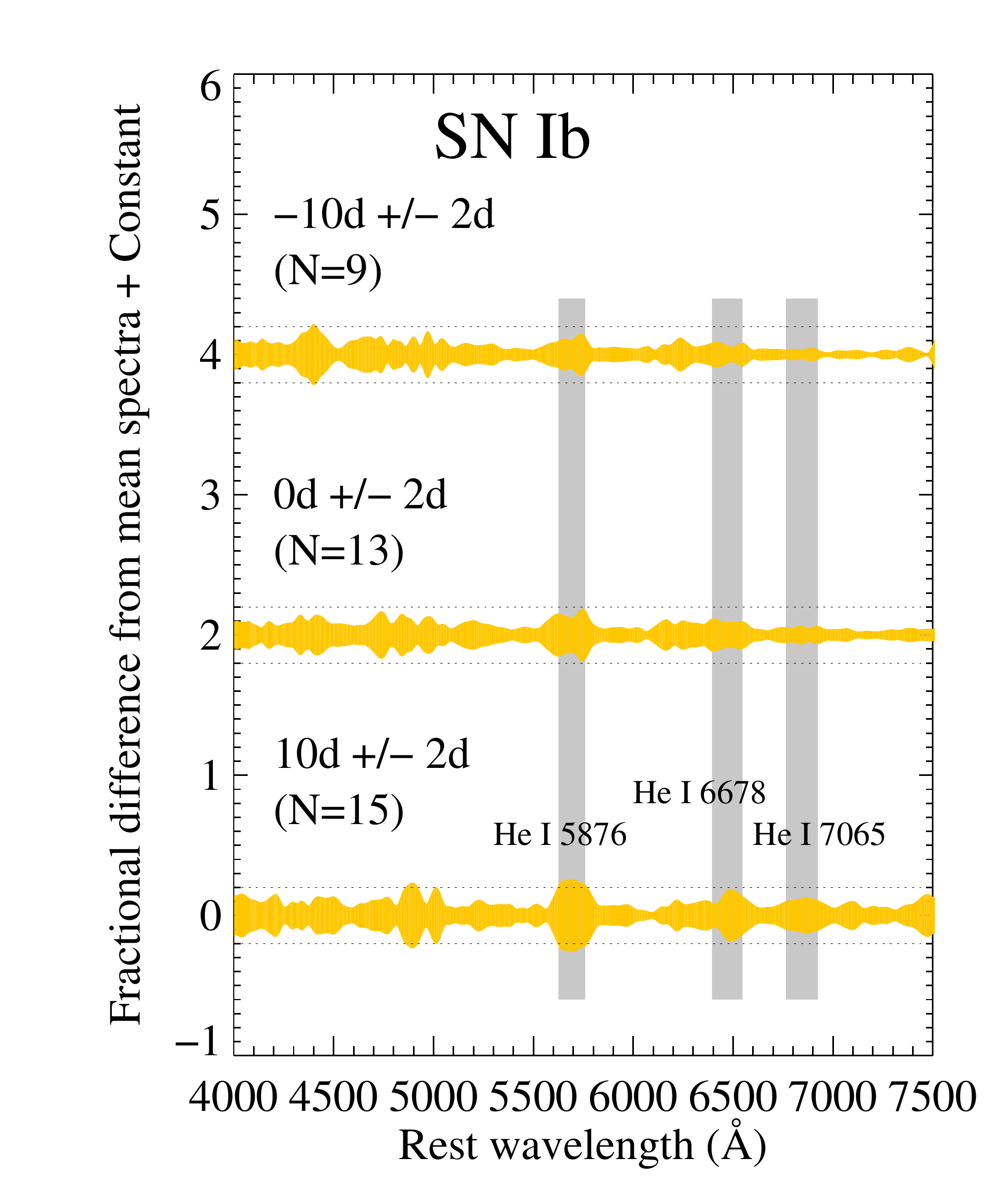}
}

\subfigure{%
\includegraphics[scale=0.5,angle=0,trim = 10mm 0mm 5mm 10mm, clip]{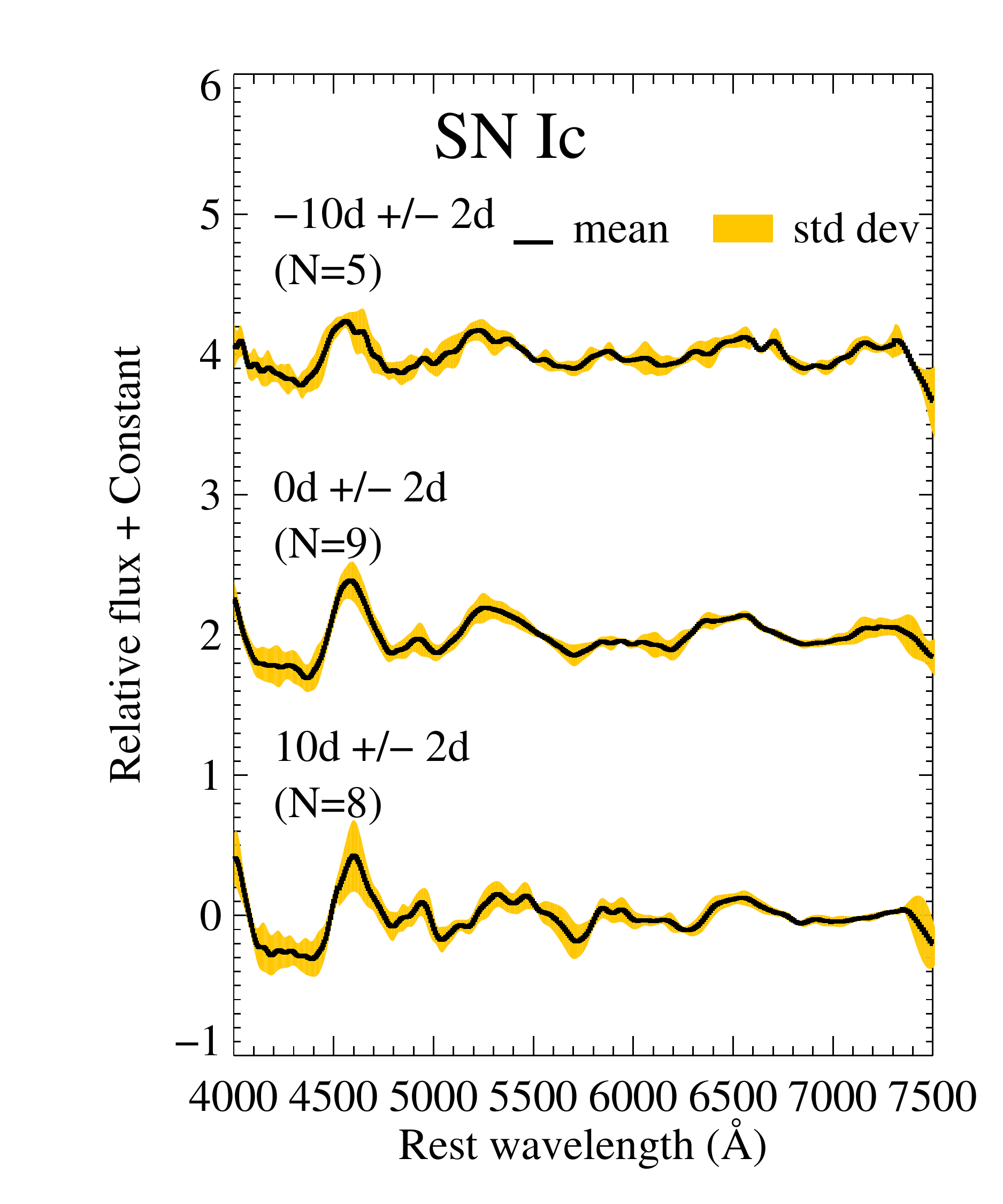}
}
\quad
\subfigure{%
\includegraphics[scale=0.5,angle=0,trim = 15mm 0mm 0mm 10mm, clip]{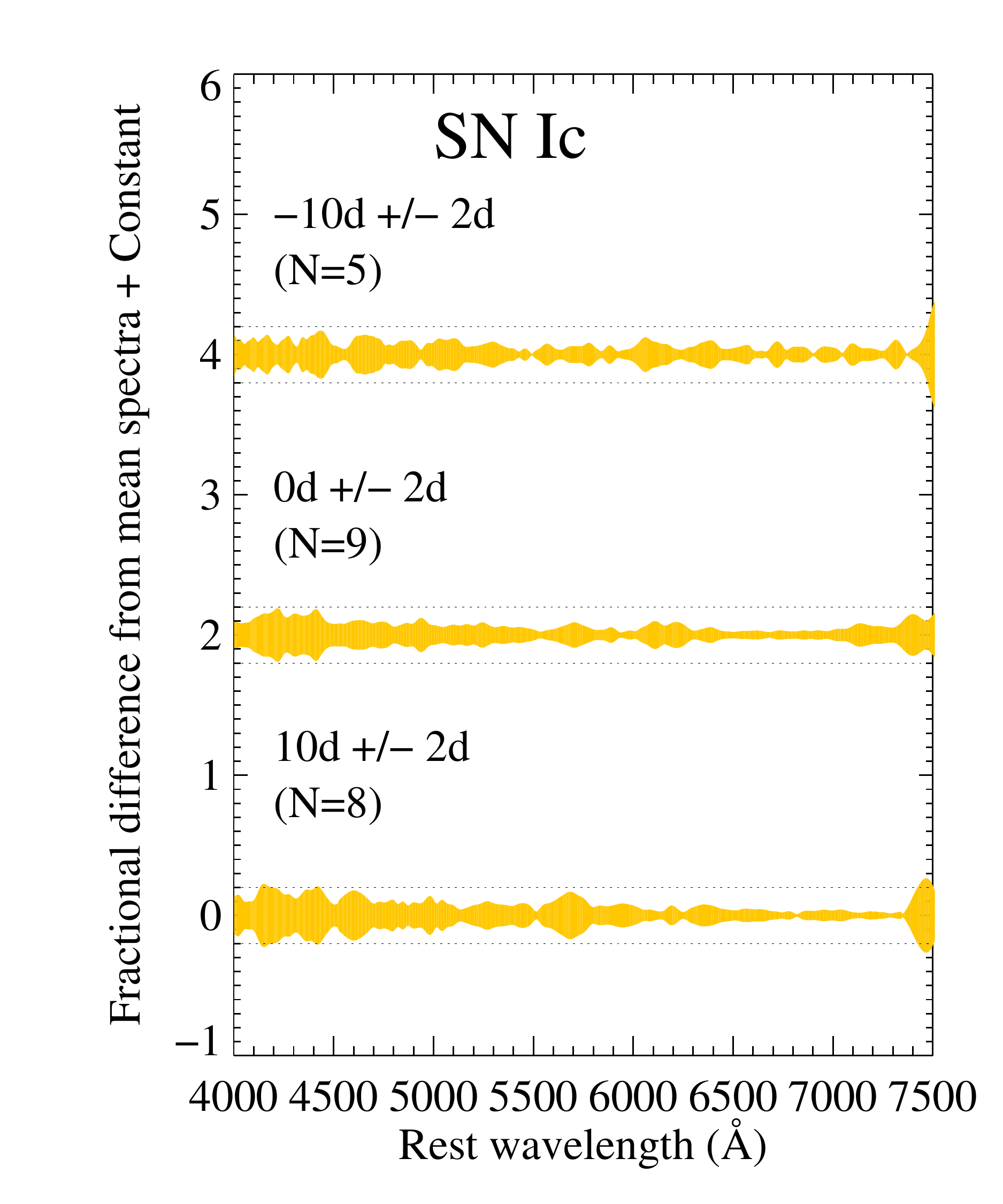}
}
\caption{Same as Figure \ref{fig_mean_residual_spec_IIb}, but for SNe Ib and Ic.
%Standard deviation spectra added to / subtracted from the corresponding mean spectra of SNe IIb (2 black lines for each phase range), at phase ranges: $-12$ to $-8$ days ($-10$d$+/-2$d; top), $-2$ to 2 days (0d$+/-2$d; middle), and 8 to 12 days (10d$+/-2$d; bottom). Red lines are the flattened Nugent's SNe Ib/c templates at phases $-10$, 0, and 10 days. \textit{Right}: Ratio of standard deviations (from mean spectra) to our mean spectra is shown in black and ratio of deviations of Nugent's templates (from mean spectra) to mean spectra is shown in red. Dotted horizontal lines represent the 0.2 ratio. Three pairs of dashed lines delimit expected positions of He I $\lambda\lambda\lambda$5876, 6678, and 7065 at velocities of $-$5000 km s$^{-1}$ to $-$14000 km s$^{-1}$.
}
\label{fig_mean_residual_spec_Ib}
\end{figure*}

\begin{figure*}[t]
\subfigure{%
\includegraphics[scale=0.5,angle=0,trim =10mm 0mm 5mm 5mm, clip]{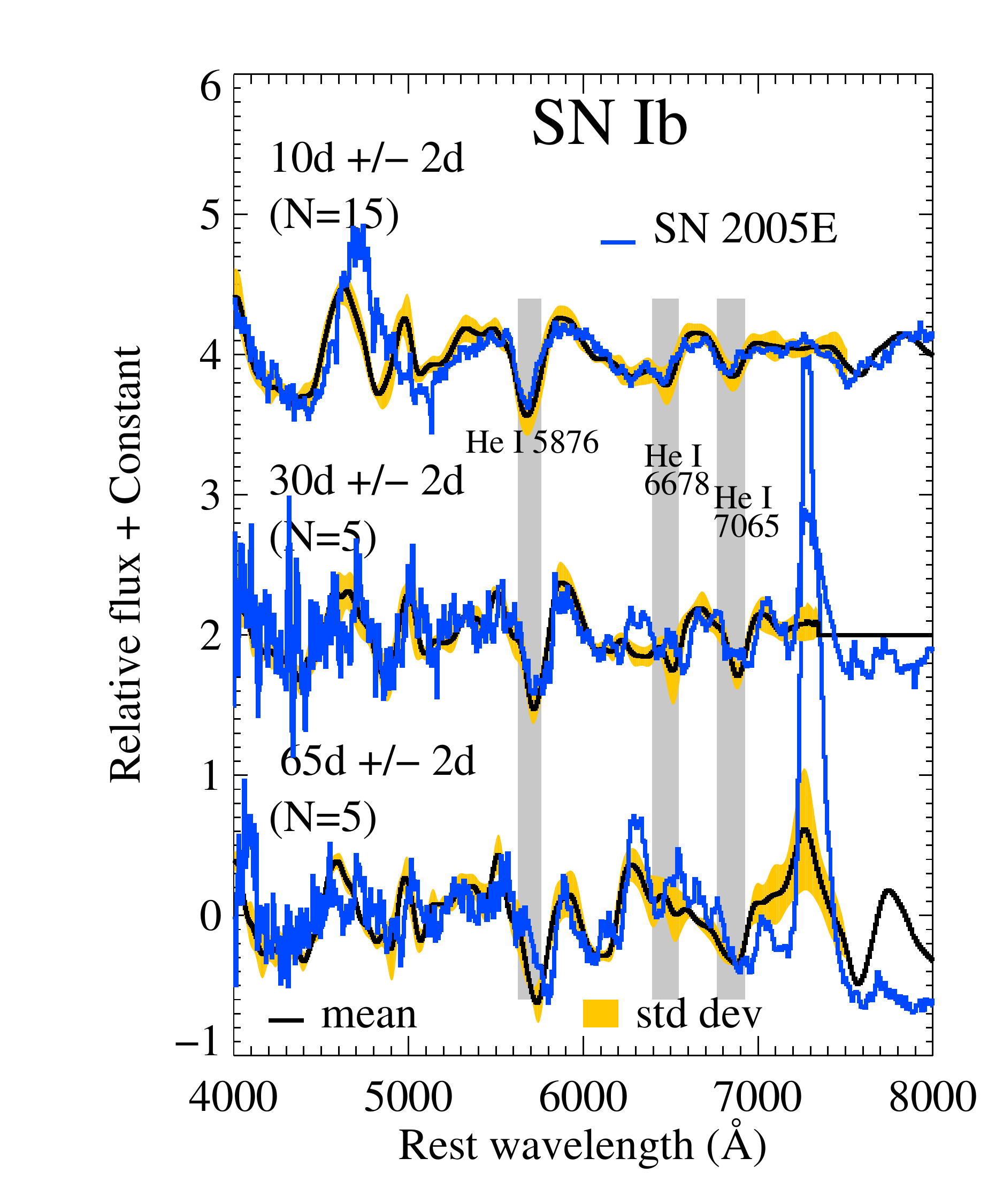}
}
\quad
\subfigure{%
\includegraphics[scale=0.5,angle=0,trim =15mm 0mm 0mm 5mm, clip]{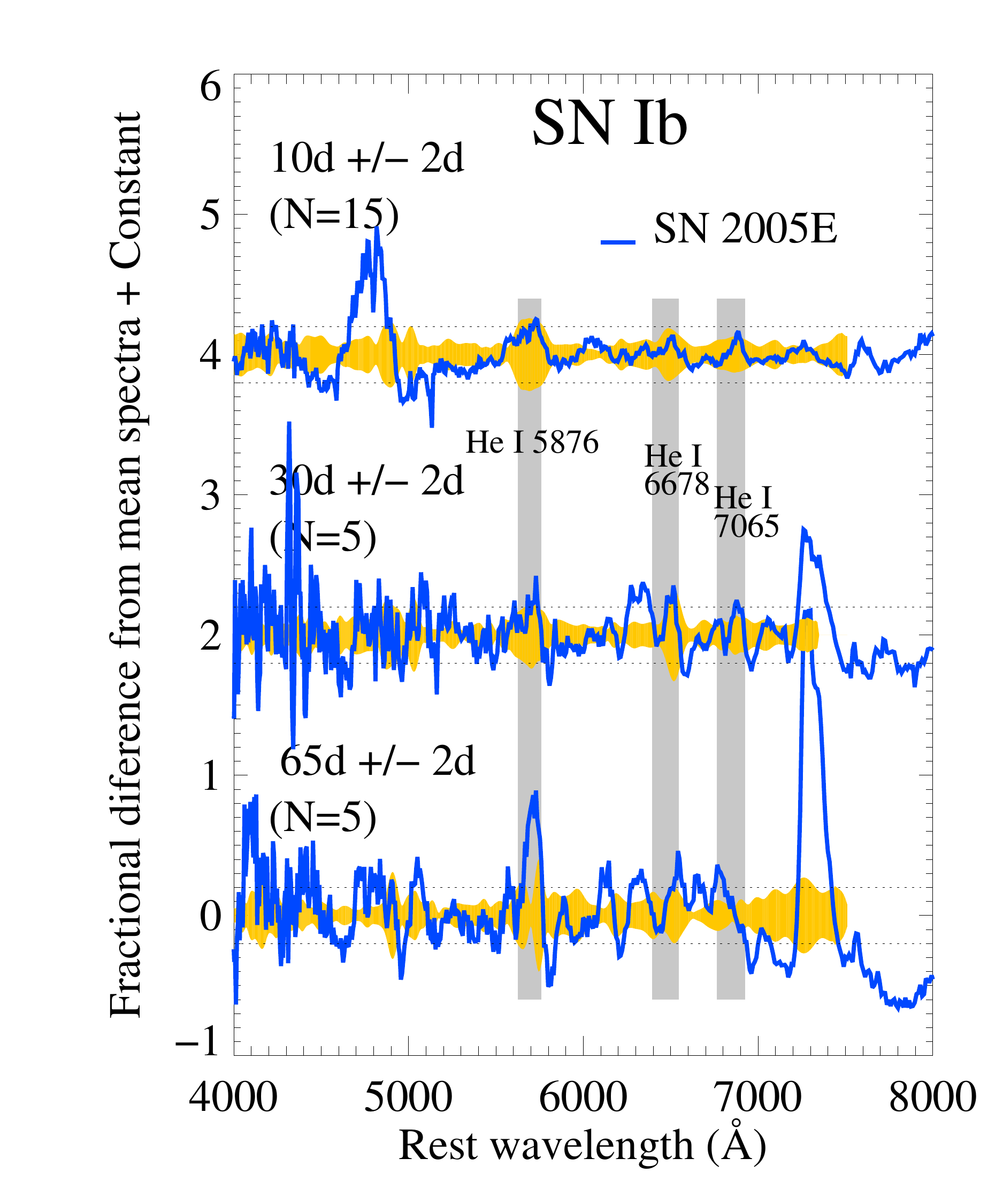}
}
\caption{The same as Figure \ref{fig_mean_residual_spec_IIb}, but for the comparison between normal SN Ib and SN Ib-Ca 2005E. The spectra of SN 2005E are at $t_{\mathrm{Vmax}}\simeq11$, 31, and 65 days.}
\label{fig_mean_residual_spec_Ib_05E}
\end{figure*}
%\begin{figure*}[t]
%\subfigure{%
%\includegraphics[scale=0.4,angle=0]{meanplotIc_oral2.pdf}
%}
%\quad
%\subfigure{%
%\includegraphics[scale=0.4,angle=0]{meanplotIc_oral2_residual.pdf}
%}
%%
%\caption{The same as Figure \ref{fig_mean_residual_spec_Ib}, but for mean spectra of SNe Ic and to compare them with the same Nugent's SNe Ib/c templates.}
%\label{fig_mean_residual_spec_Ic}
%\end{figure*}

Mean spectra and the standard deviations from the mean characterize the spectral diversity of each SN subtype, thus they can be used to determine if a newly discovered SN belongs to an already known subtype, or is a novel discovery. With the large amount of data we have, we construct mean spectra and their corresponding standard deviations as a function of subtype and phase. To reduce the effect of reddening, these mean spectra are constructed from continuum-divided spectra.

%If mean spectra are constructed from spectra in absolute flux, they can be used for photometric classification \citep[e.g.,][]{poznanski02} to construct model light curves which are the expected light curves of different types of SNe at a range of redshift and in various bands. Model light curves are compared to the observed ones in order to classify the latter. Currently, the \citet{nugent02} templates\footnote{https://c3.lbl.gov/nugent/nugent\_templates.html} are used as the representative spectra of stripped SNe. With the large amount of data we have, we construct mean spectra that are more representative than the \citet{nugent02} templates. Note that since our mean spectra are constructed from spectra that are divided by the continuum, our mean spectra cannot be used in the photometric classification yet.

In Figures~\ref{fig_mean_residual_spec_IIb} and \ref{fig_mean_residual_spec_Ib}, we show the mean spectra, one standard deviation from the mean, and the ratio of the standard deviation to the mean for SNe IIb, SNe Ib and SNe Ic at $t_{\mathrm{Vmax}}=-10$, 0, and $10$ days. The mean spectra and their corresponding standard deviation for SNe Ic-bl are published in \citet{modjaz15}. Although the spectral diversity of SN features are already shown in previous figures, such as Figure \ref{fig_vabs_Halpha_f3}, now we can directly observe the spectral diversity of different stripped SN classes via their mean spectra and the corresponding standard deviations. For the mean spectra of SNe Ib (Figure \ref{fig_mean_residual_spec_Ib}), the deepest absorption feature is due to He I $\lambda$5876 with possible contamination from Na I D. Other two obvious features are produced by He I $\lambda$$\lambda$6678 and 7065. These three lines evolve to lower velocities but grow stronger as time goes by. The above velocity evolution is expected since the velocity scales with radius in homologous expansion and the photosphere---where the absorption features are produced---recedes towards the center of the expansion as time elapses. For the mean spectra of SNe IIb (Figure \ref{fig_mean_residual_spec_IIb}), the strongest absorption feature is due to H$\alpha$, which becomes weaker with time. Other three prominent absorption troughs in SNe IIb are attributed to He I $\lambda$$\lambda$$\lambda$5876, 6678, and 7065, whose velocity evolutions and strength evolutions behave similarly to those in the mean spectra of SNe Ib. These trends are confirmed by velocity and pEW measurements of H$\alpha$ and He I lines for individual SNe in Figures \ref{fig_vabs_Halpha_f3}, \ref{fig_vabs_HeI5876}, and \ref{fig_HeI5876_pEW_IbIIb}.

We test whether the so-called ``prototypical''  SN IIb 1993J is truly representative of SNe IIb by comparing its spectra with our mean spectra of SNe IIb. As shown in Figure \ref{fig_mean_residual_spec_IIb}, SN 1993J is fully consistent with the mean spectra of SNe IIb, except for the H$\alpha$ absorption feature at phase $t_{\mathrm{Vmax}}=10$ days. Thus, SN 1993J is a typical SN IIb from a spectroscopic point of view.

The ratio between the standard deviation and the mean spectra in the right panels of Figures \ref{fig_mean_residual_spec_IIb} and \ref{fig_mean_residual_spec_Ib} show that in most cases, the spectra of SNe IIb, SNe Ib, and SNe Ic are within 20\% of the corresponding mean spectrum. The spectral variance is relatively small at $t_{\mathrm{Vmax}}=-10$ and 0 days, but relatively large around the absorption features. For example, the spectral variance around He I lines in the mean spectra of SNe IIb and SNe Ib is larger than that at other wavelengths. For the mean spectra of SNe IIb, the spectral variance is also large around the H$\alpha$ and H$\beta$ features. This indicates that there is a diversity in the thickness of either the helium or hydrogen layers (or both) of the progenitors.

In Figure \ref{fig_mean_residual_spec_Ib_05E}, we show how our average spectra can be used to discover and study relatively novel SN types such as Calcium-rich Type Ib SNe \citep[SNe Ib-Ca;][]{filippenko03_03dg, perets10, kasliwal12}. Compared to SNe Ib, whose spectroscopic hallmark is the prominent He I lines, SNe Ib-Ca also display unusually strong lines of calcium during their nebular phase. Although the origin of SNe Ib-Ca is contentious \citep{perets10, meng15}, in Figure  \ref{fig_mean_residual_spec_Ib_05E} we spectroscopically compare the SN Ib-Ca 2005E \citep{perets10} to our SN Ib average spectra. The velocities and strengths of He I lines in the spectra of SN 2005E are consistent with those of the average SN Ib spectra at comparable phases. At $t_{\mathrm{Vmax}}\simeq65$ days, SN 2005E exhibits a much stronger Ca emission line around 7300 \AA~than the SN Ib average spectrum. We also note that at $t_{\mathrm{Vmax}}\simeq10$ days, the Fe II lines around 5000 \AA~in the SN 2005E spectrum form a ``v"-like feature (which is commonly seen in SNe Ic-bl), instead of the ``w"-like feature that appears in SNe Ib (as well as in SNe IIb and SNe Ic). Thus, instead of comparing newly discovered SNe to individual SNe with a known type, comparing the new SNe to our average spectra of different SN subtypes is more powerful for detecting spectroscopically novel SNe (though they may also have photometric differences).

\section{Summary and Conclusions}
\label{conc}

\begin{deluxetable*}{cccccccc}
\tabletypesize{\scriptsize}
\tablecaption{Summary of comparisons between our observations and predictions as well as other observations in the literature\label{table_comp2}}
\tablehead{
\colhead{Absorption Feature} &
\colhead{Our Observation} &
\colhead{M01\tablenotemark{a}}  &
\colhead{D12\tablenotemark{b}} &
\colhead{F13\tablenotemark{c}}  &
}
\startdata
%H$\alpha$, He I $\lambda\lambda\lambda$5876, 6678, 7065, Fe II $\lambda$5169  & V$_{\mathrm{Ib}}$ $>$ V$_{\mathrm{IIb}}$ &    NA  &    consistent &    NA \\
%He I $\lambda$5876 & pEW$_{\mathrm{Ib}}$ $>$ pEW$_{\mathrm{IIb}}$, around $t_{\mathrm{Vmax}}\simeq0$ days &    NA  &    consistent &    NA \\
% & pEW$_{\mathrm{Ib}}$ $\simeq$ pEW$_{\mathrm{IIb}}$, after $t_{\mathrm{Vmax}}\simeq10$ days &    NA  &    inconsistent &    NA \\
%He I $\lambda$6678 & pEW$_{\mathrm{Ib}}$ $>$ pEW$_{\mathrm{IIb}}$ &    NA  &    consistent &    NA \\
%He I $\lambda$7065 & pEW$_{\mathrm{Ib}}$ $>$ pEW$_{\mathrm{IIb}}$, around $t_{\mathrm{Vmax}}\simeq0$ days &    NA  &    consistent &    NA \\
% & pEW$_{\mathrm{Ib}}$ $<$ pEW$_{\mathrm{IIb}}$, after $t_{\mathrm{Vmax}}\simeq10$ days &    NA  &    inconsistent &    NA \\[1ex]
% \hline \\[0.05ex]
O I $\lambda$7774 & pEW$_{\mathrm{Ic}}$ $>$ pEW$_{\mathrm{Ib}}$ &    consistent  &    inconsistent &    consistent \\
Fe II $\lambda$5169 & V$_{\mathrm{Ic}}$ $>$ V$_{\mathrm{Ib}}$ &  NA   &  inconsistent &  NA\\
O I $\lambda$7774 & V$_{\mathrm{Ic}}$ $>$ V$_{\mathrm{Ib}}$ &  NA   &  inconsistent &  NA
\enddata
\tablenotetext{a}{M01 = \citet{matheson01}}
\tablenotetext{b}{D12 = \citet{dessart12}}
\tablenotetext{c}{F13 = \citet{frey13}}
\end{deluxetable*}

In this paper, we have improved the SN identification scheme and constrained the progenitors of different kinds of stripped SNe by analyzing the spectra of the largest stripped SN sample (which consists of 242 spectra of 14 SNe IIb, 262 spectra of 21 SNe Ib, and 207 spectra of 17 SNe Ic) in a statistical and quantitative way. To be thorough, we have derived robust error bars for velocity and pEW measurements by constructing uncertainty arrays from the SN spectra themselves (by separating noise from SN signal in Fourier space) for the very first time and then propagating those uncertainties into our final measurements via MC simulations.

To better classify SNe IIb and Ib, we have quantified the properties of spectral features by measuring absorption velocity and pEW values of H$\alpha$, He I $\lambda\lambda\lambda$5876, 6678, 7065, and Fe II $\lambda$5169. We then compared these values for SNe IIb with those for SNe Ib in a statistical way. On the one hand, there is substantial overlap in almost all of our above measurements, suggesting an observational continuum between SNe IIb and Ib. The observational continuum is most likely due to a continuum of sizes or masses of the hydrogen or helium layers. On the other hand, the H$\alpha$ pEW values of SNe Ib are different from those of SNe IIb at all phases, which can be used to classify these two SN subtypes. Although there is no clear boundary between the pEW values of H$\alpha$ for the SN IIb sample and those for the SN Ib sample, there is no overlap at any phase. Thus if a SN is identified as a SN IIb once, it remains a SN IIb. Therefore, the H$\alpha$ pEW value can be used to differentiate SNe IIb from SNe Ib at all epochs. We suggest that to classify a new SN IIb/Ib, not only the spectroscopic SN identification tools (e.g., SNID, GELATO\footnote{https://gelato.tng.iac.es/}) should be used, but also the properties (e.g., H$\alpha$ pEW values) of the new SN IIb/Ib should be measured, announced, and compared to SNe IIb and Ib described here. 

We have addressed the question of hidden helium in SNe Ic by comparing our observations of O I $\lambda$7774 strength, Fe II $\lambda$5169 velocity, and O I $\lambda$7774 velocity for SNe Ib and Ic with theoretical predictions. Table \ref{table_comp2} summarizes these comparisons. We find that on average, SNe Ib have higher photospheric velocities (as traced by Fe II $\lambda$5169 velocities) than SNe Ic, which is inconsistent with the predictions of \citet{dessart12} based on a hidden helium model for SN Ic progenitors, \textbf{if the only difference between SN Ib and SN Ic progenitors were the level of 56Ni mixing}. Furthermore, we find that the average pEW values of O I $\lambda$7774 for SNe Ic are stronger than those for SNe Ib, which supports predictions of \citet{frey13} based on a helium-free model for SN Ic progenitors, though the new mixing algorithm used in \citet{frey13} needs to be verified. This indicates that progenitors of SNe Ic may not contain helium before the explosion. \textbf{The same conclusion has been reached in \citet{taddia15} using early time light curves of stripped SNe.}

We have constructed continuum-divided mean spectra and their corresponding standard deviation arrays for SNe IIb, SNe Ib, and SNe Ic to characterize their respective spectral diversity, which show systematic differences between SN subtypes. These mean spectra should be used in discovery of spectroscopically novel SNe and spectroscopic classification of SNe. By comparing the spectra of SN 1993J with the SN IIb mean spectra, we show that SN 1993J is a typical SN IIb from a spectroscopic point of view. We also find that the spectral diversity is relatively large around absorption features such as He I lines and H lines, which should be taken into account when progenitor models are coupled to spectral synthesis codes.
%By comparing the mean spectra of SNe Ib and SNe Ic with the \citet{nugent02} templates, we argue that our mean spectra should be used in cases where the SN absorption features are important (e.g., discovery of spectroscopically novel SNe and spectroscopic classification of SNe), since our mean spectra are constructed from flattened spectra (i.e., continuum-divided spectra), however, they include many more SNe than Nugent included, cover a larger wavelength range at some phases, and account for the spectral diversity of the SNe.

For convenience, we summarize the trends we observe in our sample, which may help constrain the progenitor models of SNe IIb, SNe Ib and SNe Ic:
\begin{itemize}
\item Average velocities of H$\alpha$, He I  $\lambda\lambda\lambda$5876, 6678, 7065, and Fe II $\lambda$5169 in SNe Ib are systematically higher than those in SNe IIb.
\item The H$\alpha$ pEW values in SNe IIb are systematically higher than those in SNe Ib at all phases.
\item The He I $\lambda\lambda$5876, 6678 velocities in most SNe IIb and Ib decrease rapidly over time while the He I $\lambda$7065 velocities only decrease slightly.
\item SNe IIb and Ib have comparable pEW values of He I $\lambda$5876. On average, SNe Ib have larger pEW values of He I $\lambda$6678 than SNe IIb. The He I $\lambda$7065 pEW values are comparable in SNe IIb and Ib before $t_{\mathrm{Vmax}}\simeq10$ days. However, SNe IIb show much stronger He I $\lambda$7065 than SNe Ib at later epochs.
\item Three SNe IIb and four SNe Ib in our sample have flat-velocity He I lines while none of them show flat-velocity Fe II $\lambda$5169 line.
\item On average, SNe Ic have stronger O I $\lambda$7774 than SNe Ib from $t_{\mathrm{Vmax}}\simeq-10$ days to $t_{\mathrm{Vmax}}\simeq25$ days.
\item The average Fe II $\lambda$5169 velocities and O I $\lambda$7774 velocities in SNe Ic are systematically higher than those in SNe Ib.
\item The average Fe II $\lambda$5169 velocities are highest for SNe Ic, followed by SNe Ib, and finally by SNe IIb, which systematically have the lowest average Fe II $\lambda$5169 velocities.
\item For all SN subtypes, the spectral variance is larger around absorption features such as H lines and He I lines than at other wavelengths.
\end{itemize}

Both asphericity and intrinsic diversity will contribute to the diversity within a subset, and the spectral diversity within the observable sample of SNe IIb is comparable to the degree of asymmetry observed in galactic SN IIb Cassiopeia A, as probed by light echo spectra \citep{finn16}. However the systematic spectral trend as a function of subtype (e.g., on average, SNe Ic having the highest Fe II velocities, vs. SNe IIb having the lowest) is then unlikely to be explained with asphericity alone. Future radiative transfer calculations that aim to elucidate the nature of SNe IIb, Ib, and Ic need to reproduce the many observed trends that have been presented here. In order to complete the sequence of massive star explosions, the measured absorption velocities and pEW values for H$\alpha$ in SNe IIb and Ib should be compared with those in SNe II. We suggest the use of realistic non-LTE codes that properly treat non-thermal excitations to explore all our SN Ic spectra for the presence of He I lines.  The spectral diversity in stripped SNe should be explored further using machine-learning methods. In order to be used in the photometric SN classifiers \citep[e.g.,][]{poznanski02}, mean spectra that retain the continuum information, include more SN spectra, sampled at a smaller phase bin, and cover a longer wavelength range should be constructed for different types of SNe. 

The mean spectra and new SNID templates produced in this study can be downloaded via our SNYU webpage.\footnote{http://cosmo.nyu.edu/SNYU/spectra}

\acknowledgments
We are grateful to Luc Dessart, Ross Fadely, Lluis Galbany, Avishay Gal-Yam, David W. Hogg, Dan Milisavljevic, Jerod Parrent, Nathan Smith, and Sung-Chul Yoon for useful discussions. We thank Stephane Blondin for making available to us the codes from \citet{Blondin2007} and \citet{blondin11_spec_distance_Ia} to measure absorption velocities and pEW, as well as to produce mean spectra.

Y.-Q. Liu is supported in part by a NYU/CCPP James Arthur Graduate Fellowship. M. Modjaz and the SNYU group are supported by the NSF CAREER award AST-1352405 and by the NSF award AST-1413260. F. Bianco is supported in part by the NYU/CCPP James Arthur Postdoctoral Fellowship. 

This research has made use of NASA's Astrophysics Data
System Bibliographic Services (ADS), the HyperLEDA database and the
NASA/IPAC Extragalactic Database (NED) that is operated by the Jet
Propulsion Laboratory, California Institute of Technology, under
contract with the National Aeronautics and Space Administration.  

\bibliographystyle{apj}
\bibliography{refs}

\appendix
\renewcommand{\thesubsection}{\Alph{section}}

\section{\\A. Spectral Pre-processing}
\label{pre-process}

We need both a flattened (i.e., continuum-divided) version and an original (i.e., continuum-included) version for all of the spectra in our SN sample, for different purposes. The former is used to construct mean spectra for different SN types as a function of phase. The latter is used to measure velocity and strength of absorption features.

Some spectra in our sample are original spectra, while others are continuum-divided spectra. The latter are the SNID templates from the SNID database templates-2.0 \citep{Blondin2007}, for which we do not possess the original spectra. The SNID templates are flattened spectra where the continuum has been divided out. We also made some phase and type modifications to these flattened spectra based on \citet{modjaz14} and summarized in \citet{liu14}. For flattened spectra in our sample, we constructed their ``original" spectra by adding the continuum flux (whose information is stored in the SNID template) back to the spectra. The original spectra in our sample contain the continuum, and thus we generated their SNID templates by following the steps in \citet{Blondin2007}, as well as the instructions on the webpage of SNID.\footnote{http://people.lam.fr/blondin.stephane/software/snid/howto.html} In this manner, we have constructed new SNID templates of all published stripped SN spectra. The SNID templates of the CfA stripped SNe were published in \citet{liu14}; additional SNID templates of SNe Ic, SNe Ic-bl, and SN-GRBs are published in \citet{modjaz15}; additional SNID templates of SNe IIb, and SNe Ib are published here. 

\section{\\B. Constructing uncertainty arrays of spectra}
\label{smooth}

\begin{figure*}[t]
\center
\includegraphics[scale=0.8,angle=90,trim = 0mm 0mm 0mm 0mm, clip]{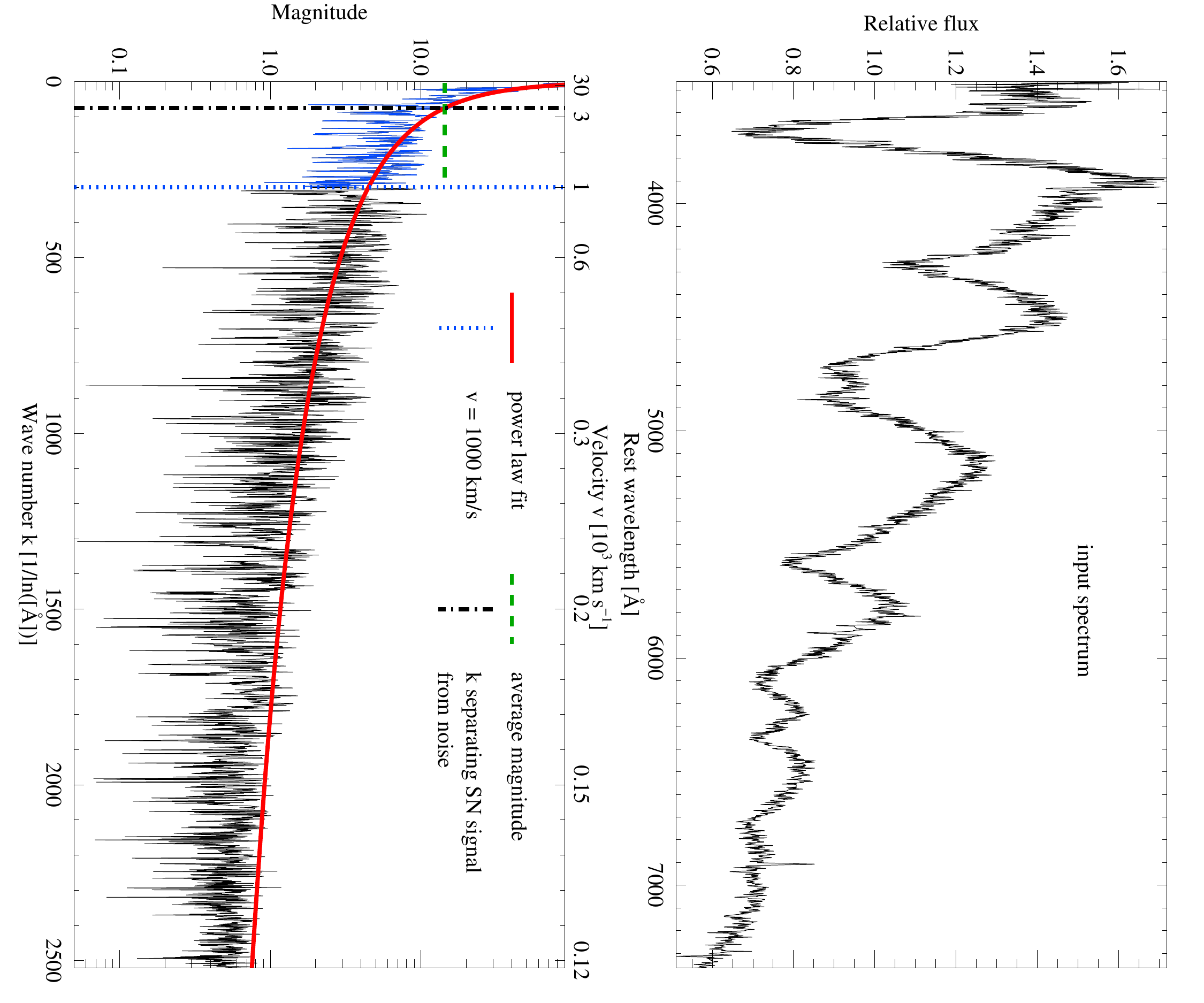}
%\hskip 
\caption{Using Fourier Transform (FT) to separate noise from SN signal in Fourier space and to filter a spectrum. \textit{Upper}: Input spectrum---spectrum of SN Ib SN 2004gq at $t_{\mathrm{Vmax}}=-9$ days. \textit{Lower}: FT of the input spectrum (black); A power law fit to magnitudes at wave numbers $k>3$ is shown in red; Magnitudes corresponding to velocities $1,000<v<100,000$ km s$^{-1}$ (or $300>k>3$) are shown in blue; Average magnitude of $3<k<300$ is shown in green; We identify as noise any FT features at $k>k_{noise}$, where $k_{noise}$ is the intersection (indicated as a black dotted-dashed line) between the above power law fit (red) and the average magnitude (green). In this case, the intersection is at $k=75$, which corresponds to $v=4000$ km s$^{-1}$. The filtered spectrum is generated by inverting the FT after setting the magnitudes for $k>k_{noise}$ to 0 (see the top panel of Figure \ref{fig_variance_our}).}
\label{fig_smooth_FFT}
\end{figure*}

\begin{figure*}[t]
\center
\includegraphics[scale=0.85,angle=0,trim = 0mm 0mm 0mm 00mm, clip]{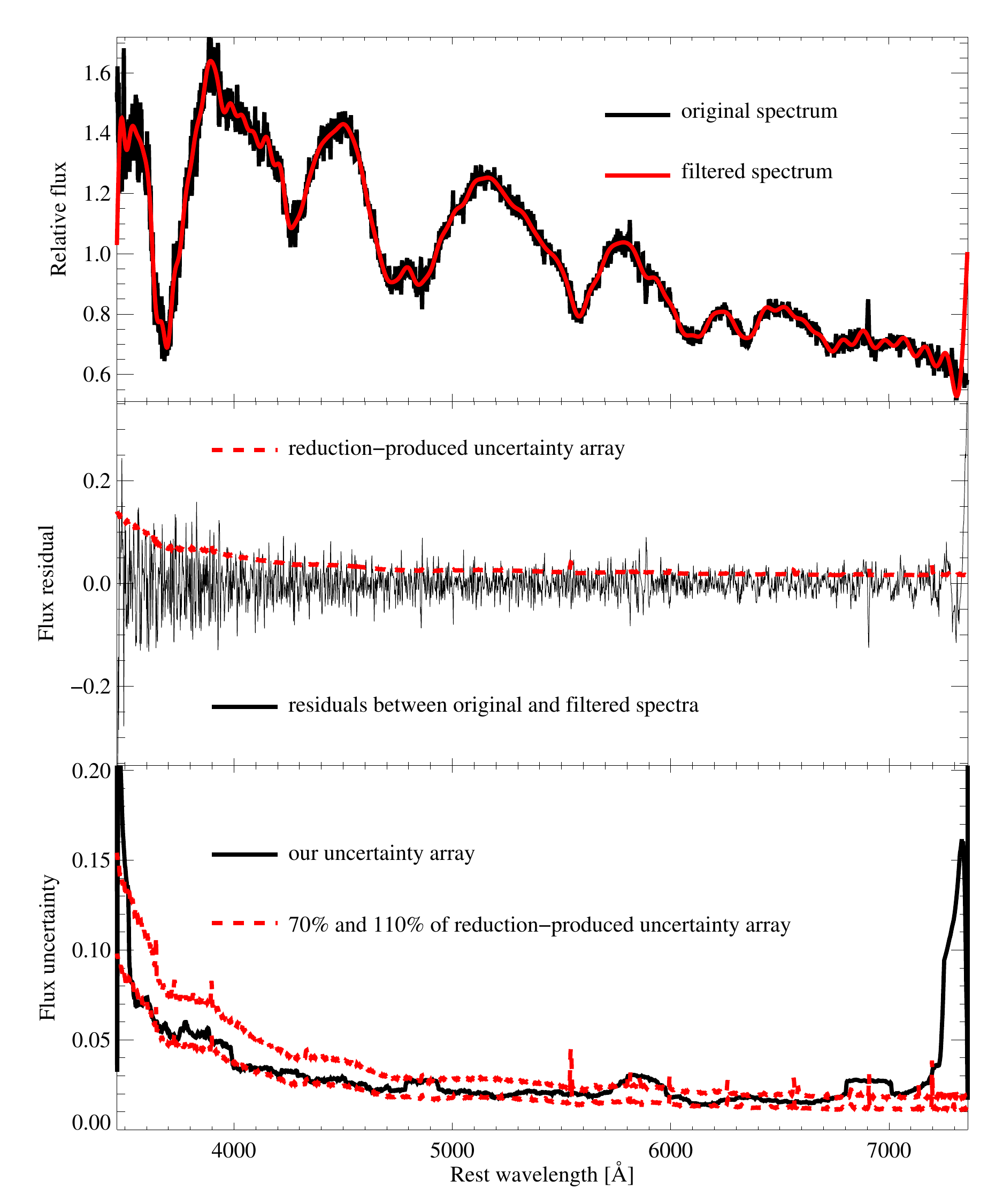}
\caption{Our uncertainty array, compared to the corresponding reduction-produced uncertainty array. \textit{Top}: spectrum of SN Ib SN 2004gq at $t_{\mathrm{Vmax}}=-9$ days (black) and the corresponding Fourier-filtered spectrum (red). \textit{Middle}: residuals between the Fourier-filtered spectrum and the original spectrum (black), i.e., noise in spectrum. For comparison, the reduction-produced uncertainty array (which is the square root of variance spectrum) is shown in red. \textit{Bottom}: our uncertainty array (black) and 70\% and 110\% of the reduction-produced uncertainty array (red). The high flux variance values at the edges of the wavelength range are due to the edge effect of FT. This will not affect our analysis, since we focus on spectral features that are not at the edges.}
\label{fig_variance_our}
\end{figure*}

We need to know the uncertainty arrays of spectra to estimate the errors on the velocity and strength measurements, since uncertainty arrays describe the noise level in spectra, which gives rise to the uncertainty in the measurements. Uncertainty arrays are ideally produced during data reduction, in particular during optimal extraction \citep{horne86}. The reduction-produced uncertainty arrays mainly include Poisson noise and the readout noise of the CCD. Other sources of noise have a negligible effect on the final uncertainty. The noise varies slowly with wavelength, except at the positions of sky lines or H II region emission lines, so the uncertainty arrays should generally change smoothly with wavelength. 

However, most spectra from the literature do not have published reduction-produced uncertainty arrays, nor could we obtain their reduction-produced uncertainty arrays by re-reducing the data, since we do not have the input files (e.g., the raw 2D frames) needed for optimal extraction. Instead, we have developed the following method to derive the uncertainty array from the reduced SN spectrum itself. 

We designed an empirical method to filter spectra in Fourier space, where we first separate noise from signal in Fourier space, and then we calculate the uncertainty array as the standard deviations of the noise over a wavelength range. We describe the two steps in detail below, and show them in Figure \ref{fig_smooth_FFT}. Methods for producing uncertainty arrays, or spectral filtering have been designed and implemented in the past \citep[e.g.,][]{blondin06}, including Fourier based filtering methods \citep[e.g.,][]{Blondin2007}. Unlike its predecessors, however, our method is fully automated: our Fourier filtering algorithm automatically finds the appropriate wave number $k_\mathrm{noise}$ to separate the signal from the noise for each individual spectrum. This is particularly useful if the goal is, for example, to produce a consistent analysis of large heterogeneous spectral datasets, as in this paper. 

Although not periodic, we expect the spectral features to show power in Fourier space at some characteristic scales. We note that in wavelength space, which is the natural space for noise in spectra and is the space for spectra when observed at telescopes, noise is white noise. By definition, white noise has the same power at all wave numbers ($k$) in Fourier space, so it should be easy to isolate the Fourier components that carry the spectral signal by identifying the wave numbers where the power begins to deviate from a constant, e.g., by looking for the intersection of a power law fit to all Fourier components of a spectrum with a flat-line fit to the ones at large wave numbers \citep{press07}. However, the natural space of spectral features is velocity space: the spectral features will have some characteristic sizes when observed in flux-velocity space, and thus some characteristic wave numbers in Fourier space.

The first step of our method is to convert the wavelength axis of a spectrum to velocity space by binning the spectrum on a logarithmic wavelength axis. Since Fourier transform (FT) can be applied to evenly sampled data only, the spectrum in velocity space has to be re-binned (we use a bin size equal to its smallest dispersion), and then the Fast FT algorithm (FFT) is applied. However, by transforming the noise from flux-wavelength space to flux-velocity space and re-binning it, we increase the power at the low wave numbers, so we can no longer treat the noise as white. The low wave numbers are also the wave numbers where the spectral signal has power. The increase in noise power at low wave numbers is marginal compared to the power of the spectral signal, thus the spectral signal still dominates the power at low wave numbers, but there will be a domain where both contribute equally. Our goal is to identify the $k_\mathrm{noise}$ where the noise begins to have a significant contribution to the power spectrum.

Let us assume that very high and very low velocities are not going to be associated with SN spectral features. Velocities can be converted to wave numbers as $k = c/v$, where $c$ is the speed of light and $v$ is the corresponding velocity. We can safely assume that velocities higher than $\sim100,000$ km s$^{-1}$ (which corresponds to $k<3$) or lower than $\sim1,000$ km s$^{-1}$ (which corresponds to $k>300$) are not consistent with the velocity of SN spectral features, since none of the SNe in our sample exhibit such extreme velocity widths. 

The second step of our method is to fit a power law to the magnitude $M$ ($=\sqrt{P}$, where $P$ is the power of spectrum in Fourier space) for $k>3$ (to avoid the divergence of $M$ for $k=0$). Then, since velocities smaller than $1,000$ km s$^{-1}$ are too low for SN features in our sample, we assume that the $M$ for $k>300$ is entirely due to noise, and calculate the average $M$ for $3<k<300$. We find that the $k_\mathrm{noise}$ corresponds to the intersection of the power law fit and the average $M$ above, which is a good estimate of the wave number that separates spectral signal from the noise. The filtered SN spectrum is then obtained by inverting the FT of the spectrum after suppressing $M$ with $k > k_\mathrm{noise}$. The filtered SN spectrum is shown in the top panel of Figure \ref{fig_variance_our}, and it is the version of the spectrum that we use in the velocity and pEW measurements.

The uncertainty array is obtained by deriving the noise spectrum as the residuals between original spectrum and filtered spectrum, and then calculating the standard deviations of this noise spectrum within a rolling window of 100 \AA, a characteristic width for SN features. An example is shown in the middle and bottom panels of Figure \ref{fig_variance_our}. Note that since using a rolling window leads to correlated uncertainties the data points in our uncertainty array are not independent.

We find that this method produces uncertainty arrays that are consistent with the reduction-produced ones, having tested this on the subset of the M14 spectra for which reduction-produced uncertainty arrays are available. This subset
consists of 28 spectra from nine SNe IIb, nine SNe Ib, and ten SNe Ic. The uncertainty arrays produced with our prescription for all SNe in the sample are within 70$\%-$110$\%$ of the ones produced from the optimal extraction (Figure \ref{fig_variance_our}, bottom panel). For consistency, all uncertainty arrays used in this study were produced using our method. To ensure the noise derived with our novel method and the Fourier-filtered spectra are reasonable, we visually inspected them as well.

As mentioned earlier, methods to estimate the uncertainty array from a processed spectrum had been developed in the past, but they either require fine tuning on a spectrum by spectrum level, or are automated through the use of parameter choices that are not appropriate for all spectra. \citet{Blondin2007} filtered spectra by applying the same bandpass filter to all spectra in Fourier space, but different spectra have different characteristic widths, especially for different subtypes, that should correspond to different bandpass filters. \citet{blondin06} used an inverse-variance-weighted Gaussian filter with the same ``smoothing factor'' for all spectra in wavelength space. The standard deviation of the Gaussian filter is the product of the wavelength and the ``smoothing factor''. Thus, the Gaussian filter is wider at longer wavelengths. This accounts for the increased noise at red wavelengths due to the  sky emission lines, but it does not take into consideration the fact that the low quantum efficiency at the edges of a CCD will increase the noise at both the blue and the red edges of a spectrum. With our method, each spectrum is filtered with a unique ``smoothing factor" that is set by the characteristics of the spectrum itself.

\section{\\C. The effect of SN continuum on the measurements} 
\label{continuum}

Ideally, we should measure the velocity and pEW of the SN features themselves, whereas what we actually measured was the velocity and pEW of SN features superimposed on the SN continuum. The continuum should not affect the pEW measurements since it was divided out from the spectra. Since the SN continuum may systematically change the velocity measurements, we checked the effect of SN continuum on velocity measurements by doing the following. We chose a subset of SN spectra and took velocity measurements in both the original spectra and the corresponding continuum-subtracted spectra. Here the continuum is a 13-point cubic spline fit to the corresponding spectrum, the same as that defined in SNID. In principle, the velocity measurements in continuum-subtracted spectra should not be biased by the SN continuum. We found that the differences between the two kinds of velocity measurements are randomly distributed within the error bars from the negative values to the positive values (Section \ref{error_bar}). Moreover, the benefits of subtracting the SN continuum, e.g., reducing the effect of reddening, are irrelevant for the velocity and pEW measurements. Thus, we measured these quantities in the original spectra, since they are less processed, and thus have less uncertainty compared to their continuum-subtracted counterparts. 

\section{\\D. Rolling Weighted Average of the Measurements} 
\label{mean_KS}

In order to explore bulk properties for different stripped SN subtypes, we calculated rolling weighted averages for each of the various SN properties. The rolling average helps to reduce the bias induced by starting at specific chosen phases, to smooth out short-term fluctuations, and to highlight long-term trends. A cumulative distribution function (CDF) plot of various SN properties or a two-sample Kolmogorov-Smirnov (K-S test) or Anderson-Darling tests (A-D test) on various SN properties would not be appropriate in this study, since the SN properties here evolve with time while the CDF and CDF-based tests cannot treat a two-dimensional dataset well \citep{babu06}. The CDF and CDF-based tests can be applied to SN properties within a narrow phase range where the time evolution is ignored, but this does not allow testing for similarities in their evolution. The latter is important in this study given that the SN spectra change significantly in the first one to three months after maximum light.   

The rolling weighted average of the measurements in individual spectra is calculated in the following way: 
\begin{enumerate}
\item We choose a bin size of five days for phases before $t_{\mathrm{Vmax}}=30$ days and of 10 days for later phases. The step size for these bins is one day, i.e., two adjacent bins are $[t_{\mathrm{begin}}, t_{\mathrm{end}}]$ and $[t_{\mathrm{begin}}+1, t_{\mathrm{end}}+1]$, where $t_{\mathrm{begin}}$ and $t_{\mathrm{end}}$ are phases in days with respect to the date of maximum light.
\item The weighted-average value of an individual SN in each bin is calculated using all measurements for that individual SN within the bin. The associated error bar is the standard deviation of this average value. Thus, each SN contributes no more than one combined value with an error bar to a bin, which mitigates the impact of a single SN with many data points. 
\item Using the above weighted average values of individual SNe, the weighted average value of a SN subtype at a given bin is calculated. The associated error bar is the standard deviation of these weighted average values of individual SNe, which reflects the variation of the measurements in the SNe. If there are fewer than three SNe of the same subtype contributing to the bin, the weighted-average value is meaningless and no value is reported for that bin. 
\end{enumerate}

%For the measurements of a specific SN feature, if two SN subtypes each have more than or equal to 4 SNe within the same phase range, a two-sample Kolmogorov-Smirnov test (K-S test) was used to check if these two underlying one-dimensional probability distributions of the measurements differ from each other. In this case, the K-S statistic quantifies the maximum distance between the cumulative distributions of two datasets. Thus, the K-S test is a nonparametric test. This is extremely useful here since we do not know the underlying distributions of our measurements. The null hypothesis of no difference between two datasets is rejected at K-S probability level $\alpha$ if: $D>c(\alpha)\times \sqrt{N_{\mathrm{eff}}}$, where $D$ is the K-S statistic, $c(\alpha)$ is a coefficient which have large values for low levels of $\alpha$ and the values can be found in published tables (e.g. tables 54 and 55 in Pearson \& Hartley 1972), and $N_\mathrm{eff}$ is the effective number of data which equals $(n1\times n2)/(n1+n2)$, where $n1$ and $n2$ represent the number of data points in each dataset, respectively. A small value of K-S probability means that the distribution of one dataset is significantly different from that of the other dataset.

\end{document}